%% file: main.tex
\documentclass[sigconf,balance=false]{acmart}
\usepackage{popets}
\setcopyright{popets}
\copyrightyear{YYYY}

\acmYear{YYYY}
\acmVolume{YYYY}
\acmNumber{X}
\acmDOI{XXXXXXX.XXXXXXX}
\acmISBN{}
\acmConference{Proceedings on Privacy Enhancing Technologies}
\input{includes/packages}

\AtBeginDocument{%
  \providecommand\BibTeX{{%
    \normalfont B\kern-0.5em{\scshape i\kern-0.25em b}\kern-0.8em\TeX}}}

\begin{document}

\title[]{Opted Out, Yet Tracked: Are Regulations Enough to \\Protect Your Privacy?}
\author{Zengrui Liu}
\affiliation{%
  \institution{Texas A\&M University}
  \city{College Station}
  \state{Texas}
  \country{USA}}
\email{lzr@tamu.edu}

\author{Umar Iqbal}
\affiliation{%
  \institution{Washington University in St. Louis}
  \city{St. Louis}
  \state{Missouri}
  \country{USA}}
\email{umar.iqbal@wustl.edu}

\author{Nitesh Saxena}
\affiliation{%
  \institution{Texas A\&M University}
  \city{College Station}
  \state{Texas}
  \country{USA}}
\email{nsaxena@tamu.edu}




\providecommand{\ione}{\emph{(i)}}
\providecommand{\itwo}{\emph{(ii)}}
\providecommand{\ithree}{\emph{(iii)}}
\providecommand{\ifour}{\emph{(iv)}}
\providecommand{\ifive}{\emph{(v)}}

\input{docs/0_abstract}
\maketitle
\input{docs/1_introduction}

\input{docs/2_background}
\input{docs/3_methodology}

\input{docs/4_evaluation}
\input{docs/5_discussion}
\input{docs/6_conclusion}
\begin{acks}

We would like to express our gratitude to the reviewers and the shepherd whose insightful comments and constructive feedback greatly contributed to the enhancement of this paper. 
The work of the second author is supported by the National Science Foundation under grant number CNS-2127309 (Computing Research Association for the CIFellows 2021 Project).

\end{acks}

\balance
\bibliographystyle{plain}

\bibliography{references}

\input{docs/7_appendix}
\input{docs/new_tables_0713}
\end{document}

%% file: includes/packages.tex
\usepackage{marvosym}
\usepackage{colortbl}

\newcolumntype{?}[1]{!{\vrule width #1}}
\usepackage{url}
\usepackage{booktabs}
\usepackage{xspace}

\usepackage{breakurl}
\usepackage{subfig}
\usepackage{multirow}
\usepackage{color,soul}
\usepackage{enumitem}
\usepackage{tikz}
\usepackage{listings}
\usepackage{adjustbox}
\usepackage[font=small]{caption}
\usepackage{balance}

\definecolor{lowbids}{HTML}{B3E5FC}
\definecolor{lowerbids}{HTML}{039BE5}
\definecolor{highbids}{HTML}{FFCDD2}
\definecolor{higherbids}{HTML}{EF5350}

%% file: docs/0_abstract.tex
\begin{abstract}
Data protection regulations, such as GDPR and CCPA, require websites and embedded third-parties, especially advertisers, to seek user consent before they can collect and process user data.
Only when the users opt in, should these entities collect, process, and share user data. 
Websites typically incorporate Consent Management Platforms (CMPs), such as OneTrust and CookieBot, to solicit and convey user consent to the embedded advertisers, with the expectation that the consent will be respected. 
However, neither the websites nor the regulators currently have any mechanism to audit advertisers' compliance with the user consent, i.e., to determine if advertisers indeed do not collect, process, and share user data when the user opts out. 

In this paper, we propose an auditing framework that leverages advertisers' bidding behavior to empirically assess the violations of data protection regulations. 
Using our framework, we conduct a measurement study to evaluate four of the most widely deployed CMPs, i.e., Didomi, Quantcast, OneTrust, and CookieBot, as well as advertiser-offered opt-out controls, i.e., National Advertising Initiative's opt-out, under GDPR and CCPA. 
Our results indicate that in many cases user data is unfortunately still being collected, processed, and shared even when users opt-out. 
We also find that some CMPs are better than the others at conveying user consent and that several ad platforms ignore user consent. 
Our results also indicate that advertiser-offered opt-out are equally ineffective at protecting user privacy.

\end{abstract}


%% file: docs/1_introduction.tex
\section{Introduction}
\label{sec:introduction}

\footnote{This paper has been accepted by The 24th Privacy Enhancing Technologies Symposium (PETs 2024). }
There has been a recent increase in the promulgation of data protection regulations, such as General Data Protection Regulation (GDPR) \cite{GDPR}, California Consumer Privacy Act (CCPA) \cite{CCPA}, and General Personal Data Protection Act (LGPD) \cite{LGPD}, across the globe. 
At a high level, data protection regulations aim to protect user privacy by mandating online services to take user consent before collection, processing, and sharing of user data. 
Because of their mass deployment, automatic enforcement, and legal binding, data protection regulations possess the potential to protect user privacy; provided that users do not consent to data collection and processing.
In fact, infringement fines have already amounted to billions. 
For example, in case of GDPR -- arguably the most mature data protection regulation -- the fines have accumulated to a total of \EUR 1.6 billion \cite{gdpr_fines_over_time}.
However, despite legal binding, prior research has found that online services often trick users into giving positive consent \cite{Matte20SPDoCookieBanners}, do not include controls to opt-out of data collection and processing \cite{dnsmpi}, or deploy user interfaces that are unintuitive to navigate in terms of providing consent \cite{alizadeh2020gdpr, habib19optout}.
In cases where users are indeed able to exercise their rights, user data is poorly handled. 
For example, online services often ignore or respond late to data access requests \cite{urban2019study} and even leak sensitive user data to unauthorized users because of weak authentication mechanisms \cite{bufalieri2020gdpr, di2019personal}.
In some cases, the existence of these issues could be attributed to the complexity of the regulations, unpreparedness, or oversights of online services. 
In other cases, it could be attributed to inconsideration of online services towards data protection regulations.


%
Regulators have mostly focused on auditing compliance of large well-known corporations, such as Amazon \cite{amazon_fine} and Google \cite{CNIL_google_fine}, perhaps because of the lack of systematic mechanisms to automatically detect infringements at scale \cite{google_amazon_fine}.
Prior research \cite{Matte20SPDoCookieBanners,dnsmpi, alizadeh2020gdpr, habib19optout} has focused on auditing the implementation deficiencies in consent management platforms/tools but it has largely ignored the instances where \textit{compliance is correctly conveyed but online services fail to comply}. 
Though, negligence in implementation raises doubts on the seriousness of online services in protecting users' privacy, it does not by itself imply non-compliance.

In this paper, we set out to fill this gap in the state-of-the-art research and deployed practice by regulatory bodies in assessing whether online services are actually compliant with the data regulations or not. To this end, we propose a framework to automatically audit regulatory compliance. 
We focus on cases where user consent is correctly conveyed but online services may not necessarily comply. 
We evaluate our auditing framework on the web, whereby websites typically record user consent using consent management platforms (CMPs), e.g., OneTrust \cite{OneTrust}, and convey it to advertisers under GDPR and CCPA.
Our key idea is to leak user interest data in controlled A/B experiments, opt-out/in of processing and selling through CMPs, and leverage advertisers \textit{ bidding behavior} as a side channel in the advertising ecosystem to infer the processing and selling of user information.
Since the bidding behavior of advertisers is shaped up by their pre-existing knowledge of the user, we expect to receive higher bids when advertisers process or sell leaked user interest data, i.e., are non-compliant with the law, despite the user choosing to opt-out.

%

We summarize our key contributions as follows:

\begin{enumerate}
	\item We propose a framework to automatically audit regulatory compliance of online services. We implement our framework by extending OpenWPM \cite{Englehardt16MillionSiteMeasurementCCS}. The framework allows us to imitate real user, automatically opt-out/opt-in of data processing and selling, and capture advertisers bidding by advertisers. 
    
    \item As a case study, we use our proposed framework to audit regulatory compliance of online services under GDPR and CCPA with four consent management platforms, i.e, Didomi\cite{Didomi}, Quantcast\cite{Quantcast}, OneTrust \cite{OneTrust}, and CookieBot \cite{Cookiebot}. Our results indicate that in many cases the advertisers do not necessarily comply with the user consent to opt-out of data processing and selling. Some CMPs perform better than the others, though. For example, when consent is conveyed through Didomi, advertisers bidding behavior significantly changes under CCPA.

    \item We also pursue a comparative analysis between state-enforced regulations and advertiser-offered controls, i.e. National Advertising Initiative's (NAI) central opt-out \cite{naa_central_optout}, in reduction of collection and selling of user data. Our results indicate that the advertiser-offered NAI's opt-out controls might be equally ineffective at protecting user privacy. 
    
\end{enumerate}

\textit{Paper Organization:} 
The rest of the paper is outlined as follows. Section \ref{sec:background} presents an overview of online privacy threats and protection mechanisms. 
Section \ref{sec:methodology} describes the design of our framework to audit regulatory compliance of online services. 
Section \ref{sec:evaluation} presents the results of our auditing.
Section \ref{sec:discussion} presents discussion and limitations of our proposed auditing framework. 
Section \ref{sec:conclusion} offers the main conclusions from our work.

%% file: docs/2_background.tex
\section{Background \& Related Work}
\label{sec:background}

\subsection{Online Tracking}
Online trackers capture users browsing histories and activities across the web to facilitate online behavioral advertising, among other use cases \cite{FTCDataBrokers}.
%
Online tracking is typically conducted through cookies that are set by third party resources loaded on websites, with the key idea being third parties having cross-site access to their cookies. 
Since most third parties are present on a limited number of websites, they often partner with each other to increase their coverage. 
Prior research has shown that trackers engage in data sharing partnerships and exchange cookies with \textit{as much as 118} other third parties \cite{Englehardt16MillionSiteMeasurementCCS}, which allows them to increase their coverage by as much as \textit{7 times} \cite{PapadopoulosCookieSynchronization19WWW}.  

Online tracking, and especially tracking driven advertising, poses a serious threat to users' privacy both at the individual and the societal level. 
At the individual level, trackers collect sensitive personal information, for example, about health and sexual orientation, which is then used to hyper-target the individuals, for instance, through personalized ads \cite{targeted_ads_wp,targeted_ads_vox}.
At the societal level, tracking driven advertising has been leveraged to conduct mass surveillance \cite{mass_surveillance_ads}, increase political polarization \cite{AliAdDeliveryPolarization21WSDM}, spread misinformation \cite{misinformation_ads}, and discriminate \cite{discrimination_ads}.
Overall, people are frustrated by the privacy harms facilitated by online tracking.

\subsection{Protection Mechanisms}
\subsubsection{Self-Regulations}
To tackle user privacy concerns and pressure from the regulatory bodies, such as the Federal Trade Commission (FTC), the online advertising industry has responded with self-regulations \cite{FTC_self_regulation_principles, FTC_congress}. 
However, prominent self-regulatory actions, such as the ones facilitated by browsers, for example, Platform for Privacy Preferences (P3P) \cite{p3p_w3c} and Do Not Track (DNT) \cite{dnt_w3c}, and the ones offered by the advertisers, for example, Digital Advertising Alliance's (DAA) AdChoices \cite{adchoices} and National Advertising Initiative's (NAI) central opt-out \cite{naa_central_optout}, are either not respected by majority of the vendors or they are too convoluted to be used or understood by the lay users.

\textit{Browser-Facilitated Automated Controls.}
Browsers provide several mechanisms that advertisers can leverage to enforce self-regulatory measures in an automated manner. 
P3P and DNT request headers are two such mechanisms.
P3P, now discontinued, was an automated mechanism for online services (e.g., website and third-party vendors) to communicate their privacy policies to web browsers.
It was implemented by major web browsers, such as Internet Explorer and Firefox \cite{p3p_internet_explorer,p3p_mozilla}, and supported by thousands of websites \cite{cranor2008p3pdeployment}.
However, P3P was often misrepresented by online services \cite{leon2010token,RaeyP3PTWEB09} likely because it was not enforced under any state regulation. 
Similarly, DNT was proposed to convey user's privacy preferences to the online services in an automated manner.
However, it also enjoyed limited adoption and it had practically no impact in limiting tracking. 
Libert et al. \cite{libert2018automated} reported that only 7\% of the websites mentioned DNT in their privacy policies, and in majority of those cases specified that the DNT signal is not respected.
Miguel et al. \cite{Miguel15CoNEXTOBA}, conducted an A/B study and identified that the DNT signal essentially had no impact on ad targeting, experienced by users.

\textit{Advertiser-Offered Manual Controls.}
In response to the concerns from FTC, advertising networks formed National Advertising Initiative (NAI), which provides a central interface for users to opt-out from targeted advertising, i.e., if users opt-out through NAI's central opt-out interface, they will (supposedly) no longer be tracked for online advertising \cite{nai_opt_out}.
McDonald and Cranor \cite{mcdonald2010americans} conducted a user study and found that only 11\% of respondents understood NAI's opt-out mechanism, which indicates that its adoption is perhaps low.
Similarly, taking a step forward in self-regulations, several of the advertising consortiums, created Digital Advertising Alliance (DAA) with an aim to provide easy to access user transparency and control, with ``AdChoices'' icon, to opt-out of targeted advertisements \cite{adchoices}.
Unfortunately, similar to NAI's opt-out, only 9.9\% of ads shown on top websites had AdChoices icon \cite{tracking_the_trackers_adchoices}.

\subsubsection{User-Managed Privacy Protections}
Currently, the most effective way for users to self-protect their privacy is to rely on off-the-shelf privacy-enhancing tools, such as AdBlock Plus \cite{adblockplusweb}, Privacy Badger \cite{privacybadger_web}, and Disconnect \cite{disconnect_me}.
However, privacy-enhancing tools are not available by default in browsers and need to be installed separately; which limits their adoptability to mostly tech-savvy users. 
Further, trackers engage in an arms-race with privacy-enhancing tools and try to come up with evasive tactics, for example, bounce tracking \cite{bounce_tracking_definition} and CNAME cloaking \cite{Dao20CNAME}, to evade privacy protections.

The other likely more feasible alternative is to rely on default privacy protections offered by the mainstream browsers, which are available to a larger population.
However, these protections are too weak to completely protect user privacy. 
For example, some main-stream browsers block third-party cookies, which makes them susceptible to new and sophisticated ways of tracking, such as browser fingerprinting \cite{Englehardt16MillionSiteMeasurementCCS,Iqbal21FPInspectorSP}.
Further, some browsers, such as Google Chrome, are too cautious even in blocking third-party cookies because of website breakage concerns \cite{chrome_3p_cookie_blocking_delayed}.

\subsubsection{State-Enforced Regulations: Focus of Our Work}
Both self-regulations and user-managed privacy protections do not have any legal binding and are thus blatantly bypassed by the advertisers and trackers. 
Only recently, legislators have promulgated regulations, such as General Data Protection Regulation (GDPR) \cite{GDPR} in EU and California Consumer Privacy Act (CCPA) \cite{CCPA} in California, that have potential to rein-in online advertising and tracking ecosystem. 
These regulations have clearly-stated frameworks that define protocols to collect, share, and use personal user information.
Most importantly, their infringements can be prosecuted; which can lead to heavy fines \cite{gdpr_faq, CCPA}.
For example, both Amazon and Google were recently fined for \EUR 746 \cite{amazon_fine, LuxembourgAmazonFine} and \EUR 50 millions \cite{CNIL_google_fine} under GDPR, respectively.
Essentially, these regulations possess the ability to keep advertising and tracking ecosystem in check. 


Both GDPR and CCPA guarantee a right for individuals to opt-out of processing and selling of their data. 
Under GDPR, online services need to take user consent (Articles 4 (11)) before they can process user data (Article 6 (1) (a)).
GDPR has a broad definition of data processing, that includes \textit{collection, recording, organization, structuring, storage, adaptation or alteration, retrieval, consultation, use, disclosure by transmission, dissemination or otherwise making available, alignment or combination, restriction, erasure or destruction} of user data (Article 4 (2)). 
Under CCPA, online services need to provide user control to opt-out of sale of personal user data (Section 1798 (a) (1)). 
CCPA has a broad definition of personal data selling, that includes \textit{selling, renting, releasing, disclosing, disseminating, making available, and transferring} data to another business or a third party for monetary or other valuable consideration (Section 178.140 (t) (1)).
Both GDPR's and CCPA's data processing and selling definition covers routine data exchanges, such as processing user data to serve personalized ads (e.g., through Real-Time Bidding (RTB) protocol \cite{RTB_protocol}), and sharing data with advertising partners under data sharing partnerships (e.g., with cookie syncing \cite{google_rtb_docs}).
In fact, The Office of the California's Attorney General explicitly lists several such examples as violations of CCPA \cite{ccpa_enforcemnt_exampels_jdsupra, ccpa_enforcemnt_exampels_oag}. 
It is noteworthy that GDPR requires to obtain consent beforehand (Article 6 (1)(a)): \textit{Processing shall be lawful only if and to the extent that at least one of the following applies: (a) the data subject has given consent to the processing of his or her personal data for one or more specific purposes.}
Whereas, CCPA requires to provide tools to opt-out later (Section 1798.120 (a)): \textit{A consumer shall have the right, at any time, to direct a business that sells or shares personal information about the consumer to third parties not to sell or share the consumer's personal information. This right may be referred to as the right to opt-out of sale or sharing.}
CCPA does not require a beforehand consent because it only restricts the selling/sharing of personal data and not its collection.

Both GDPR and CCPA require websites to provide \textit{privacy notices} with information and controls to opt-in/out of personal information collection and/or processing. 
To obtain user consent, websites typically embed off-the-shelf consent management platforms (CMPs), e.g., OneTrust \cite{OneTrust} and Cookiebot \cite{Cookiebot}. 
CMPs scan websites and identify all cookies set by the HTTP headers and scripts, from both the first and third party resources. 
In case of GDPR, CMPs should ensure that only strictly necessary cookies are shared and consent is obtained before non-essential cookies, such as for advertising and analytics, are shared. 
In case of CCPA, CMPs should ensure that they provide controls to users to opt-out to sell their personal information.
\begin{figure}[!h]
   \centering
       \subfloat[Consent management dialog for GDPR.]{
       \includegraphics[width=0.45\columnwidth]{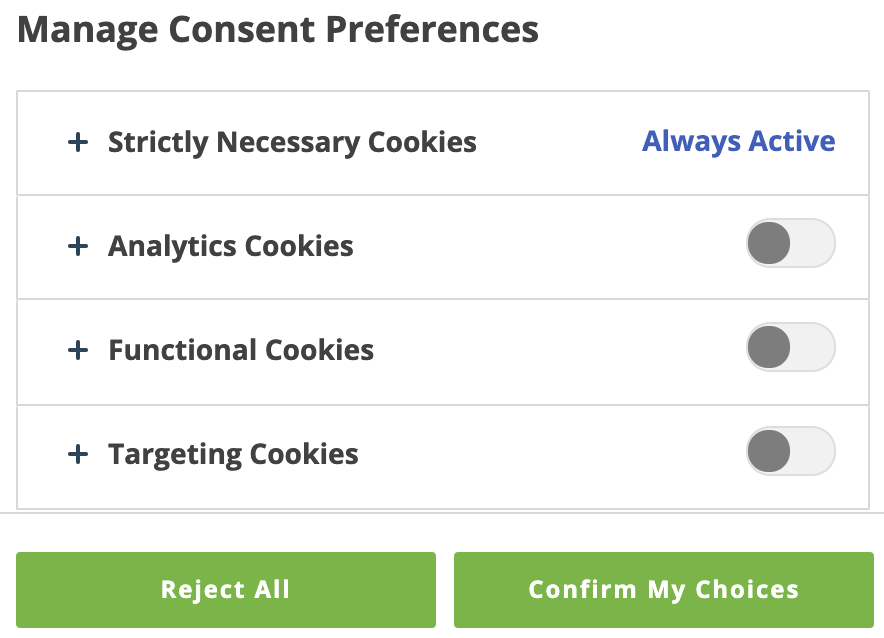}
       \label{fig:GDPR-OneTrust}
       }
       \hfill
        \subfloat[Consent management dialog for CCPA.]{
       \includegraphics[width=0.45\columnwidth]{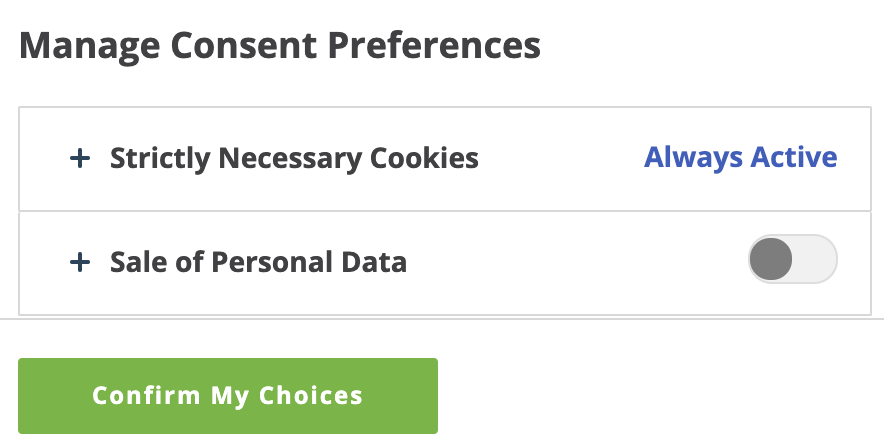}
       \label{fig:CCPA-OneTrust}
       }
       \caption{OneTrust's consent management dialog for GDPR and CCPA.}
   \label{fig:OneTrust-CMPs}
\end{figure}

Figure \ref{fig:GDPR-OneTrust} shows an example consent dialog displayed under GDPR, and Figure \ref{fig:CCPA-OneTrust} shows an example consent dialog displayed under CCPA. 

\subsection{Header bidding}
%

Header bidding~\cite{HB_protocol} is a strategy in which websites, known as publishers, allocate various advertising slots for advertisers. The advertiser with the most competitive bid wins the opportunity to showcase their ads within the relevant ad slots. In the context of client-side header bidding, users can conveniently view all the bids directly from their web browser. A prominent illustration of this approach is demonstrated by \texttt{prebid.js}~\cite{prebid}. We've identified the Alexa top 100K websites, encompassing those with advertisements (e.g., cnn.com) and those without (e.g., google.com), to verify their utilization of \texttt{prebid.js}. Our findings revealed that 5421 websites are employing \texttt{prebid.js} with the standard API label \texttt{pbjs} (Personalized custom API labels are also utilized within \texttt{prebid.js}, and these are excluded from our tally of detections.). This signifies that over 5\% of websites continue to employ \texttt{prebid.js}.

%

In contrast, server-side header bidding involves conducting the bidding auction on the ad server instead of the user's browser. Consequently, the bids from participating advertisers remain concealed from the web user's perspective. Noteworthy instances of server-side header bidding include Google Open Bidding, Amazon TAM, and Prebid Server. Comparing to server-side header bidding, the client side header bidding has the following advantages:

\begin{enumerate}

\item Through header bidding, publishers retain the ability to select buyers via header bidding wrappers. Additionally, publishers have the authority to set the minimum price for each ad unit. Consequently, the entire auction process becomes visible and clear for both publishers and advertisers. However, such transparency is not as pronounced with server-side header bidding. While publishers still determine the floor price, they lack visibility into the buyers participating in the auction, resulting in a more concealed auction process.

\item The primary rationale for favoring client-side header bidding over server-side header bidding lies in auction management. Header bidding wrappers empower publishers to oversee and manage the auction. Publishers can add buyers, establish timeout configurations, and ensure simultaneous bid requests to all buyers using these wrappers. In contrast, server-side header bidding involves the server reaching out to buyers and initiating bid requests, hence the management is primarily executed by the server.

\item Client-side header bidding enables advertisers to directly access ad units from publishers' web pages using wrappers, thereby gaining access to user cookie data. This data can be further employed for targeted advertising. Conversely, server-side header bidding encounters limitations in cookie matching.

\end{enumerate}

Since the server-side header bidding does not expose auction at the client side, we do not consider it in our measurements.

\subsection{Statistical Analysis}
To evaluate if there are significant differences in advertisers bidding behavior when users opt-out under GDPR and CCPA, we conduct Mann-Whitney U test of statistical significance \cite{cohen1977statistical}.
Mann-Whitney U test is a nonparametric test to compare the differences between two distributions.
Since we perform multiple comparisons, i.e., compare bid values for all 16 personas, we also conduct Bonferroni correction on the statistical test. 
Our null hypothesis is that the bid distributions for opt-in and opt-out are similar to each other. 
We reject the null hypothesis, when the p-value (after correction, i.e., original value multiplied by 16) is less than 0.05 (reflecting a 95\% confidence interval), i.e., the distributions are statistically different. 
%
We also measure the magnitude of the difference between bid values by calculating the effect size \cite{cohen1977statistical}. 
%
Effect size less than 0.3, between 0.3 and 0.5, and greater than 0.5 is considered small, medium, and large, respectively. Effect sizes are reported only in cases where statistically significant differences are observed.
In instances where no bids are accumulated under either opt-out or opt-in conditions, the calculation of p-value and effect size becomes unfeasible, as these measures necessitate two datasets for meaningful comparison.

\subsection{Related Work}
Prior research has identified that online services design unintuitive and hard to navigate data access interfaces \cite{alizadeh2020gdpr, habib19optout}, trick users into giving positive consent \cite{Matte20SPDoCookieBanners}, and do not include controls to opt-out of data selling \cite{dnsmpi}.
Alizadeh et al. \cite{alizadeh2020gdpr} conducted a user study to understand data rights under GDPR and identified that the participants find data access interfaces unintuitive and hard to navigate. 
Specifically, users prefer structured and easy-to-navigate data usage reports in contrast to data dumps, that are hard to explore. 
Habib et al. \cite{alizadeh2020gdpr} conducted a measurement study of 150 websites and identified that the privacy controls were hard to locate on the majority of websites.
Furthermore, in several instances, links to privacy control did not lead to stated choices.
Matte et al. \cite{Matte20SPDoCookieBanners} investigated CMPs and identified that the consent is often incorrectly conveyed.
Specifically, websites often register consent before the user has made any choice, register positive consent regardless of user's choice, or nudge users to give pre-selected positive consent. 
Urban et al. \cite{urban2020beyond} discovered that 93\% of the examined websites included third-party elements originating from regions that potentially do not conform to the prevailing legal framework.
More recently, Nortwick and Wilson \cite{dnsmpi}, conducted a measurement study of top 500K English websites and identified that only 2\% of the websites provided controls to users to opt-out of data selling, i.e., ``Do Not Sell My Personal Information'' (DNSMPI), under CCPA. 
The study by Toth et al. \cite{toth-pets22} found that CMPs themselves may exhibit dark patterns and could track users' data to some extent by investigating 10 consent services from 5 CMPs deployed on different blank websites.
They also identified that default configurations of consent pop-ups often violate regulations and that their configuration options may lead to non-compliance.
Recently, Nguyen et al. \cite{nguyen-ccs22} studied the implementation of consent notices specifically on Android apps and identified that about 20\% of these apps violate at least one GDPR consent. 
In the study conducted by Demir et al. \cite{demir2022reproducibility}, a thorough examination of 114 prior research papers on web measurement was undertaken. The findings revealed a significant trend, with the majority (72.6\%) of these papers lacking a comprehensive inventory of the pages they had analyzed. This has substantial implications for experiment reproducibility, as it highlights a prevalent issue where the majority of experiments cannot be replicated in terms of the specific sites and pages that were studied.
%


Though negligence in obtaining consent and not providing easy-to-navigate opt-out controls raises doubts on online services' seriousness in protecting users' data and respecting their consent, it does not automatically imply non-compliance. 
Prior work, to the best of our knowledge, has not directly measured non-compliance through consent notices on traditional web browsers, especially for the cases where consent is properly conveyed to the online services. 
To bridge that gap, in our work, we set out to audit the usage and selling of personal user data, where the user has directed online services to cease the processing and selling of their data, and their consent is properly recorded by the CMPs.

\subsection{Inferring Non-Compliance}
\label{subsection:non-compliance}
Online services, including publishers, advertisers, and trackers, do not offer much transparency in the usage and sharing of collected data, which makes it challenging to directly assess non-compliance.
Though prior work has not directly measured advertisers and trackers non-compliance, they have relied on side channel information to infer the usage and sharing of user data \cite{Olejnik14SellingPrivacyNDSS,lecuyer2015sunlight,Bashir16TrackingFlowsUsenix,Papadopoulos17IMCYouAreTheProduct,Cook20HeaderBiddingPETS}.

A series of studies \cite{Cook20HeaderBiddingPETS,Olejnik14SellingPrivacyNDSS,Papadopoulos17IMCYouAreTheProduct} leaked user interest data, in controlled experiments, and leveraged advertisers bidding behavior as a side channel to infer the usage and sharing of user data. 
Especially in the study \cite{Olejnik14SellingPrivacyNDSS} the author mentioned that the prices for profiles ``Only category'' are about 40\% higher than those for ``New user''.
Their main insight is that the advertisers bidding behavior is shaped by their pre-existing knowledge of the user, which typically results in higher bid values, as compared to bid values for users for which advertisers do not have knowledge.
Specifically, higher bids made by the advertiser to which the data was leaked indicates the usage of the leaked data for ad targeting.
Whereas, higher bids from the advertiser to which data was not leaked indicates the sharing of data from advertisers to which the data was leaked.

Data sharing is an essential component of online advertising ecosystem and it is baked into ad delivery protocols, such as RTB \cite{RTB_protocol} and HB \cite{HB_protocol} protocols. 
Prior work \cite{Englehardt16MillionSiteMeasurementCCS,PapadopoulosCookieSynchronization19WWW} has identified that advertisers and trackers use ad delivery protocols, to directly share user data with each other at the client side, e.g., by cookie syncing \cite{cookie_syncing}. 
%
Thus, client side data sharing can be directly inferred by analyzing network requests (e.g., redirects), between advertising and tracking services.

We argue that analyzing advertisers bidding behavior and network traffic should suffice in establishing whether advertisers comply with the user consent, when they opt-out of processing and selling of their data under GDPR and CCPA.
%
%
Thus, in this study, we leverage advertisers bidding behavior and network traffic to audit regulatory compliance of advertisers under GDPR and CCPA.


%% file: docs/3_methodology.tex
\section{Methodology}
\label{sec:methodology}

\begin{figure}
    \centering

        \includegraphics[width=0.47\textwidth]{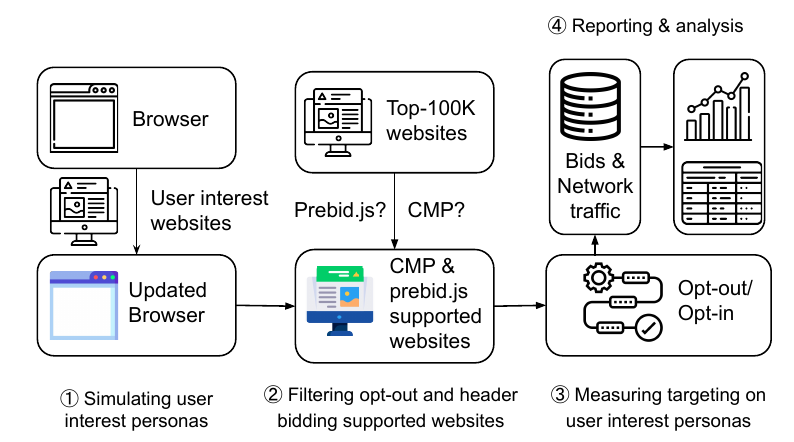}
     
        \caption{High level overview of our framework to audit regulatory compliance. (1) We use OpenWPM \cite{Englehardt16MillionSiteMeasurementCCS} to automatically visit top-50 websites from 16 different interest categories to simulate 16 user interest personas. (2) We filter top websites that support opt-outs through Didomi, Quantcast, OneTrust, and CookieBot under GDPR and CCPA and also support header bidding through \texttt{prebid.js} \cite{prebid}. (3) We then visit the filtered websites with user interest personas, opt-out of data processing and selling, and collect bids and network requests from advertisers. (4) We then analyze the collected bids and network requests to infer data processing and selling from advertisers.} 
        \vspace{-1.2em}
        \label{fig:CMP}
\end{figure}

In this section, we describe our framework to audit advertising and tracking services under GDPR and CCPA. 
At a high level, we simulate synthetic user personas (profiles) with specific interests, intentionally leak those interests to advertisers and trackers, opt-out of processing and selling of user data, and analyze advertisers bidding behavior and network traffic to infer the usage and selling of leaked data. 
Figure \ref{fig:CMP} gives an overview of our approach. 

\subsection{Pre-Conditions: Crawling Under GDPR \& CCPA} 
\subsubsection{Web Crawling}
We rely on OpenWPM \cite{Englehardt16MillionSiteMeasurementCCS} to set up our auditing framework. 
OpenWPM is a widely used Firefox-based, open-source web measurement tool that has been used in numerous research studies \cite{openwpm_usage_stats}. 
OpenWPM by default provides functionality to automatically visit websites and capture network requests, among other things.
To adapt to our needs, we further extend OpenWPM to automatically opt-out of processing and selling of user data, and to capture advertisers bids on ad slots. In the opt-out process, we perform either the JavaScript opt-out event activation or engage the opt-out button provided by the CMP. We describe comprehensive information in Section \ref{susub:opt-out}. The JavaScript opt-out function is accomplished using a browser extension, while the opt-out button is clicked through the utilization of Selenium. It's worth noting that OpenWPM employs Selenium for browser handling, allowing functions compatible with Selenium to also function within OpenWPM. Throughout the phase of recording advertisers' bidding behaviors, the \texttt{getBidResponses} method, supported by \texttt{prebid.js}, is employed.

\subsubsection{Simulating Measurements under GDPR \& CCPA}
We conduct our measurements from EU and California to audit third-party compliance under both GDPR and CCPA. 
We choose Frankfurt (Germany) and Northern California as representative jurisdictions under GDPR and CCPA, respectively.
We rely on Amazon EC2 to simulate web crawls through the respective locations.
We setup a new EC2 node, i.e., with a unique IP address, for each OpenWPM instance.

\subsection{	Simulating User Interest Personas} 

Recognizing the distinct bidding patterns exhibited by advertisers based on varying user interests, we undertake the simulation of 16 unique user interest personas. These personas are shaped by the categories of the top 16 Alexa-listed websites\footnote{Adult, Art, Business, Computers, Games, Health, Home, Kids, News, Recreation, Reference, Regional, Science, Shopping, Society, and Sports}\cite{alexa}. Notably, our selection approach compensates for Alexa's shift in offerings after September 2020\cite{16_lists_github}.The method of persona simulation involves initializing a fresh browser profile within an OpenWPM instance. This occurs on a pristine EC2 node, each furnished with an exclusive IP address. The process entails iterative visits to the top 50 websites within each category, with the browser profile being updated following each visit. Importantly, it is pertinent to highlight that our interactions were limited to the main pages of these websites. Our rationale for crafting these simulated personas rests upon their potential to persuade advertisers and trackers regarding the interests of each persona. The goal is to incentivize advertisers to submit higher bids when tailoring personalized ads for each persona. Inclusive of the aforementioned 16 personas, our study also introduces a control persona, represented by an empty browser profile. This control persona functions as a benchmark, facilitating the measurement of disparities in bidding behavior across personas. Noteworthy measures undertaken in our methodology encompass the activation of OpenWPM's bot mitigation features and the introduction of random delays ranging between 10 to 30 seconds subsequent to loading each website. These steps serve to enhance the authenticity of user behavior emulation.

We deployed a total of 34 instances: 17 in Germany (16 personas + 1 Control) using 17 different IPs, and 17 in California (16 personas + 1 Control) using 17 different IPs.

We automated the loading of selected web pages and simulated various user interactions, including mouse movements and scrolls, with random time gaps between actions. The mouse cursor is moved by varying amounts in different directions, scrolling occurs at unpredictable intervals down the page, and waiting periods are introduced randomly to create irregular patterns in page visits. This approach is consistent with previous research in web measurement \cite{Iqbal22USENIXKhaleesi,Cook20HeaderBiddingPETS}.

\subsection{Filtering Opt-Out and Header Bidding Supported Websites} 
\label{subsection:filtering-opt-out}

\input{docs/number-of-testing-sites.tex}

We shortlist websites that support opt-out through CMPs and also implement header bidding through \texttt{prebid.js}.
We identify such websites, by crawling Alexa top-100K websites, using OpenWPM, and probing for the presence of CMPs and \texttt{prebid.js} (as described in Section \ref{susub:opt-out} and \ref{susub:bid-collection}).
%
%

Table \ref{table:number-of-testing-websites} lists the presence of CMPs and \texttt{prebid.js} on Alexa top-100K websites. 
We note that a large number of websites deploy CMPs but not all of them deploy \texttt{prebid.js}.
However, scanning top-100K websites allows us to filter a meaningful number (i.e., 352) of websites that deploy CMPs and \texttt{prebid.js} under both GDPR and CCPA.

\subsubsection{Opting-out of Processing \& Selling of User Data} 
\label{susub:opt-out}
\begin{figure}[!h]
    \centering
        \subfloat[Main page of consent dialog.]{
        \includegraphics[width=0.9\columnwidth]{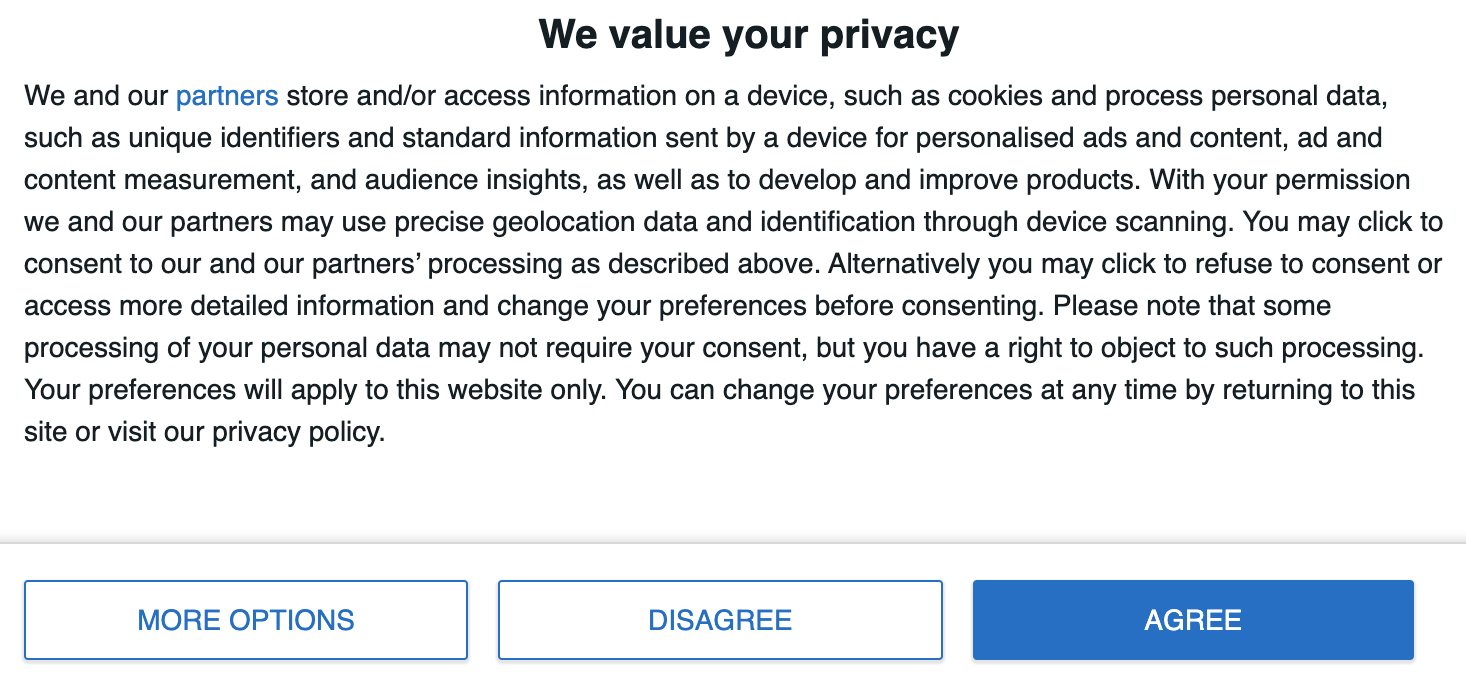}
        \label{fig:main-Quantcast}
        }
        \hfill
         \subfloat[Expanded consent dialog with all options.]{
        \includegraphics[width=0.9\columnwidth]{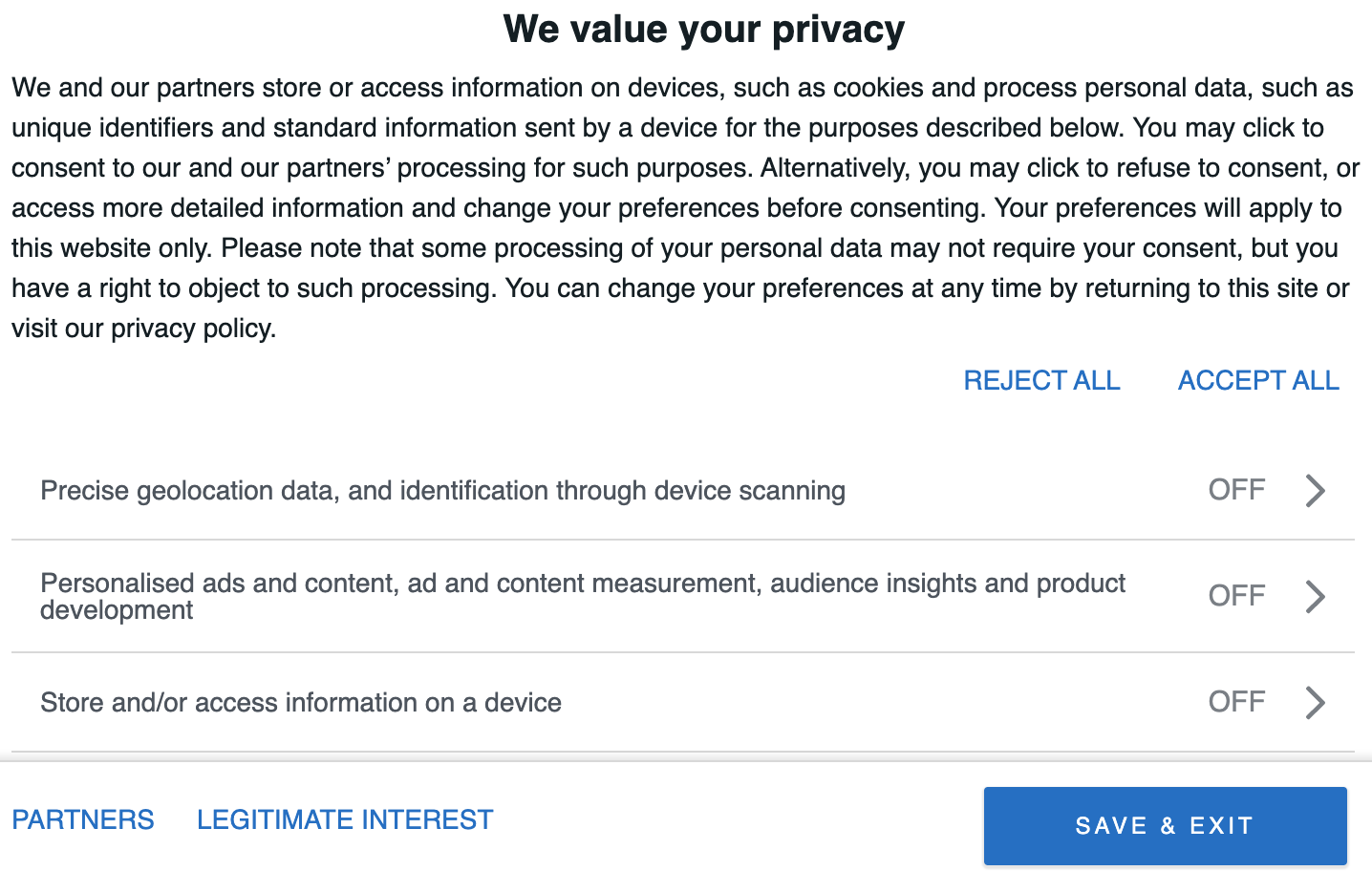}
        \label{fig:selection-Quantcast}
        }
        \caption{Quantcast's consent management dialog for GDPR and CCPA.}
    \label{fig:Quantcast-CMPs}
\end{figure}

We extend OpenWPM to programmatically opt-out of processing and selling of user data from Didomi, \cite{Didomi}, Quantcast \cite{Quantcast}, OneTrust \cite{OneTrust}, and CookieBot \cite{Cookiebot}, four of the widely used consent management platforms (CMPs) \cite{WebAlmanacStateofTheWebCMPs, hils2020measuring}. 
We conducted assessments on alternative CMPs as well. Our evaluations encompassed TrustArc, Crownpeak, LiveRamp, CookieYes, Osano, AdRoll, iubenda, and Usercentric. Employing Frankfurt, GE and Los Angeles, US IPs, we filtered websites to verify their compatibility with Opt-out, Opt-in, and prebid.js. Notably, CookieBot, Didomi, OneTrust, and Quantcast emerged as the four CMPs with the greatest prevalence among the top 100K websites listed on Alexa.
At a high level, we either trigger the JavaScript opt-out event or click the opt-out button of the CMP. 
Specifically, for Didomi, we check for the presence of consent dialog with \texttt{Didomi.notice.isVisible}, trigger \texttt{Didomi.setUserDisag-\\reeToAll} method to opt-out, and then hide the consent dialog by setting the display attributes of consent dialog markup to \texttt{none} \cite{didomi_api_docs}.
For OneTrust, we check for the presence of consent dialog with \texttt{window.OneTrust}, trigger \texttt{window.OneTrust.RejectAll} method to opt-out and hide the consent dialog \cite{OneTrust_api_docs}.
For CookieBot, we check for the presence of consent dialog with \texttt{window.Cookiebot}, traverse the DOM to find the opt-out button with id \texttt{CybotCookiebotD-\\ialogBodyButtonDecline} and click it. 
For Quantcast, we check for the presence of consent dialog by traversing the DOM to find the dialog with \texttt{qc-cmp2-summary-buttons} class name and click the button with \textit{Reject} or similar text.\footnote{Similar text candidates are manually compiled from the list of button text for all Quantcast consent dialogs.}
If the reject button is not present on the first page of consent dialog, we expand the dialog by clicking the button with \textit{more options} text and then click the \textit{Reject All} button. 
Figure \ref{fig:Quantcast-CMPs} shows the Quantcast dialog. 
The full lists of websites used to test Opt-out/Opt-in are listed on Github \cite{16_lists_github}

\subsubsection{Ensuring the Functionality of Selected CMPs}

\input{tables/4CMPs_requests_cookies/summary.tex}
We must ensure a comprehensive understanding of how each selected CMP interacts with advertisers both prior to and following user consent. CookieBot \cite{Cookiebot_blocking}, Didomi \cite{Didomi_blocking}, and OneTrust \cite{Onetrust_blocking} have official documentation outlining their procedures for managing 3rd party scripts that load prior to user consent. However, there is no available information pertaining to Quantcast. We conducted manual testing on all four CMPs using German IPs, calculating the quantity of HTTP requests generated by 3rd parties and the number of 3rd party cookies. Employing a Macbook device equipped with a Chrome browser, all data was monitored through Chrome DevTools. The exact counts of 3rd party HTTP requests and cookies are presented in Table \ref{table:requests-cookies}, depicting the figures before consent is given, after consent is granted, or after consent is denied. From the standpoint of request and cookie numbers, it appears that Cookiebot, Didomi, and Quantcast prevent 3rd parties from setting trackers in cookies before user consent. Additionally, the number of requests is lower in the Opt-out scenario compared to Opt-in. Conversely, such a trend is not evident in the case of OneTrust. Notably, cookies from domains ``\textbf{googleads.g.doubleclick.net}" and ``\textbf{agead2.googlesyndication.com}" consistently appear both before and after user consent or rejection.

Each of the four CMPs also establishes consent cookies following user decisions to either grant or deny consent. In the case of Cookiebot, a new cookie named ``CookieConsent" is created once the user makes a choice regarding consent. Didomi employs two cookies, namely ``didomi\_token" and ``euconsent-v2," to store consent-related information. OneTrust uses the ``OptanonConsent" cookie to retain the user's consent preference. In the context of Quantcast, three cookies are generated upon user consent: ``euconsent-v2," ``\_pbjs\_userid\_consent\_data," and ``addtl\_consent." If users opt to decline consent, only the ``euconsent-v2" cookie is present. It's important to note that all these cookies are designated as 1st party cookies and are not affiliated with the CMPs themselves.

\subsection{Measuring Targeting on User Interest Personas} 




\subsubsection{Measuring Targeting on Personas}
Next, we measure targeting received by our personas to infer compliance (or lack thereof) under GDPR and CCPA.
As mentioned earlier, we register negative user consent, i.e., opt-out of processing and selling of user data, through Didomi, Quantcast, OneTrust, and CookieBot and capture bids through \texttt{prebid.js}.
After filtering the websites, we iteratively visit each website nine times from each of our 16 (+ control) personas under both GDPR and CCPA.
We visit the websites first time to opt-out of processing or selling of data and the next eight times to collect bids.

Occasionally, the internet connection might experience instability, causing the browser to miss out on receiving bids. The quantity of bids received by the browser during distinct visits can also vary. It is unfeasible to ensure that the framework or a typical browser captures bids from all potential advertisers within a single visit. Therefore, we configured our framework to access the website on 8 occasions. The choice of the number 8 isn't dictated by any particular rationale; rather, we opted for a relatively sizable number due to necessity.
Specifically, additional factors, e.g., day/week and website popularity, may influence the bids \cite{Cook20HeaderBiddingPETS,Olejnik14SellingPrivacyNDSS,Papadopoulos17IMCYouAreTheProduct}. 
In addition, we also use identical hardware/software, collect bids at the same time, from the same location, and on the same websites, across all personas.
Overall, we expect that crawling websites several times and keeping conditions consistent will minimize the variability in bids.

\subsubsection{Capturing Bidding by Advertisers}
\label{susub:bid-collection}
We treat advertisers bidding behavior as an indication of advertisers and trackers non-compliance with the user consent (as discussed in $\S$ \ref{subsection:non-compliance}).
To this end, we audit advertisers and trackers on websites that support header bidding, more specifically \texttt{prebid.js} --- a widely used implementation of header bidding protocol \cite{HB_usage}, primarily because header bidding occurs at the client side and allows us to intercept the bidding process \cite{HB_protocol}. 
Additional header bidding solutions such as Amazon Transparent Ad Marketplace and Google Ad Manager (GAM) featuring EBDA (Exchange Bidding Dynamic Allocation) operate through server-side mechanisms, preventing us from accessing bid data on the client side. Previously, Google offered a client-side header bidding service. Nevertheless, the current unavailability of the API googletag.pubads().getBidsReceived() has impacted this functionality.
To capture the bids, we first identify header bidding supported websites.
We identify such websites by injecting a script on the webpage that probes the \texttt{prebid.js} version; if we receive a response, we consider the website as \texttt{prebid.js} supported website.
Note that we do not consider personalized custom API labels (i.e., other than \texttt{pbjs}) in our measurements.
After identification, we capture the bids by calling the \texttt{getBidResponses} method which returns the bids CPMs\footnote{The bid values are expressed in CPM (cost per mille); which is the amount that an advertiser pays a website per one thousand visitors who see its advertisements.}. 
In case, we do not get any bids, we request the bids ourselves by calling the \texttt{requestBids} method.

\subsubsection{Capturing Cookie Syncing by Advertisers}
Client side data sharing is a standard practice in the online advertising ecosystem. 
Advertisers most commonly share data through cookie syncing \cite{google_rtb_docs}.
Specifically, advertisers read their cookies (or other identifiers) from the browser and embed them in redirect requests which force the browser to send the embedded identifiers to the redirected advertiser. 
Since cookie syncing involves redirects from the browser, network traffic can be analyzed to detect cookie syncing events. 
To evaluate advertisers compliance, we measure whether opt-outs, under GDPR and CCPA, reduce cookie syncing.
We use the heuristic from prior work \cite{Iqbal22USENIXKhaleesi} to detect cookie syncing in network traffic when users opt-out and opt-in using CMPs.

\subsubsection{Baseline Comparison with Opt-in}
To understand the impact of regulations, we also establish a baseline by opting in to the processing and selling of data. 
Our rationale for opting in is to get the upper bound on processing and selling of data, as inferred by advertisers bidding behavior.
To opt-in, we enable all cookie controls mentioned in Section \ref{susub:opt-out}.
For Didomi we call the \texttt{Didomi.setUserAgreeToAll}, for OneTrust we call \texttt{window. OneTrust.AllowAll}, for CookieBot we click the opt-in button with id \texttt{CybotCookiebotDialogBodyLevelButtonLevelOptinAl lowAll}, and for Quantcast we click the button with \textit{Accept} or similar text.

\subsubsection{Comparison With Advertisers Self Regulations}
We also compare state-enforced regulations, i.e., GDPR and CCPA, with advertisers offered controls, i.e., with NAI's central opt-out \cite{nai_opt_out} in curbing the processing and selling of data.
We opt-out of NAI's offered controls by automatically navigating to their opt-out webpage \cite{nai_opt_out} and by clicking \textit{OPT OUT OF ALL} button.
To evaluate advertisers offered controls, we select a different set of websites that support \texttt{prebid.js} but do not support CMPs.
Specifically, we filter Alexa top-50 websites and identify 28 websites that support \texttt{prebid.js} but do not support any CMPs under both GDRP and CCPA.
It is important to select websites that do not support CMPs because otherwise we cannot distinguish between the impact of advertisers offered controls and state-enforced regulations. 

%% file: docs/number-of-testing-sites.tex
%


\begin{table}
\centering
\scriptsize
\caption{CMP and prebid.js deployment on Alexa top-100K websites under GDPR and CCPA. +PB represents the count of websites for each CMP that also deploy \texttt{prebid.js}. Common websites column represents the count of websites that deploy both CMPs and \texttt{prebid.js} and are common across GDPR and CCPA.}
\adjustbox{max width=\columnwidth}{
\begin{tabular}{l|rr|rr|c}
\hline
\multirow{3}{*}{\textbf{CMP}} & \multicolumn{2}{c|}{\textbf{GDPR}} & \multicolumn{2}{c|}{\textbf{CCPA}} & \multirow{2}{*}{\textbf{Common}}   \\

\multicolumn{1}{c|}{\textbf{}} & \multicolumn{1}{c}{\textbf{CMP}} & \multicolumn{1}{c|}{\textbf{+PB}} & \multicolumn{1}{c}{\textbf{CMP}} & \multicolumn{1}{c|}{\textbf{+PB}} & \multirow{2}{*}{\textbf{websites}} \\

\multicolumn{1}{c|}{\textbf{}} & \multicolumn{1}{c}{\textbf{count}} & \multicolumn{1}{c|}{\textbf{count}} & \multicolumn{1}{c}{\textbf{count}} & \multicolumn{1}{c|}{\textbf{count}} &   \\

\hline
\textbf{Didomi}               & 528      & 238      & 709      & 209    & 192   \\
\textbf{Quantcast}            & 2,563    & 234      & 2,875    & 94     & 77    \\
\textbf{Onetrust}             & 3,046    & 96       & 3,222    & 183    & 71    \\
\textbf{Cookiebot}            & 756      & 22       & 599      & 15     & 12    \\
\midrule
\textbf{Total}                & 6,893        & 590        & 7,405        & 501      & 352 \\ 
\bottomrule
\end{tabular}
}
\label{table:number-of-testing-websites}
\end{table}

%% file: tables/4CMPs_requests_cookies/summary.tex
\begin{table}
\centering

\caption{Number of 3rd party requests and cookies set during manual test on 4 CMPs: Cookiebot, Didomi, Onetrust and Cookiebot. Column ``3rd R'' represents the number of 3rd party HTTP requests, and column ``3rd C'' represents the number of 3rd party cookies.}

\resizebox{\linewidth}{!}{

    \begin{tabular}{c?{0.3mm}cc?{0.3mm}cc?{0.3mm}cc?{0.3mm}cc} 
\toprule

\multicolumn{1}{c?{0.3mm}}{\textbf{CMP}  }                 & \multicolumn{2}{c?{0.3mm}}{\textbf{Cookiebot}}      & \multicolumn{2}{c?{0.3mm}}{\textbf{Didomi}}         & \multicolumn{2}{c?{0.3mm}}{\textbf{Onetrust}}       & \multicolumn{2}{c}{\textbf{Quantcast}}       \\ 

\multicolumn{1}{c?{0.3mm}}{\textbf{Tested Website} }       & \multicolumn{2}{c?{0.3mm}}{\textbf{sammobile.com}}  & \multicolumn{2}{c?{0.3mm}}{\textbf{marca.com}}      & \multicolumn{2}{c?{0.3mm}}{\textbf{breitbart.com}}  & \multicolumn{2}{c}{\textbf{foreca.fi}}       \\ 

  \multicolumn{1}{c?{0.3mm}}{}                             & \textbf{3rd R} & \textbf{3rd C} & \textbf{3rd R} & \textbf{3rd C} & \textbf{3rd R} & \textbf{3rd C} & \textbf{3rd R} & \textbf{3rd C}  \\

\toprule

\textbf{Before Choice}         & 103                    & 0                    & 223                   & 0                    & 251                    & 3                    & 27                    & 0                    \\
\textbf{Opt-in}                & 210                    & 11                   & 648                   & 10                   & 11                     & 4                    & 173                   & 4                    \\
\textbf{Refresh after Opt-in}  & 537                    & 11                   & 645                   & 3                    & 159                    & 3                    & 90                    & 1                    \\
\textbf{Opt-out}               & 161                    & 0                    & 18                    & 0                    & 8                    & 3                    & 87                    & 1                    \\

\textbf{Refresh after Opt-out} & 112                    & 1                    & 208                   & 0                    & 171                    & 2                    & 83                    & 0           \\

\bottomrule             
\end{tabular}
}
\label{table:requests-cookies}
\end{table}

%% file: docs/4_evaluation.tex
\section{Results and Comprehensives Analysis}
\label{sec:evaluation}








In this section, we analyze advertisers compliance when users opt-out of data processing and selling. 
We compare and contrast targeting by advertisers across personas and configurations, make statistical observations, and draw conclusions from those observations about advertisers' compliance under GDPR and CCPA.
We present our findings at the granularity of individual CMP because some CMPs might better handle user consent than the others.
We measure advertisers compliance as follows:

\begin{enumerate}[leftmargin=0.5cm]
    \item \textit{\textbf{Data usage.}} Opting out should lead to lower bid values, in interest personas as compared to the control, from advertisers. If advertisers continue to bid higher, they may still be using user data when users opt out of data processing and selling. 
    \item \textit{\textbf{Server-side data sharing.}} Opting out should lead to lower bid values, in interest personas as compared to the control, from advertisers to whom data is not directly leaked. If advertisers to which data is not leaked bid higher, advertisers might still be sharing data when users opt out of data processing and selling. 
    \item \textit{\textbf{Client-side data sharing.}} Opting out should eliminate or significantly reduce cookie syncing events, in interest personas as compared to the control, from advertisers. If advertisers continue to sync cookie with each other, they may be sharing/selling user data when users opt out of data processing and selling. 
\end{enumerate}

As the steps of analysis in each CMP and NAI are similar, we only listed Cookiebot in this section,  put the rest in Appendix \ref{appendix:analysis_cmps}, and only remain the takeaway of each CMPs and NAI except Cookiebot.
\subsection{Cookiebot}

\noindent
\textit{\textbf{Data usage.}} 
%
We evaluate reduction in data usage by analyzing advertisers bidding behavior. 
Table \ref{table:cookiebot-table1} presents advertisers bidding on personas when users opt-out and opt-in through Cookiebot under GDPR and CCPA.
We note that all personas, with the exception of the Shopping where the bid value is same as the control under CCPA, receive higher bids as compared to the control when users opt-out under both GDPR and CCPA.
%
%
%

Next, we analyze if there is statistically significant difference between advertisers bidding patterns when users opt-out or opt-in under GDPR and CCPA. 
%
%
%
It can be seen in Table \ref{table:cookiebot-table1} that advertisers bidding behavior does not significantly changes regardless of whether users opt-out or opt-in under both GDPR and CCPA.

\input{tables/new/cookiebot-1.tex}

\noindent
\textit{\textbf{Server-side data sharing.}}
We evaluate reduction in server-side data sharing by analyzing  bidding from advertisers to which we do not leak data. 
%
%
Table \ref{table:cookiebot-table2} presents bids from advertisers to which we did not explicitly leak data. 
It can be seen that all personas, with the exception of Shopping for CCPA, receive higher bids on average than the control persona. 
Even in the case of Shopping persona, the bid value is only 0.01 less than the control. 

\input{tables/new/cookiebot-2.tex}
\input{docs/advertisers_cookie_syncing.tex}

\noindent
\textit{\textbf{Client-side data sharing.}} 
We evaluate reduction in client-side data sharing by measuring advertiser cookie syncing in network traffic.
Table \ref{table:cookiebot-table2} presents advertiser cookie syncing behavior.
Under GDPR, we note that there is substantial difference between advertisers cookie syncing behavior for opt-out and opt-in. 
Specifically, we only experience cookie syncing events in one persona (i.e., News) when we opt-out but we experience substantial more cookie syncing when we opt-in. 
Under CCPA, however, advertisers engage in cookie syncing events on 12 personas when we opt-out and all 16 personas when we opt-in. 
The total number of cookie syncing events on average in both opt-out and opt-in remains the same. 

We further investigate cookie syncing frequency of individual advertisers.
Table \ref{tbl:Advertisers_cookie_syncing} presents the top 5 most prevalent advertisers that participate in cookie syncing, when we opt-out under both GDPR and CCPA.
It can be seen from the table that advertisers participate in as many as 3 and 128 cookie syncing events when we opt-out under GDPR and CCPA with Cookiebot, respectively.

\noindent
\textbf{\textit{Takeaway.}} 
%
%
The effectiveness of opt-out under CCPA appears limited, with average bids comparable between opt-out and opt-in across all 16 personas settings. Additionally, the syncing events in opt-out are not significantly fewer than in opt-in for most personas settings. Conversely, under GDPR, bid differences between opt-out and opt-in suggest opt-out might not be effective. Nonetheless, the number of syncing events is consistently lower in opt-out than in opt-in, indicating its effectiveness. This means both GDPR and CCPA compliance may not be ensured on the same websites.

\subsection{Didomi, Onetrust, Quantcast and NAI}

For Didomi, Onetrust, Quantcast and NAI, we did similar client-side data sharing analysis and server-side data sharing analysis, as what we did for Cookiebot.
The comprehensive analysis details can be found in Appendix \ref{appendix:analysis_cmps}.

In Didomi, significant decreases in data utilization and sharing are observed when users choose to opt-out under the guidelines of GDPR and CCPA. The decline in data usage is more pronounced under CCPA than under GDPR. Conversely, the decrease in client-side data sharing is more prominent under GDPR compared to CCPA. While the utilization of Didomi for obtaining consent noticeably minimizes targeted activities, it does not entirely eradicate them. This is evident in the continued heightened bidding for certain user profiles and the involvement of advertisements in cookie synchronization. Consequently, achieving GDPR compliance might be feasible through Didomi, but ensuring CCPA compliance might not be guaranteed for the same set of websites.

In OneTrust, differences in advertiser behavior between GDPR and CCPA were observed when users opted out. Specifically, opting out did not result in a statistically significant difference in data usage under GDPR, but it did under CCPA. The prevalence of both server-side and client-side data sharing was higher under CCPA compared to GDPR. Surprisingly, advertisers synchronized more cookies (meaning they shared more data on the client side) under CCPA compared to GDPR. This implies that compliance with GDPR and CCPA might not be guaranteed on the same websites.

In Quantcast, significant decreases in data usage and sharing aren't observed when users choose to opt-out. Advertiser bidding behavior undergoes notable changes for 5 user personas under GDPR, albeit with a minor impact. When users opt-in under GDPR, there's a noticeable increase in cookie syncing events. This suggests that ensuring compliance with both GDPR and CCPA might not be assured on identical websites.

In NAI, the utilization of advertisers' data remains relatively stable. Nevertheless, advertisers tend to offer lower bids under CCPA compared to GDPR. Likewise, we have observed a marked decrease in data sharing, both on the server-side and client-side, under CCPA. As a result, it is possible that websites might not be concurrently compliant with both GDPR and CCPA regulations.

In the majority of cases within four CMPs and NAI, compliance with GDPR and CCPA regulations is lacking. There is little distinction in bids between opt-out and opt-in scenarios, and the volume of syncing events remains comparable regardless of opt-out or opt-in choices. It is plausible that some advertisers adhered to GDPR and CCPA regulations, and the CMP or NAI systems effectively functioned. Nonetheless, these advertisers were not extensively involved in most bidding events and cookie syncing occurrences. To validate this presumption, our focus shifted to advertisers, and a more extensive analysis is presented in Section \ref{sec:aac}.

\subsection{Advertisers Bidding Behavior with pre-opt-out}
Under GDPR processing personal data is prohibited, unless the data subject has consented to the processing (Article 6). 
However, under CCPA, data selling and sharing should be stopped immediately stop once consumers opt-out (Section 798.120 (a), Section 7013 (a)).
Thus to eliminate the impact of data collection and sharing prior to opting-out, we conduct additional experiments where we opt-out prior to simulating personas. 
Similar to post opt-out, we note that under both GDPR and CCPA advertisers continue to use data even when we opt-out prior to collecting bids. 
We discuss advertisers bidding behavior with pre-opt-out in detail in Appendix \ref{appendix:pre-opt-out}.


\input{docs/tp6_advertisers_bids.tex}
\subsection{Advertisers across CMPs}
\label{sec:aac}

\subsubsection{Cookie Syncing Advertisers}

We have compiled a list of the top five advertisers from a pool of 53, based on the highest number of syncing events during the Opt-out phase (Table \ref{tbl:Advertisers_cookie_syncing}). Additionally, we have tabulated the event counts during the Opt-in phase in Table \ref{tbl:Advertisers_cookie_syncing_opt_in}. A noticeable trend emerges: under GDPR regulations, CookieBot, Didomi, and Quantcast have effectively curbed a substantial number of syncing events. However, in the case of OneTrust, the disparity in the number of syncing events between Opt-out and Opt-in appears minimal. When considering CCPA, Didomi and NAI emerge as more successful than Cookiebot, OneTrust, and Quantcast in preventing syncing events. Nevertheless, it's worth noting that the counts of syncing events originating from websites utilizing Didomi and NAI are already quite high during the Opt-out phase.

\textbf{\textit{Takeaway.}} In terms of obstructing cookie syncing, CMPs exhibit superior performance under GDPR compared to CCPA. Notably, cookie events persist post Opt-out in CCPA scenarios. Among all CMPs, OneTrust shows the weakest performance, as the event counts between Opt-out and Opt-in remain comparable.

\subsubsection{Analysis of Bids from Syncing Advertisers}

\input{docs/tp6_advertisers_bids_numbers.tex}

In both Table \ref{tbl:Advertisers_cookie_syncing} and Table \ref{tbl:Advertisers_cookie_syncing_opt_in}, there are four advertisers that consistently appear, occupying the top four positions in both instances. Consequently, we extracted bids data from these shared four advertisers and the two unique advertisers from each table. This data was then employed to compile Table \ref{tbl:Bids_of_advertisers_syncing_GDPR} and Table \ref{tbl:Bids_of_advertisers_syncing_CCPA}. Notably, certain scenarios, such as the "Arts" persona under Cookiebot, exhibit significantly higher bid counts in Opt-out compared to Opt-in. To provide a comprehensive overview, we calculated the bid quantities under similar circumstances, as detailed in Table \ref{tbl:Number_Bids_of_advertisers_syncing_GDPR} and Table \ref{tbl:Number_Bids_of_advertisers_syncing_CCPA}.

Under GDPR regulations, both CookieBot and Didomi show reduced bid counts in Opt-out compared to Opt-in, whereas the figures remain relatively consistent for OneTrust, Quantcast, and NAI. Furthermore, the averages between Opt-out and Opt-in exhibit similarity for OneTrust, Quantcast, and NAI. In the context of CCPA, more bids emerges in Opt-out compared to GDPR, accompanied by greater bid values. The trend in bid counts between Opt-out and Opt-in under CCPA resembles that observed under GDPR.

\textbf{\textit{Takeaway.}} A comparative analysis between GDPR and CCPA reveals that all four CMPs and NAI demonstrate superior bid-blocking performance for the six selected advertisers under GDPR. A CMP-to-CMP comparison under GDPR reveals the superior performance of Cookiebot and Didomi compared to OneTrust, Quantcast, and NAI. Bid value and bid count patterns remain consistent between Opt-out and Opt-in for OneTrust, Quantcast, and NAI. When contrasting CMPs under CCPA, Cookiebot and Didomi still outperform OneTrust and Quantcast. Bid values and counts remain comparable between Opt-out and Opt-in for OneTrust, Quantcast, and NAI, with substantially higher values observed in Opt-out scenarios.

\subsubsection{Syncing versus Non-Syncing}
\input{docs/bids_from_syncing_vs_not.tex}

The subsequent analysis compares advertisers engaged in cookie syncing to those not involved, specifically focusing on Opt-out settings. The averages of bid values for both cookie syncing and non-participating advertisers under GDPR and CCPA are presented in Table \ref{tbl:Bids_of_advertisers_syncing_vs_not_GDPR} and Table \ref{tbl:Bids_of_advertisers_syncing_vs_not_CCPA}. Additionally, Table \ref{tbl:Number_Bids_of_advertisers_syncing_vs_not_GDPR} and Table \ref{tbl:Number_Bids_of_advertisers_syncing_vs_not_CCPA} display the bid count originating from advertisers who either participated or did not participate in cookie syncing under both GDPR and CCPA regulations.

\begin{enumerate}[leftmargin=0.5cm]





\item In Cookiebot, bids from syncing advertisers are significantly fewer than those from non-syncing counterparts. Notably, for personas such as Arts, Health, Reference, and Shopping, the average bid from syncing advertisers surpasses that from non-syncing advertisers under CCPA. Remarkably, the bid count remains comparable between GDPR and CCPA scenarios.

\item In Didomi, the number of bids is notably limited under GDPR. Conversely, under CCPA, an overwhelming majority of bids originate from syncing advertisers, with the average bid value showing parity between syncing and non-syncing participants.

\item OneTrust showcases a trend where bids predominantly stem from non-syncing advertisers under both GDPR and CCPA, with a notable shift towards syncing advertisers under CCPA compared to GDPR. Additionally, the bid values from syncing advertisers considerably surpass those from non-syncing counterparts under both GDPR and CCPA.

\item Quantcast exhibits a situation where nearly all bids arise from syncing advertisers under GDPR. Meanwhile, under CCPA, bids are sourced from both syncing and non-syncing advertisers, with syncing advertisers accounting for a larger share. Interestingly, under CCPA, the bid values from non-syncing advertisers are notably greater than those from syncing participants.

\item For NAI, the bid count from non-syncing advertisers surpasses that from syncing advertisers. Remarkably, the bid count maintains similarity between GDPR and CCPA contexts. In the context of GDPR, syncing advertisers yield lower values compared to non-syncing ones, whereas the reverse trend is observed under CCPA. Syncing advertisers exhibit lower values under GDPR compared to CCPA, while non-syncing advertisers display the opposite trend.

\end{enumerate}
\input{docs/bids_from_syncing_vs_not_numbers.tex}
\textbf{\textit{Takeaway.}} All 4 CMPs have a very small amount of bids from advertisers which participated in cookie syncing under GDPR, except NAI. There are still bids from no-syncing advertisers in CookieBot and Onetrust under GDPR. There are more bids under CCPA, especially in Didomi(Syncing), Onetrust(Syncing), and Quantcast(Both Syncing and no-Syncing), except NAI. NAI has a similar number of bids under GDPR and CCPA. The bid value is higher under CCPA than under GDPR, except Cookiebot and NAI. Cookiebot has similar average bids from no-syncing under GDPR and CCPA. NAI has opposite value comparison trends between syncing advertisers and no-syncing ones under both GDPR and CCPA.

%% file: tables/new/cookiebot-1.tex
\begin{table}[htpb]

\centering
\scriptsize
\caption{Ad bidding under GDPR and CCPA after opt-out (Out) and opt-in (In) with Cookiebot. Avg. column represents the mean of all bid value. \textcolor{highbids}{Light red} and \textcolor{lowbids}{Light blue} indicate bid values that are higher and lower than Control's avg., respectively. \textcolor{higherbids}{Dark red} and \textcolor{lowerbids}{Dark blue} indicate bid values that are Control's avg. $\pm$ std., respectively. Column p-val. and Eff. represent p-value and effect size, respectively}
\adjustbox{max width=\columnwidth}{
\begin{tabular}{l?{0.3mm}cc|cc?{0.3mm}cc|cc} 
    \toprule
                    & \multicolumn{4}{c?{0.3mm}}{\textbf{GDPR}}                                                  & \multicolumn{4}{c}{\textbf{CCPA}}                                                   \\ 

                    & \textbf{Out} & \textbf{In} & \multicolumn{2}{c?{0.3mm}}{\textbf{Stat. Test}} & \textbf{Out} & \textbf{In} & \multicolumn{2}{c}{\textbf{Stat. Test}}  \\ 

\textbf{Persona}    & \textbf{Avg.}    & \textbf{Avg.}   & \textbf{p-val.} & \textbf{Eff.}        & \textbf{Avg.}    & \textbf{Avg.}   & \textbf{p-val.} & \textbf{Eff.}         \\ 
    \toprule
\textbf{Adult}      & 0.37 {\cellcolor{highbids}}  & 0.33 {\cellcolor{highbids}}  & 1.85 & 0.08     & 0.37 {\cellcolor{highbids}}  & 0.34 {\cellcolor{highbids}}  & 2.64 & 0.06  \\
\textbf{Arts}       & 0.35 {\cellcolor{highbids}}  & 0.32 {\cellcolor{highbids}}  & 2.65 & 0.06     & 0.38 {\cellcolor{highbids}}  & 0.31 {\cellcolor{highbids}}  & 0.64 & 0.11  \\
\textbf{Business}   & 0.35 {\cellcolor{highbids}}  & 0.35 {\cellcolor{highbids}}  & 7.82 & 2E-03 & 0.36 {\cellcolor{highbids}}  & 0.32 {\cellcolor{highbids}}  & 3.08 & 0.06  \\
\textbf{Computers}  & 0.35 {\cellcolor{highbids}}  & 0.36 {\cellcolor{highbids}}  & 3.71 & 0.05     & 0.37 {\cellcolor{highbids}}  & 0.30 & 0.51 & 0.12  \\
\textbf{Games}      & 0.38 {\cellcolor{highbids}}  & 0.38 {\cellcolor{highbids}}  & 7.42 & 0.01     & 0.36 {\cellcolor{highbids}}  & 0.28 {\cellcolor{lowbids}}  & 0.14 & 0.15  \\
\textbf{Health}     & 0.35 {\cellcolor{highbids}}  & 0.34 {\cellcolor{highbids}}  & 6.38 & 0.02     & 0.38 {\cellcolor{highbids}}  & 0.38 {\cellcolor{highbids}}  & 5.36 & 0.03  \\
\textbf{Home}       & 0.38 {\cellcolor{highbids}}  & 0.31 {\cellcolor{highbids}}  & 0.75 & 0.11     & 0.36 {\cellcolor{highbids}}  & 0.34 {\cellcolor{highbids}}  & 4.53 & 0.04  \\
\textbf{Kids}       & 0.37 {\cellcolor{highbids}}  & 0.34 {\cellcolor{highbids}}  & 2.33 & 0.07     & 0.35 {\cellcolor{highbids}}  & 0.29 {\cellcolor{lowbids}}  & 0.32 & 0.13  \\
\textbf{News}       & 0.35 {\cellcolor{highbids}}  & 0.39 {\cellcolor{highbids}}  & 2.31 & 0.07     & 0.37 {\cellcolor{highbids}}  & 0.33 {\cellcolor{highbids}}  & 2.12 & 0.07  \\
\textbf{Recreation} & 0.37 {\cellcolor{highbids}}  & 0.35 {\cellcolor{highbids}}  & 4.43 & 0.04     & 0.36 {\cellcolor{highbids}}  & 0.31 {\cellcolor{highbids}}  & 1.71 & 0.08 \\
\textbf{Reference}  & 0.37 {\cellcolor{highbids}}  & 0.33 {\cellcolor{highbids}}  & 1.85 & 0.08     & 0.37 {\cellcolor{highbids}}  & 0.32 {\cellcolor{highbids}}  & 0.89 & 0.10  \\
\textbf{Regional}   & 0.38 {\cellcolor{highbids}}  & 0.34 {\cellcolor{highbids}}  & 2.34 & 0.07     & 0.35 {\cellcolor{highbids}}  & 0.30 & 0.93 & 0.10  \\
\textbf{Science}    & 0.35 {\cellcolor{highbids}}  & 0.39 {\cellcolor{highbids}}  & 1.48 & 0.09     & 0.36 {\cellcolor{highbids}}  & 0.32 {\cellcolor{highbids}}  & 2.22 & 0.07  \\
\textbf{Shopping}   & 0.37 {\cellcolor{highbids}}  & 0.35 {\cellcolor{highbids}}  & 7.33 & 0.01     & 0.32 & 0.32 {\cellcolor{highbids}}  & 5.08 & 0.03  \\
\textbf{Society}    & 0.37 {\cellcolor{highbids}}  & 0.37 {\cellcolor{highbids}}  & 7.93 & 1E-03 & 0.37 {\cellcolor{highbids}}  & 0.33 {\cellcolor{highbids}}  & 1.31 & 0.09  \\
\textbf{Sports}     & 0.37 {\cellcolor{highbids}}  & 0.35 {\cellcolor{highbids}}  & 4.39 & 0.04     & 0.33 {\cellcolor{highbids}}  & 0.31 {\cellcolor{highbids}}  & 7.78 & 2E-3  \\
\midrule
\textbf{Control}    & 0.33 & 0.26 & 0.53 & 0.11     & 0.32 & 0.30 & 6.77 & 0.01                 \\
\bottomrule
\end{tabular}
}
\vspace{-1em}
\label{table:cookiebot-table1}
\end{table}

%% file: tables/new/cookiebot-2.tex
\begin{table}[htpb]
    \scriptsize
\centering
\caption{Ad bidding and cookie syncing under GDPR and CCPA after opt-out (Out) and opt-in (In) with Cookiebot. Avg. column represents the mean of all bid value from advertisers who did not bid or appear when we simulated personas but appeared and bid after we opt-out. Out and In under C-Sync. represent number of cookie syncing events after opt-out and opt-in, respectively.}
\begin{tabular}{l?{0.3mm}c|cc?{0.3mm}c|cc} 
\toprule
                    & \multicolumn{3}{c?{0.3mm}}{\textbf{GDPR}}                              & \multicolumn{3}{c}{\textbf{CCPA}}                               \\ 

                    & \textbf{Out} & \multicolumn{2}{c?{0.3mm}}{\textbf{C-Sync.}} & \textbf{Out} & \multicolumn{2}{l}{\textbf{C-Sync.}}  \\ 

\textbf{Persona}    & \textbf{Avg.}    & \textbf{Out} & \textbf{In}           & \textbf{Avg.}    & \textbf{Out} & \textbf{In}            \\ 
\toprule
\textbf{Adult}      & 0.37 {\cellcolor{highbids}}& 0  & 25 & 0.37 {\cellcolor{highbids}}& 0  & 11  \\
\textbf{Arts}       & 0.35 {\cellcolor{highbids}}& 0  & 7  & 0.37 {\cellcolor{highbids}}& 4  & 3   \\
\textbf{Business}   & 0.35 {\cellcolor{highbids}}& 0  & 44 & 0.37 {\cellcolor{highbids}}& 3  & 3   \\
\textbf{Computers}  & 0.35 {\cellcolor{highbids}}& 0  & 28 & 0.37 {\cellcolor{highbids}}& 0  & 4   \\
\textbf{Games}      & 0.38 {\cellcolor{highbids}}& 0  & 27 & 0.36 {\cellcolor{highbids}}& 0  & 4   \\
\textbf{Health}     & 0.35 {\cellcolor{highbids}}& 0  & 36 & 0.37 {\cellcolor{highbids}}& 28 & 25  \\
\textbf{Home}       & 0.38 {\cellcolor{highbids}}& 0  & 3  & 0.36 {\cellcolor{highbids}}& 6  & 14  \\
\textbf{Kids}       & 0.37 {\cellcolor{highbids}}& 0  & 41 & 0.36 {\cellcolor{highbids}}& 5  & 26  \\
\textbf{News}       & 0.35 {\cellcolor{highbids}}& 3  & 22 & 0.37 {\cellcolor{highbids}}& 3  & 3   \\
\textbf{Recreation} & 0.37 {\cellcolor{highbids}}& 0  & 6  & 0.36 {\cellcolor{highbids}}& 21 & 13  \\
\textbf{Reference}  & 0.37 {\cellcolor{highbids}}& 0  & 24 & 0.37 {\cellcolor{highbids}}& 14 & 5   \\
\textbf{Regional}   & 0.38 {\cellcolor{highbids}}& 0  & 52 & 0.36 {\cellcolor{highbids}}& 28 & 16  \\
\textbf{Science}    & 0.35 {\cellcolor{highbids}}& 0  & 26 & 0.37 {\cellcolor{highbids}}& 29 & 18  \\
\textbf{Shopping}   & 0.37 {\cellcolor{highbids}}& 0  & 50 & 0.31 {\cellcolor{lowbids}}& 37 & 12  \\
\textbf{Society}    & 0.37 {\cellcolor{highbids}}& 0  & 5  & 0.37 {\cellcolor{highbids}}& 0  & 5   \\
\textbf{Sports}     & 0.37 {\cellcolor{highbids}}& 0  & 43 & 0.33 {\cellcolor{highbids}}& 5  & 7   \\
\midrule
\textbf{Control}    & 0.33 & -- & -- & 0.32 & -- & --                 \\
\midrule
\textbf{Average} & -- & 1 & 27 & -- & 11 & 11 \\
\bottomrule
\end{tabular}
\vspace{-1em}
\label{table:cookiebot-table2}
\end{table}


%% file: docs/advertisers_cookie_syncing.tex
\begin{table*}[!h]
    \scriptsize
\centering
\caption{Most prevalent advertisers that participate in cookie syncing, when we opt-out under GDPR and CCPA. These advertiser appear in all personas across CookieBot, Didomi, Onetrust, Quantcast and NAI configurations. CB, DM, OT, QC and NAI columns represent the count of cookie syncing events under  CookieBot, Didomi, Onetrust, Quantcast and NAI for each advertiser.}
\adjustbox{max width=\textwidth}{
    \begin{tabular}{l?{0.3mm}ccccc?{0.3mm}ccccc?{0.3mm}c} 
    \toprule
     & \multicolumn{5}{c?{0.3mm}}{\textbf{GDPR}} & \multicolumn{5}{c?{0.3mm}}{\textbf{CCPA}}  &   \\ 
     \textbf{Advertisers}& \multicolumn{1}{c}{\textbf{Cookiebot}} & \multicolumn{1}{c}{\textbf{Didomi}}&\multicolumn{1}{c}{\textbf{OneTrust}}&\multicolumn{1}{c}{\textbf{Quantcast}}& \multicolumn{1}{c?{0.3mm}}{\textbf{NAI}}& \multicolumn{1}{c}{\textbf{Cookiebot}} & \multicolumn{1}{c}{\textbf{Didomi}}& \multicolumn{1}{c}{\textbf{OneTrust}}& \multicolumn{1}{c}{\textbf{Quantcast}}& \multicolumn{1}{c?{0.3mm}}{\textbf{NAI}} & \multicolumn{1}{c}{\textbf{Total}}   \\ 
    \toprule

    \textbf{PubMatic}       & 0 & 31 & 101 & 149 & 107 & 128 & 211 & 306 & 299 & 190 & 1522  \\
    \textbf{AT\&T}          & 3 & 0  & 40  & 170 & 0   & 25  & 39  & 324 & 317 & 0   & 918   \\
    \textbf{VerizonMedia}   & 0 & 9  & 12  & 0   & 0   & 23  & 43  & 419 & 76  & 49  & 631   \\
    \textbf{RubiconProject} & 0 & 0  & 0   & 148 & 22  & 0   & 17  & 88  & 225 & 58  & 558   \\
    \textbf{GumGum}         & 0 & 0  & 0   & 42  & 0   & 0   & 0   & 0   & 286 & 0   & 328  \\       
\bottomrule
\end{tabular}
}
\label{tbl:Advertisers_cookie_syncing}
\end{table*}

\begin{table*}[!h]
    \scriptsize
\centering
\caption{Most prevalent advertisers that participate in cookie syncing, when we opt-in under GDPR and CCPA. These advertiser appear in all personas across CookieBot, Didomi, Onetrust, Quantcast and NAI configurations. CB, DM, OT, QC and NAI columns represent the count of cookie syncing events under  CookieBot, Didomi, Onetrust, Quantcast and NAI for each advertiser.}
\adjustbox{max width=\textwidth}{
    \begin{tabular}{l?{0.3mm}ccccc?{0.3mm}ccccc?{0.3mm}c} 
    \toprule
     & \multicolumn{5}{c?{0.3mm}}{\textbf{GDPR}} & \multicolumn{5}{c?{0.3mm}}{\textbf{CCPA}}  &   \\ 
     \textbf{Advertisers}& \multicolumn{1}{c}{\textbf{Cookiebot}} & \multicolumn{1}{c}{\textbf{Didomi}}&\multicolumn{1}{c}{\textbf{OneTrust}}&\multicolumn{1}{c}{\textbf{Quantcast}}& \multicolumn{1}{c?{0.3mm}}{\textbf{NAI}}& \multicolumn{1}{c}{\textbf{Cookiebot}} & \multicolumn{1}{c}{\textbf{Didomi}}& \multicolumn{1}{c}{\textbf{OneTrust}}& \multicolumn{1}{c}{\textbf{Quantcast}}& \multicolumn{1}{c?{0.3mm}}{\textbf{NAI}} & \multicolumn{1}{c}{\textbf{Total}}   \\ 
    \toprule

\textbf{PubMatic}       & 325 & 375 & 131 & 294 & 95 & 125 & 353 & 365 & 323 & 255 & 2641 \\
\textbf{AT\&T}          & 27  & 624 & 94  & 350 & 0  & 22  & 575 & 332 & 405 & 167  & 2596\\
\textbf{VerizonMedia}   & 0   & 725 & 20  & 228 & 2  & 8   & 559 & 412 & 425 & 134  & 2513\\
\textbf{RubiconProject} & 20  & 651 & 17  & 422 & 0  & 0   & 470 & 36  & 162 & 60  &1838 \\
\textbf{RichAudience}   & 0   & 221 & 0   & 211 & 0  & 0   & 224 & 67  & 212 & 25  &960   \\
\bottomrule
\end{tabular}
}
\label{tbl:Advertisers_cookie_syncing_opt_in}
\end{table*}

%% file: docs/tp6_advertisers_bids.tex
\begin{table}[!h]
    \scriptsize
\centering
\caption{Average of bids data from 6 advertisers in cookie syncing, when we opt-out and opt-in under GDPR. CB, DM, OT, QC and NAI columns represent the count of cookie syncing events under CookieBot, Didomi, Onetrust, Quantcast and NAI for each advertiser.}
\adjustbox{max width=\columnwidth}{
    \scalebox{1}{ \begin{tabular}{l?{0.3mm}ll?{0.3mm}ll?{0.3mm}ll?{0.3mm}ll?{0.3mm}ll} 
    \toprule
    & \multicolumn{10}{c}{\textbf{GDPR}} \\
     & \multicolumn{2}{c?{0.3mm}}{\textbf{CB}} & \multicolumn{2}{c?{0.3mm}}{\textbf{DM}}  & \multicolumn{2}{c?{0.3mm}}{\textbf{OT}}& \multicolumn{2}{c?{0.3mm}}{\textbf{QC}}& \multicolumn{2}{c}{\textbf{NAI}}   \\ 
& \multicolumn{1}{c}{\textbf{Out}} & \multicolumn{1}{c?{0.3mm}}{\textbf{In}}  & \multicolumn{1}{c}{\textbf{Out}} & \multicolumn{1}{c?{0.3mm}}{\textbf{In}} & \multicolumn{1}{c}{\textbf{Out}} & \multicolumn{1}{c?{0.3mm}}{\textbf{In}} & \multicolumn{1}{c}{\textbf{Out}} & \multicolumn{1}{c?{0.3mm}}{\textbf{In}} & \multicolumn{1}{c}{\textbf{Out}} & \multicolumn{1}{c}{\textbf{In}}   \\      
     
     \textbf{Persona}& \multicolumn{1}{c}{\textbf{Avg.}} & \multicolumn{1}{c?{0.3mm}}{\textbf{Avg.}} & \multicolumn{1}{c}{\textbf{Avg.}} & \multicolumn{1}{c?{0.3mm}}{\textbf{Avg.}} & \multicolumn{1}{c}{\textbf{Avg.}} & \multicolumn{1}{c?{0.3mm}}{\textbf{Avg.}} & \multicolumn{1}{c}{\textbf{Avg.}} & \multicolumn{1}{c?{0.3mm}}{\textbf{Avg.}} & \multicolumn{1}{c}{\textbf{Avg.}} & \multicolumn{1}{c}{\textbf{Avg.}}   \\ 
    \toprule

\textbf{Adult}      & --   & 0.08 & --   & 0.06 & 0.02 & 0.01 & 0.11 & 0.09 & --   & --    \\
\textbf{Arts}       & --   & 0.11 & 0.05 & 0.06 & --   & 0.05 & 0.08 & 0.11 & 0.02 & --    \\
\textbf{Business}   & --   & 0.09 & --   & 0.02 & 0.02 & 0.10 & 0.10 & 0.06 & 0.05 & 0.03  \\
\textbf{Computers}  & --   & 0.38 & 0.05 & 0.05 & 0.04 & 0.01 & 0.08 & 0.09 & 0.18 & 0.17  \\
\textbf{Games}      & --   & 0.40 & 0.03 & 0.07 & 0.02 & 0.03 & 0.10 & 0.09 & 0.02 & --    \\
\textbf{Health}     & --   & 0.24 & --   & 0.05 & 0.03 & 0.15 & 0.12 & 0.08 & 0.02 & 0.12  \\
\textbf{Home}       & --   & 0.08 & 0.05 & 0.06 & 0.31 & 0.05 & 0.16 & 0.08 & --   & --    \\
\textbf{Kids}       & --   & 0.13 & 0.06 & 0.04 & --   & 0.02 & 0.07 & 0.09 & 0.02 & 0.01  \\
\textbf{News}       & 0.00 & 0.46 & --   & 0.04 & 0.03 & 0.55 & 0.09 & 0.04 & 0.02 & 0.05  \\
\textbf{Recreation} & --   & 0.13 & 0.06 & 0.02 & 0.02 & 0.13 & 0.11 & 0.07 & 0.01 & --    \\
\textbf{Reference}  & --   & 0.04 & --   & 0.03 & 0.03 & 0.20 & 0.09 & 0.06 & 0.02 & 0.01  \\
\textbf{Regional}   & --   & 0.01 & --   & 0.03 & 0.01 & 0.02 & 0.18 & 0.10 & 0.03 & 0.01  \\
\textbf{Science}    & --   & 0.49 & --   & 0.06 & 0.03 & 0.09 & 0.08 & 0.10 & 0.02 & --    \\
\textbf{Shopping}   & --   & 0.08 & --   & 0.05 & 1.10 & 0.13 & 0.06 & 0.03 & 0.02 & 0.02  \\
\textbf{Society}    & --   & 0.37 & 0.05 & 0.03 & 0.02 & 0.02 & 0.07 & 0.07 & --   & --    \\
\textbf{Sports}     & --   & 0.23 & 0.05 & 0.05 & 0.02 & 0.16 & 0.18 & 0.17 & 0.15 & --   \\

\bottomrule
\end{tabular}
}}
\label{tbl:Bids_of_advertisers_syncing_GDPR}
\end{table}

\begin{table}[!h]
    \scriptsize
\centering
\caption{Average of bids data from 6 advertisers in cookie syncing, when we opt-out and opt-in under CCPA. CB, DM, OT, QC and NAI columns represent the count of cookie syncing events under CookieBot, Didomi, Onetrust, Quantcast and NAI for each advertiser.}
\adjustbox{max width=\columnwidth}{
   \scalebox{1}{ \begin{tabular}{l?{0.3mm}ll?{0.3mm}ll?{0.3mm}ll?{0.3mm}ll?{0.3mm}ll} 
    \toprule
    & \multicolumn{10}{c}{\textbf{CCPA}} \\
     & \multicolumn{2}{c?{0.3mm}}{\textbf{CB}} & \multicolumn{2}{c?{0.3mm}}{\textbf{DM}}  & \multicolumn{2}{c?{0.3mm}}{\textbf{OT}}& \multicolumn{2}{c?{0.3mm}}{\textbf{QC}}& \multicolumn{2}{c}{\textbf{NAI}}   \\ 
& \multicolumn{1}{c}{\textbf{Out}} & \multicolumn{1}{c?{0.3mm}}{\textbf{In}}  & \multicolumn{1}{c}{\textbf{Out}} & \multicolumn{1}{c?{0.3mm}}{\textbf{In}} & \multicolumn{1}{c}{\textbf{Out}} & \multicolumn{1}{c?{0.3mm}}{\textbf{In}} & \multicolumn{1}{c}{\textbf{Out}} & \multicolumn{1}{c?{0.3mm}}{\textbf{In}} & \multicolumn{1}{c}{\textbf{Out}} & \multicolumn{1}{c}{\textbf{In}}   \\      
     
     \textbf{Persona}& \multicolumn{1}{c}{\textbf{Avg.}} & \multicolumn{1}{c?{0.3mm}}{\textbf{Avg.}} & \multicolumn{1}{c}{\textbf{Avg.}} & \multicolumn{1}{c?{0.3mm}}{\textbf{Avg.}} & \multicolumn{1}{c}{\textbf{Avg.}} & \multicolumn{1}{c?{0.3mm}}{\textbf{Avg.}} & \multicolumn{1}{c}{\textbf{Avg.}} & \multicolumn{1}{c?{0.3mm}}{\textbf{Avg.}} & \multicolumn{1}{c}{\textbf{Avg.}} & \multicolumn{1}{c}{\textbf{Avg.}}   \\ 
    \toprule

\textbf{Adult}      & --   & 0.10 & 0.34 & 0.04 & 0.45 & 0.23 & 0.26 & 0.14 & 0.53 & 0.13  \\
\textbf{Arts}       & 0.53 & 0.05 & 0.11 & 0.08 & 0.20 & 0.38 & 0.36 & 0.16 & 0.26 & 0.18  \\
\textbf{Business}   & 0.24 & 0.09 & 0.14 & 0.07 & 0.44 & 0.19 & 0.14 & 0.18 & 0.10 & 0.24  \\
\textbf{Computers}  & --   & 0.05 & 0.11 & 0.06 & 0.27 & 0.38 & 0.10 & 0.14 & 0.19 & 0.14  \\
\textbf{Games}      & --   & 0.07 & 0.08 & 0.05 & 0.16 & 0.23 & 0.17 & 0.06 & 0.31 & 0.18  \\
\textbf{Health}     & 0.44 & 0.41 & 0.46 & 0.12 & 0.34 & 0.18 & 0.67 & 0.48 & 0.28 & 0.47  \\
\textbf{Home}       & --   & 0.08 & 0.13 & 0.04 & 0.28 & 0.27 & 0.13 & 0.10 & 0.12 & 0.14  \\
\textbf{Kids}       & 0.03 & 0.09 & 0.14 & 0.08 & 0.46 & 0.24 & 0.20 & 0.18 & 0.24 & 0.22  \\
\textbf{News}       & 0.01 & 0.12 & 0.06 & 0.08 & 0.42 & 0.27 & 0.10 & 0.21 & 0.10 & 0.17  \\
\textbf{Recreation} & 0.14 & 0.07 & 0.13 & 0.11 & 0.47 & 0.22 & 0.36 & 0.26 & 0.09 & 0.14  \\
\textbf{Reference}  & 0.52 & 0.11 & 0.10 & 0.05 & 0.38 & 0.18 & 0.13 & 0.16 & 0.06 & 0.18  \\
\textbf{Regional}   & 0.03 & 0.11 & 0.10 & 0.07 & 0.37 & 0.21 & 0.18 & 0.11 & 0.57 & 0.23  \\
\textbf{Science}    & 0.11 & 0.06 & 0.14 & 0.16 & 0.38 & 0.29 & 0.13 & 0.08 & 0.17 & 0.18  \\
\textbf{Shopping}   & 0.39 & 0.10 & 0.09 & 0.09 & 0.42 & 0.23 & 0.19 & 0.12 & 0.20 & 0.21  \\
\textbf{Society}    & --   & 0.09 & 0.12 & 0.08 & 0.35 & 0.19 & 0.21 & 0.07 & 0.18 & 0.27  \\
\textbf{Sports}     & 0.32 & 0.13 & 0.11 & 0.09 & 0.28 & 0.24 & 0.16 & 0.20 & 0.26 & 0.26 \\

\bottomrule
\end{tabular}
}}
\label{tbl:Bids_of_advertisers_syncing_CCPA}
\end{table}

%% file: docs/tp6_advertisers_bids_numbers.tex
\begin{table}[!h]
    \scriptsize
\centering
\caption{Number of bids data from 6 advertisers in cookie syncing, when we opt-out and opt-in under GDPR. CB, DM, OT, QC and NAI columns represent the count of cookie syncing events under CookieBot, Didomi, Onetrust, Quantcast and NAI for each advertiser.}
\adjustbox{max width=\textwidth}{
    \scalebox{1}{ \begin{tabular}{l?{0.3mm}ll?{0.3mm}ll?{0.3mm}ll?{0.3mm}ll?{0.3mm}ll} 
    \toprule
    & \multicolumn{10}{c}{\textbf{GDPR}} \\
     & \multicolumn{2}{c?{0.3mm}}{\textbf{CB}} & \multicolumn{2}{c?{0.3mm}}{\textbf{DM}}  & \multicolumn{2}{c?{0.3mm}}{\textbf{OT}}& \multicolumn{2}{c?{0.3mm}}{\textbf{QC}}& \multicolumn{2}{c}{\textbf{NAI}}   \\ 
& \multicolumn{1}{c}{\textbf{Out}} & \multicolumn{1}{c?{0.3mm}}{\textbf{In}}  & \multicolumn{1}{c}{\textbf{Out}} & \multicolumn{1}{c?{0.3mm}}{\textbf{In}} & \multicolumn{1}{c}{\textbf{Out}} & \multicolumn{1}{c?{0.3mm}}{\textbf{In}} & \multicolumn{1}{c}{\textbf{Out}} & \multicolumn{1}{c?{0.3mm}}{\textbf{In}} & \multicolumn{1}{c}{\textbf{Out}} & \multicolumn{1}{c}{\textbf{In}}   \\      
     
     \textbf{Persona}& \multicolumn{1}{c}{\textbf{\#.}} & \multicolumn{1}{c?{0.3mm}}{\textbf{\#.}} & \multicolumn{1}{c}{\textbf{\#.}} & \multicolumn{1}{c?{0.3mm}}{\textbf{\#.}} & \multicolumn{1}{c}{\textbf{\#.}} & \multicolumn{1}{c?{0.3mm}}{\textbf{\#.}} & \multicolumn{1}{c}{\textbf{\#.}} & \multicolumn{1}{c?{0.3mm}}{\textbf{\#.}} & \multicolumn{1}{c}{\textbf{\#.}} & \multicolumn{1}{c}{\textbf{\#.}}   \\ 
    \toprule

\textbf{Adult}      & 0 & 15 & 0 & 524 & 6 & 1 & 144 & 84  & 0 & 0  \\
\textbf{Arts}       & 0 & 11 & 1 & 234 & 0 & 2 & 158 & 161 & 3 & 0  \\
\textbf{Business}   & 0 & 5  & 0 & 408 & 3 & 3 & 148 & 68  & 2 & 1  \\
\textbf{Computers}  & 0 & 22 & 2 & 627 & 2 & 3 & 149 & 79  & 1 & 2  \\
\textbf{Games}      & 0 & 13 & 2 & 533 & 1 & 3 & 150 & 70  & 1 & 0  \\
\textbf{Health}     & 0 & 28 & 0 & 499 & 4 & 6 & 134 & 74  & 2 & 1  \\
\textbf{Home}       & 0 & 21 & 1 & 337 & 2 & 5 & 120 & 108 & 0 & 0  \\
\textbf{Kids}       & 0 & 8  & 1 & 248 & 0 & 1 & 152 & 75  & 2 & 1  \\
\textbf{News}       & 1 & 21 & 0 & 741 & 3 & 5 & 57  & 121 & 1 & 1  \\
\textbf{Recreation} & 0 & 8  & 1 & 432 & 6 & 5 & 139 & 84  & 1 & 0  \\
\textbf{Reference}  & 0 & 15 & 0 & 429 & 4 & 4 & 156 & 121 & 2 & 1  \\
\textbf{Regional}   & 0 & 10 & 0 & 239 & 6 & 2 & 178 & 72  & 3 & 1  \\
\textbf{Science}    & 0 & 14 & 0 & 694 & 5 & 3 & 153 & 110 & 1 & 0  \\
\textbf{Shopping}   & 0 & 10 & 0 & 626 & 1 & 2 & 46  & 104 & 2 & 3  \\
\textbf{Society}    & 0 & 11 & 1 & 371 & 5 & 1 & 109 & 76  & 0 & 0  \\
\textbf{Sports}     & 0 & 11 & 1 & 365 & 2 & 6 & 133 & 151 & 2 & 0 \\

\bottomrule
\end{tabular}
}}
\label{tbl:Number_Bids_of_advertisers_syncing_GDPR}
\end{table}

\begin{table}[!h]
    \scriptsize
\centering
\caption{Number of bids data from 6 advertisers in cookie syncing, when we opt-out and opt-in under CCPA. CB, DM, OT, QC and NAI columns represent the count of cookie syncing events under CookieBot, Didomi, Onetrust, Quantcast and NAI for each advertiser.}
\adjustbox{max width=\textwidth}{
   \scalebox{1}{ \begin{tabular}{l?{0.3mm}ll?{0.3mm}ll?{0.3mm}ll?{0.3mm}ll?{0.3mm}ll} 
    \toprule
    & \multicolumn{10}{c}{\textbf{CCPA}} \\
     & \multicolumn{2}{c?{0.3mm}}{\textbf{CB}} & \multicolumn{2}{c?{0.3mm}}{\textbf{DM}}  & \multicolumn{2}{c?{0.3mm}}{\textbf{OT}}& \multicolumn{2}{c?{0.3mm}}{\textbf{QC}}& \multicolumn{2}{c}{\textbf{NAI}}   \\ 
& \multicolumn{1}{c}{\textbf{Out}} & \multicolumn{1}{c?{0.3mm}}{\textbf{In}}  & \multicolumn{1}{c}{\textbf{Out}} & \multicolumn{1}{c?{0.3mm}}{\textbf{In}} & \multicolumn{1}{c}{\textbf{Out}} & \multicolumn{1}{c?{0.3mm}}{\textbf{In}} & \multicolumn{1}{c}{\textbf{Out}} & \multicolumn{1}{c?{0.3mm}}{\textbf{In}} & \multicolumn{1}{c}{\textbf{Out}} & \multicolumn{1}{c}{\textbf{In}}   \\      
     
     \textbf{Persona}& \multicolumn{1}{c}{\textbf{\#.}} & \multicolumn{1}{c?{0.3mm}}{\textbf{\#.}} & \multicolumn{1}{c}{\textbf{\#.}} & \multicolumn{1}{c?{0.3mm}}{\textbf{\#.}} & \multicolumn{1}{c}{\textbf{\#.}} & \multicolumn{1}{c?{0.3mm}}{\textbf{\#.}} & \multicolumn{1}{c}{\textbf{\#.}} & \multicolumn{1}{c?{0.3mm}}{\textbf{\#.}} & \multicolumn{1}{c}{\textbf{\#.}} & \multicolumn{1}{c}{\textbf{\#.}}   \\ 
    \toprule

\textbf{Adult}      & 0 & 12 & 18 & 288 & 42 & 54  & 67  & 77  & 12 & 20  \\
\textbf{Arts}       & 2 & 22 & 28 & 427 & 34 & 23  & 62  & 46  & 13 & 44  \\
\textbf{Business}   & 5 & 17 & 16 & 369 & 88 & 66  & 60  & 102 & 12 & 36  \\
\textbf{Computers}  & 0 & 26 & 21 & 484 & 41 & 16  & 62  & 94  & 9  & 19  \\
\textbf{Games}      & 0 & 31 & 16 & 325 & 33 & 131 & 87  & 205 & 21 & 12  \\
\textbf{Health}     & 3 & 19 & 27 & 435 & 51 & 76  & 79  & 109 & 20 & 42  \\
\textbf{Home}       & 0 & 13 & 8  & 301 & 35 & 21  & 63  & 156 & 19 & 17  \\
\textbf{Kids}       & 2 & 14 & 15 & 401 & 44 & 100 & 117 & 67  & 14 & 26  \\
\textbf{News}       & 2 & 20 & 23 & 434 & 46 & 115 & 69  & 49  & 17 & 36  \\
\textbf{Recreation} & 1 & 16 & 24 & 412 & 54 & 130 & 67  & 100 & 12 & 24  \\
\textbf{Reference}  & 1 & 22 & 15 & 329 & 47 & 109 & 69  & 83  & 7  & 38  \\
\textbf{Regional}   & 2 & 30 & 17 & 506 & 43 & 102 & 82  & 85  & 28 & 34  \\
\textbf{Science}    & 4 & 15 & 31 & 456 & 52 & 82  & 63  & 60  & 6  & 12  \\
\textbf{Shopping}   & 3 & 21 & 19 & 450 & 58 & 99  & 87  & 95  & 26 & 27  \\
\textbf{Society}    & 0 & 16 & 11 & 407 & 45 & 95  & 71  & 77  & 17 & 44  \\
\textbf{Sports}     & 3 & 22 & 18 & 477 & 37 & 102 & 93  & 109 & 23 & 41 \\

\bottomrule
\end{tabular}
}}
\label{tbl:Number_Bids_of_advertisers_syncing_CCPA}
\end{table}

%% file: docs/bids_from_syncing_vs_not.tex
\begin{table}[!h]
    \scriptsize
\centering
\caption{Average of bids data from advertisers in cookie syncing and not in cookie syncing under GDPR. CB, DM, OT, QC and NAI columns represent the count of cookie syncing events under CookieBot, Didomi, Onetrust, Quantcast and NAI for each advertiser.}
\adjustbox{max width=\textwidth}{
    \scalebox{0.9}{ \begin{tabular}{l?{0.3mm}ll?{0.3mm}ll?{0.3mm}ll?{0.3mm}ll?{0.3mm}ll} 
    \toprule
    & \multicolumn{10}{c}{\textbf{GDPR}} \\
     & \multicolumn{2}{c?{0.3mm}}{\textbf{CB}} & \multicolumn{2}{c?{0.3mm}}{\textbf{DM}}  & \multicolumn{2}{c?{0.3mm}}{\textbf{OT}}& \multicolumn{2}{c?{0.3mm}}{\textbf{QC}}& \multicolumn{2}{c}{\textbf{NAI}}   \\ 
& \multicolumn{1}{c}{\textbf{Sync}} & \multicolumn{1}{c?{0.3mm}}{\textbf{Not}}  & \multicolumn{1}{c}{\textbf{Sync}} & \multicolumn{1}{c?{0.3mm}}{\textbf{Not}} & \multicolumn{1}{c}{\textbf{Sync}} & \multicolumn{1}{c?{0.3mm}}{\textbf{Not}} & \multicolumn{1}{c}{\textbf{Sync}} & \multicolumn{1}{c?{0.3mm}}{\textbf{Not}} & \multicolumn{1}{c}{\textbf{Sync}} & \multicolumn{1}{c}{\textbf{Not}}   \\      
     
     \textbf{Persona}& \multicolumn{1}{c}{\textbf{Avg.}} & \multicolumn{1}{c?{0.3mm}}{\textbf{Avg.}} & \multicolumn{1}{c}{\textbf{Avg.}} & \multicolumn{1}{c?{0.3mm}}{\textbf{Avg.}} & \multicolumn{1}{c}{\textbf{Avg.}} & \multicolumn{1}{c?{0.3mm}}{\textbf{Avg.}} & \multicolumn{1}{c}{\textbf{Avg.}} & \multicolumn{1}{c?{0.3mm}}{\textbf{Avg.}} & \multicolumn{1}{c}{\textbf{Avg.}} & \multicolumn{1}{c}{\textbf{Avg.}}   \\ 
    \toprule

\textbf{Adult}      & --   & 0.37 & --   & --   & 0.06 & 0.00 & 0.02 & --   & 0.13 & 0.29  \\
\textbf{Arts}       & --   & 0.35 & 0.05 & --   & 0.02 & 0.00 & --   & --   & 0.13 & 0.29  \\
\textbf{Business}   & --   & 0.35 & --   & --   & 0.05 & 0.00 & 0.01 & --   & 0.16 & 0.21  \\
\textbf{Computers}  & --   & 0.35 & 0.05 & --   & 0.18 & 0.00 & 0.02 & --   & 0.12 & 0.37  \\
\textbf{Games}      & --   & 0.38 & 0.03 & --   & 0.02 & 0.00 & 0.02 & --   & 0.15 & 0.58  \\
\textbf{Health}     & --   & 0.35 & --   & --   & 0.02 & 0.00 & 0.03 & --   & 0.18 & 0.53  \\
\textbf{Home}       & --   & 0.38 & --   & 0.05 & --   & 0.00 & 0.31 & --   & 0.20 & 0.25  \\
\textbf{Kids}       & --   & 0.37 & 0.06 & --   & 0.02 & 0.00 & 0.01 & 0.02 & 0.14 & 0.39  \\
\textbf{News}       & 0.00 & 0.35 & --   & --   & 0.02 & 0.00 & 0.03 & --   & 0.13 & 0.25  \\
\textbf{Recreation} & --   & 0.37 & 0.06 & --   & 0.02 & 0.00 & 0.02 & --   & 0.17 & 0.41  \\
\textbf{Reference}  & --   & 0.37 & --   & --   & 0.03 & 0.00 & 0.03 & --   & 0.13 & 0.35  \\
\textbf{Regional}   & --   & 0.38 & --   & --   & 0.03 & 0.00 & 0.01 & --   & 0.19 & 0.32  \\
\textbf{Science}    & --   & 0.35 & --   & 0.41 & 0.02 & 0.00 & 0.03 & --   & 0.14 & 0.49  \\
\textbf{Shopping}   & --   & 0.37 & --   & 0.03 & 0.03 & 0.00 & 1.10 & --   & 0.10 & 0.20  \\
\textbf{Society}    & --   & 0.37 & --   & 0.05 & --   & 0.00 & 0.01 & --   & 0.09 & 0.38  \\
\textbf{Sports}     & --   & 0.37 & 0.05 & --   & 0.15 & 0.00 & 0.01 & --   & 0.25 & 0.19   \\

\bottomrule
\end{tabular}
}}
\label{tbl:Bids_of_advertisers_syncing_vs_not_GDPR}
\end{table}

\begin{table}[!h]
    \scriptsize
\centering
\caption{Average of bids data from advertisers in cookie syncing and not in cookie syncing under CCPA. CB, DM, OT, QC and NAI columns represent the count of cookie syncing events under CookieBot, Didomi, Onetrust, Quantcast and NAI for each advertiser.}
\adjustbox{max width=\textwidth}{
   \scalebox{0.9}{ \begin{tabular}{l?{0.3mm}ll?{0.3mm}ll?{0.3mm}ll?{0.3mm}ll?{0.3mm}ll} 
    \toprule
    & \multicolumn{10}{c}{\textbf{CCPA}} \\
     & \multicolumn{2}{c?{0.3mm}}{\textbf{CB}} & \multicolumn{2}{c?{0.3mm}}{\textbf{DM}}  & \multicolumn{2}{c?{0.3mm}}{\textbf{OT}}& \multicolumn{2}{c?{0.3mm}}{\textbf{QC}}& \multicolumn{2}{c}{\textbf{NAI}}   \\ 
& \multicolumn{1}{c}{\textbf{Sync}} & \multicolumn{1}{c?{0.3mm}}{\textbf{Not}}  & \multicolumn{1}{c}{\textbf{Sync}} & \multicolumn{1}{c?{0.3mm}}{\textbf{Not}} & \multicolumn{1}{c}{\textbf{Sync}} & \multicolumn{1}{c?{0.3mm}}{\textbf{Not}} & \multicolumn{1}{c}{\textbf{Sync}} & \multicolumn{1}{c?{0.3mm}}{\textbf{Not}} & \multicolumn{1}{c}{\textbf{Sync}} & \multicolumn{1}{c}{\textbf{Not}}   \\      
     
     \textbf{Persona}& \multicolumn{1}{c}{\textbf{Avg.}} & \multicolumn{1}{c?{0.3mm}}{\textbf{Avg.}} & \multicolumn{1}{c}{\textbf{Avg.}} & \multicolumn{1}{c?{0.3mm}}{\textbf{Avg.}} & \multicolumn{1}{c}{\textbf{Avg.}} & \multicolumn{1}{c?{0.3mm}}{\textbf{Avg.}} & \multicolumn{1}{c}{\textbf{Avg.}} & \multicolumn{1}{c?{0.3mm}}{\textbf{Avg.}} & \multicolumn{1}{c}{\textbf{Avg.}} & \multicolumn{1}{c}{\textbf{Avg.}}   \\ 
    \toprule

\textbf{Adult}      & --   & 0.37 & 0.16 & 0.19 & 0.32 & 0.00 & 0.44 & 1.64 & 0.40 & 0.16  \\
\textbf{Arts}       & 0.53 & 0.37 & 0.39 & 0.51 & 0.26 & 0.00 & 0.20 & 1.76 & 0.58 & 0.17  \\
\textbf{Business}   & 0.21 & 0.37 & 0.16 & 0.20 & 0.10 & 0.00 & 0.42 & 1.81 & 0.28 & 0.16  \\
\textbf{Computers}  & --   & 0.37 & 0.45 & --   & 0.18 & 0.00 & 0.26 & 1.87 & 0.64 & 0.18  \\
\textbf{Games}      & --   & 0.36 & 0.15 & 0.38 & 0.31 & 0.00 & 0.16 & 1.58 & 0.34 & 0.18  \\
\textbf{Health}     & 0.44 & 0.37 & 0.22 & 0.22 & 0.43 & 0.00 & 0.49 & 1.42 & 0.82 & 0.32  \\
\textbf{Home}       & 0.02 & 0.36 & 0.19 & 0.45 & 0.11 & 0.00 & 0.26 & 1.76 & 0.29 & 0.18  \\
\textbf{Kids}       & 0.03 & 0.36 & 0.21 & 0.03 & 0.17 & 0.00 & 0.46 & 1.95 & 0.35 & 0.15  \\
\textbf{News}       & 0.01 & 0.37 & 0.16 & 0.29 & 0.10 & 0.00 & 0.41 & 1.89 & 0.29 & 0.15  \\
\textbf{Recreation} & 0.14 & 0.36 & 0.15 & 0.33 & 0.09 & 0.00 & 0.49 & 1.55 & 0.47 & 0.16  \\
\textbf{Reference}  & 0.52 & 0.37 & 0.17 & 0.22 & 0.09 & 0.00 & 0.35 & 1.86 & 0.31 & 0.17  \\
\textbf{Regional}   & 0.03 & 0.36 & 0.22 & 0.27 & 0.57 & 0.03 & 0.36 & 1.41 & 0.38 & 0.16  \\
\textbf{Science}    & 0.11 & 0.37 & 0.18 & 0.22 & 0.17 & 0.00 & 0.37 & 2.17 & 0.36 & 0.16  \\
\textbf{Shopping}   & 0.39 & 0.31 & 0.19 & 0.26 & 0.20 & 0.00 & 0.40 & 1.89 & 0.31 & 0.17  \\
\textbf{Society}    & --   & 0.37 & 0.24 & 0.22 & 0.17 & 0.00 & 0.33 & 2.09 & 0.27 & 0.14  \\
\textbf{Sports}     & 0.32 & 0.33 & 0.17 & 0.18 & 0.26 & 0.00 & 0.23 & 1.64 & 0.34 & 0.17  \\

\bottomrule
\end{tabular}
}}
\label{tbl:Bids_of_advertisers_syncing_vs_not_CCPA}
\end{table}

%% file: docs/bids_from_syncing_vs_not_numbers.tex
\begin{table}[!h]
    \scriptsize
\centering
\caption{Number of bids data from advertisers in cookie syncing and not in cookie syncing under GDPR. CB, DM, OT, QC and NAI columns represent the count of cookie syncing events under CookieBot, Didomi, Onetrust, Quantcast and NAI for each advertiser.}
\adjustbox{max width=\textwidth}{
    \scalebox{0.9}{ \begin{tabular}{l?{0.3mm}ll?{0.3mm}ll?{0.3mm}ll?{0.3mm}ll?{0.3mm}ll} 
    \toprule
    & \multicolumn{10}{c}{\textbf{GDPR}} \\
     & \multicolumn{2}{c?{0.3mm}}{\textbf{CB}} & \multicolumn{2}{c?{0.3mm}}{\textbf{DM}}  & \multicolumn{2}{c?{0.3mm}}{\textbf{OT}}& \multicolumn{2}{c?{0.3mm}}{\textbf{QC}}& \multicolumn{2}{c}{\textbf{NAI}}   \\ 
& \multicolumn{1}{c}{\textbf{Sync}} & \multicolumn{1}{c?{0.3mm}}{\textbf{Not}}  & \multicolumn{1}{c}{\textbf{Sync}} & \multicolumn{1}{c?{0.3mm}}{\textbf{Not}} & \multicolumn{1}{c}{\textbf{Sync}} & \multicolumn{1}{c?{0.3mm}}{\textbf{Not}} & \multicolumn{1}{c}{\textbf{Sync}} & \multicolumn{1}{c?{0.3mm}}{\textbf{Not}} & \multicolumn{1}{c}{\textbf{Sync}} & \multicolumn{1}{c}{\textbf{Not}}   \\      
     
     \textbf{Persona}& \multicolumn{1}{c}{\textbf{\#.}} & \multicolumn{1}{c?{0.3mm}}{\textbf{\#.}} & \multicolumn{1}{c}{\textbf{\#.}} & \multicolumn{1}{c?{0.3mm}}{\textbf{\#.}} & \multicolumn{1}{c}{\textbf{\#.}} & \multicolumn{1}{c?{0.3mm}}{\textbf{\#.}} & \multicolumn{1}{c}{\textbf{\#.}} & \multicolumn{1}{c?{0.3mm}}{\textbf{\#.}} & \multicolumn{1}{c}{\textbf{\#.}} & \multicolumn{1}{c}{\textbf{\#.}}   \\ 
    \toprule

\textbf{Adult}      & 0 & 96  & 0 & 0 & 1 & 41 & 7  & 0 & 152 & 235  \\
\textbf{Arts}       & 0 & 101 & 1 & 0 & 3 & 42 & 0  & 0 & 172 & 253  \\
\textbf{Business}   & 0 & 96  & 0 & 0 & 2 & 47 & 16 & 0 & 163 & 301  \\
\textbf{Computers}  & 0 & 102 & 2 & 0 & 1 & 42 & 9  & 0 & 161 & 312  \\
\textbf{Games}      & 0 & 94  & 2 & 0 & 1 & 44 & 1  & 0 & 169 & 302  \\
\textbf{Health}     & 0 & 102 & 0 & 0 & 2 & 38 & 4  & 0 & 149 & 288  \\
\textbf{Home}       & 0 & 95  & 0 & 1 & 0 & 38 & 2  & 0 & 138 & 225  \\
\textbf{Kids}       & 0 & 96  & 1 & 0 & 2 & 39 & 10 & 2 & 168 & 298  \\
\textbf{News}       & 1 & 102 & 0 & 0 & 1 & 48 & 4  & 0 & 66  & 222  \\
\textbf{Recreation} & 0 & 96  & 1 & 0 & 2 & 42 & 6  & 0 & 160 & 240  \\
\textbf{Reference}  & 0 & 96  & 0 & 0 & 3 & 38 & 4  & 0 & 173 & 250  \\
\textbf{Regional}   & 0 & 94  & 0 & 0 & 3 & 40 & 14 & 0 & 186 & 274  \\
\textbf{Science}    & 0 & 102 & 0 & 1 & 1 & 39 & 5  & 0 & 168 & 287  \\
\textbf{Shopping}   & 0 & 89  & 0 & 1 & 3 & 37 & 1  & 0 & 54  & 221  \\
\textbf{Society}    & 0 & 94  & 0 & 1 & 0 & 40 & 16 & 0 & 121 & 307  \\
\textbf{Sports}     & 0 & 89  & 1 & 0 & 2 & 48 & 15 & 0 & 149 & 252   \\

\bottomrule
\end{tabular}
}}
\label{tbl:Number_Bids_of_advertisers_syncing_vs_not_GDPR}
\end{table}

\begin{table}[!h]
    \scriptsize
\centering
\caption{Number of bids data from advertisers in cookie syncing and not in cookie syncing under CCPA. CB, DM, OT, QC and NAI columns represent the count of cookie syncing events under CookieBot, Didomi, Onetrust, Quantcast and NAI for each advertiser.}
\adjustbox{max width=\textwidth}{
   \scalebox{0.9}{ \begin{tabular}{l?{0.3mm}ll?{0.3mm}ll?{0.3mm}ll?{0.3mm}ll?{0.3mm}ll} 
    \toprule
    & \multicolumn{10}{c}{\textbf{CCPA}} \\
     & \multicolumn{2}{c?{0.3mm}}{\textbf{CB}} & \multicolumn{2}{c?{0.3mm}}{\textbf{DM}}  & \multicolumn{2}{c?{0.3mm}}{\textbf{OT}}& \multicolumn{2}{c?{0.3mm}}{\textbf{QC}}& \multicolumn{2}{c}{\textbf{NAI}}   \\ 
& \multicolumn{1}{c}{\textbf{Sync}} & \multicolumn{1}{c?{0.3mm}}{\textbf{Not}}  & \multicolumn{1}{c}{\textbf{Sync}} & \multicolumn{1}{c?{0.3mm}}{\textbf{Not}} & \multicolumn{1}{c}{\textbf{Sync}} & \multicolumn{1}{c?{0.3mm}}{\textbf{Not}} & \multicolumn{1}{c}{\textbf{Sync}} & \multicolumn{1}{c?{0.3mm}}{\textbf{Not}} & \multicolumn{1}{c}{\textbf{Sync}} & \multicolumn{1}{c}{\textbf{Not}}   \\      
     
     \textbf{Persona}& \multicolumn{1}{c}{\textbf{\#.}} & \multicolumn{1}{c?{0.3mm}}{\textbf{\#.}} & \multicolumn{1}{c}{\textbf{\#.}} & \multicolumn{1}{c?{0.3mm}}{\textbf{\#.}} & \multicolumn{1}{c}{\textbf{\#.}} & \multicolumn{1}{c?{0.3mm}}{\textbf{\#.}} & \multicolumn{1}{c}{\textbf{\#.}} & \multicolumn{1}{c?{0.3mm}}{\textbf{\#.}} & \multicolumn{1}{c}{\textbf{\#.}} & \multicolumn{1}{c}{\textbf{\#.}}   \\ 
    \toprule

\textbf{Adult}      & 0 & 96  & 122 & 9  & 21 & 50 & 43  & 50 & 146 & 253  \\
\textbf{Arts}       & 2 & 96  & 147 & 4  & 13 & 49 & 34  & 42 & 126 & 273  \\
\textbf{Business}   & 6 & 97  & 149 & 1  & 12 & 48 & 105 & 17 & 132 & 268  \\
\textbf{Computers}  & 0 & 90  & 142 & 0  & 10 & 48 & 43  & 39 & 114 & 298  \\
\textbf{Games}      & 0 & 97  & 155 & 4  & 21 & 48 & 36  & 50 & 154 & 270  \\
\textbf{Health}     & 3 & 94  & 160 & 1  & 28 & 48 & 66  & 75 & 179 & 274  \\
\textbf{Home}       & 2 & 99  & 109 & 18 & 21 & 48 & 40  & 42 & 119 & 275  \\
\textbf{Kids}       & 2 & 98  & 147 & 25 & 29 & 48 & 98  & 42 & 187 & 320  \\
\textbf{News}       & 2 & 96  & 157 & 10 & 18 & 48 & 48  & 28 & 126 & 235  \\
\textbf{Recreation} & 1 & 95  & 126 & 7  & 16 & 46 & 56  & 58 & 128 & 238  \\
\textbf{Reference}  & 1 & 97  & 143 & 20 & 9  & 51 & 55  & 54 & 130 & 252  \\
\textbf{Regional}   & 2 & 100 & 165 & 13 & 28 & 50 & 45  & 50 & 138 & 270  \\
\textbf{Science}    & 4 & 96  & 165 & 2  & 6  & 48 & 54  & 33 & 134 & 267  \\
\textbf{Shopping}   & 3 & 107 & 142 & 10 & 26 & 48 & 62  & 42 & 178 & 258  \\
\textbf{Society}    & 0 & 96  & 150 & 1  & 20 & 46 & 48  & 20 & 127 & 264  \\
\textbf{Sports}     & 3 & 108 & 149 & 44 & 24 & 48 & 46  & 47 & 144 & 280   \\

\bottomrule
\end{tabular}
}}
\label{tbl:Number_Bids_of_advertisers_syncing_vs_not_CCPA}
\end{table}

%% file: docs/5_discussion.tex
\section{Discussion \& Limitations}
\label{sec:discussion} 

\subsection{Overall Trends Across CMPs}
Overall we note that under CMPs most personas receive higher bids compared to control when users opt-out of data processing and selling under GDPR and CCPA. 
The variability in bid values, particularly higher bids as compared to control, indicates that the leaked user interests are used to target ads to users, despite users' consent to opt-out of  data processing as part of the regulations. 
We also note that opt-out is not statistically different from opt-in.
The similarity in bid values for opt-in and opt-out indicates that the user consent in most cases does not have any effect on processing and selling of data. 
However some CMPs perform better than the others. 
For example, advertisers bidding behavior significantly changes under CCPA when the consent is conveyed through Didomi.

We note that advertisers participate in data sharing activities both at the server and the client side without user consent.
At the server side, we received higher bid values from advertisers, who we did not explicitly leak user interests; which indicates potential selling and sharing from advertisers who we leaked user data. 
%
%
%
At the client side, we notice that the advertisers share unique user identifiers in plain sight and share their data with as many other advertisers.

Advertiser-offered opt-out controls are also ineffective in curbing the processing and selling of user data despite user consent to opt-out. 
While advertisers at large do not honor their own opt-out controls, they slightly share less data as compared to the state-enforced regulations.
The following is one example set of client-side data sharing.

\lstset{basicstyle=\small\ttfamily,columns=fullflexible,escapeinside={<@}{@>}}
\begin{lstlisting}[linewidth=\columnwidth,breaklines=true]
Persona: Games, 									CMP: Onetrust, 
Location: California, 			Action: After Opt-out.
<@\textcolor{blue}{Example 1:}@>
Current domain: https://d.agkn.com/pixel!t=711!?dpids=130278&che=1660055860221&sk=209610804238003026648
Shared data: 1660055860221
Next URL: https://sync.mathtag.com/sync/img?redir=https%3A%2F%2Fd.agkn.com%2Fpixel%2F2618%2F%3Fche%3D1660055860221%26partner_id%3D[MM_UUID]
<@\textcolor{blue}{Example 2: }@>
Current domain: https://pixel.mathtag.com/sync/img?redir=https%3A%2F%2Ftoken.rubiconproject.com%2Ftoken%3Fpid%3D35912%26puid%3D%5BMM_UUID%5D'
Shared data: pid=35912
Next URL: https://token.rubiconproject.com/token?pid=35912&puid=229362f2-6b26-4c00-9d4c-bff76b65b98c
<@\textcolor{blue}{Example 3: }@>
Current domain: https://image4.pubmatic.com/AdServer/SPug?partnerID=161235&pmc=1&pr=https%3A%2F%2Fmedia.grid.bidswitch.net%2Fsync%3Ftp_id%3D27%26tp_uid%3D3B55B7EA-493D-4230-AF83-29D023DAA37C
Shared data: tp_uid=3B55B7EA-493D-4230-AF83-29D023DAA37C
Next URL: https://media.grid.bidswitch.net/sync?tp_id=27&tp_uid=3B55B7EA-493D-4230-AF83-29D023DAA37C

\end{lstlisting}

\subsection{Consent Handling by CMPs}
At a high level, CMPs block or allow cookies to enforce user consent \cite{cookiebot_functions,  didomi_cookies_storage}. 
As a first step, CMPs scan the website and identify all first and third-party cookies.
After identifying the cookies, CMPs classify them into essential (i.e., necessary for websites to operate) and non-essential (e.g., advertising, tracking, marketing, etc.) cookies. 
To identify necessary cookies, CMPs rely on information from the website developers 
To identify non-essential cookies, CMPs do not clearly disclose their techniques, but they might just be relying on information shared by advertising and tracking services about the purpose of their cookies (e.g., Google declares the purpose of their cookies \cite{google_ads_cookies}).
%
%
Many CMPs, such as OneTrust and Cookiebot, consolidate the information across websites and maintain database of cookies and their purposes \cite{onetrust_zero_code_cookie_block, cookiebot_functions}. 
Consolidating information allows CMPs to automatically identify essential and non-essential cookies on new websites.

CMPs typically take user consent and store it at the client side in first-party cookies.
In addition to blocking cookies, CMPs also block execution of elements (e.g., scripts, iframes, videos, images) that might exfiltrate non-essential cookies before user consent is stored. 
To give website developers more control in order to accurately enforce user consent and avoid breakage by blocking essential cookies, CMPs allow website developers to block or allow cookies.

There are two main ways in which advertisers might be able to process and share user information despite negative consent. 
One, website developers may inaccurately deploy CMPs. 
For example, tracking code may execute first before CMPs even have a chance to block cookies or website developers may inaccurately list non-essential cookies as essential. 
Two, advertisers may rely on side channel information to circumvent enforcement by CMPs. 
For example, advertisers may routinely change their cookies to avoid detection or they may rely on browser fingerprinting to track users \cite{Iqbal21FPInspectorSP}.
Recently, Toth et al. \cite{toth-pets22} found that CMPs themselves may violate regulations and that their configuration options may lead to non-compliance.

\subsection{Possible Recommendations}
Our findings in general cast a serious doubt on the effectiveness of regulations as a sole means of privacy protection.
Specifically, even after users opt-out through CMPs, their data may still be used and shared by advertiser.
%
%
Unfortunately, in order to fully protect privacy, users still need to rely on privacy-enhancing tools, such as ad/tracker blocking browser extensions and privacy-focused browsers (e.g., Brave Browser). 
However, not all users may utilize privacy-enhancing tools to protect their privacy. 

%
%
%
%

Website developers have an important role in enforcement of regulations. 
Specifically, they could deploy CMPs that are better at conveying and enforcing user consent. 
For example, research like ours could help inform the effectiveness of consent conveyance by different CMPs. 
Moving forward, we also recommended that CMPs, advertisers, website developers, and regulators should work together to define protocols for conveying and enforcing consent. 




\subsection{Limitations}

\textit{Main page and subpages:}
In our experiment, the framework exclusively accessed the main pages of each training website. Discrepancies may arise between visits to main pages and the inclusion of subpages \cite{aqeel2020landing}. Our analysis indicates that opting out on testing websites does not prevent tracking when solely accessing their main pages. Our primary aim was to illustrate that opting out does not guarantee evasion of tracking. The possibility of training personas by exploring subpages of training websites remains a potential avenue for future exploration. Notably, the focus of visiting testing websites was to gather bid data, rather than persona training, which is why we limited our visits to main pages.

\textit{CCPA applicability criteria:} 
CCPA applies to online services that meet its broad applicability criteria. 
Specifically, as per Section 1798.140 (c) (1), CCPA applies to online services, that have an annual revenue of more than \$ 25 million, annually sell data of more than 50K California residents, or earn more than 50\% of their revenue from the sale of personal data of California residents. 
Since most information required to determine applicability is not publicly available, it is challenging to determine the applicability criteria at scale \cite{dnsmpi}. 
Thus, for our study, we did not strictly follow the CCPA applicability criteria. 
However, it is noteworthy that the prevalent advertisers (Table \ref{tbl:Advertisers_cookie_syncing}) in our dataset are mostly large corporates with revenue exceeding hundreds of millions \cite{appnexux_revenue,pubmatic_revenue}.

\textit{Sample size:} 
In comparison to prior work that analyzed ad bidding (e.g., Cook et al. \cite{Cook20HeaderBiddingPETS} analyzed 25 websites), we analyze a substantially large number of websites (i.e., 352 that support Didomi, Quantcast, OneTrust, and CookieBot).  
We also repeat our measurements several times (i.e., 8 times) to reduce the chance of the sample size biasing our results. 
In future, researchers could further increase the sample size by incorporating websites that support various CMPs. 
%
%
We leave the non-trivial task of automating opt-outs from different CMPs at scale as future work. 
In future, researchers could also rely on alternative methodologies that use ad content e.g., \cite{musa2022atom}, to eliminate the need to rely on ad bidding altogether for inference of data usage and sharing.
Such techniques might allows researchers to audit online services at a much larger scale.

\textit{Server-side data sharing:} 
We rely on the insight, also leveraged by prior research \cite{Cook20HeaderBiddingPETS,Bashir16TrackingFlowsUsenix}, that the advertisers behavior is shaped by their pre-existing knowledge of the user.
Using that insight, we make an inference that higher bids from advertisers to which data was not leaked indicates the sharing of data from advertisers to which the data was leaked.
However, there may be other additional uncontrolled factors that might impact the bids. 

\textit{Automated data collection:} 
We rely on OpenWPM to automatically collect bids and use Amazon's EC2 cloud platform to simulate crawls from Germany and California. 
In order to more accurately simulate real users, we enable bot mitigation in OpenWPM and also randomly wait between 10--30 seconds after loading each website.
We also refrain from using public proxy servers, which may be black listed, and instead rely on Amazon EC2. 

%% file: docs/6_conclusion.tex
\section{Conclusions}

In this paper, we proposed a framework to audit regulatory compliance of online services at scale. 
We used the proposed framework to audit online advertising services on popular websites under GDPR and CCPA.
Despite users exercising their rights under GDPR and CCPA to opt-out of processing and selling of their data using CMPs, we find that advertisers process user data to possibly target them and also share it with their partners both at the server and the client side. 
However, we find that some CMPs perform better than the others, i.e., advertisers bidding behavior significantly changes when the consent is conveyed.
We also audited advertisers' self-proposed opt-out controls, i.e., NAI's opt-out, and found that they might be equally ineffective at curbing processing and selling of user data. 
Overall, our measurements sadly indicate that the regulations may not protect user privacy, and advertisers might be in potential violation of GDPR and CCPA. 
To foster follow-up research, we will also release our code and data set at the time of publication.

\label{sec:conclusion}

%% file: docs/7_appendix.tex

\appendix


%

\section{Analysis of Didomi, OneTrust, Quantcast and NAI}
\label{appendix:analysis_cmps}
\subsection{Didomi}
\label{appendix:didomi}

\input{tables/new/didomi-1.tex}

\input{tables/new/didomi-2.tex}

\noindent
\textit{\textbf{Data usage.}} 
We evaluate reduction in data usage by analyzing advertisers bidding behavior. 
Table \ref{table:didomi-table1} presents advertisers bidding on personas when users opt-out and opt-in through Didomi under GDPR and CCPA.
We note that under GDPR, 3 personas bid higher than the control despite users opting out. 
In all 3 instances the bid values are even higher than the sum of average and standard deviation of bid values in control (i.e., 3E-3 std.), exceeding to as much as 8.2 times higher for the Science persona. 
However, it is important to note that other than the Science persona the differences for other personas is only 0.01. 
We also note that advertisers did not return any bids for 6 personas and in 2 personas the bid values are less than that of the control (with the difference of 0.02). 
Similarly, under CCPA, 7 personas bid higher than the control and for 2 personas, i.e., Arts and Computer, bid values are higher than the sum of average and standard deviation (i.e., 0.16). 
We also note that for 7 personas average bid value is less than that of the control, however, the difference is at most 0.03.

Next, we analyze if there is statistically significant difference between advertisers bidding patterns when users opt-out or opt-in under GDPR and CCPA. 
%
%
It can be seen in Table \ref{table:didomi-table1} that under GDPR, advertisers bidding behavior does not significantly change when users opt-out, except for the personas where we did not receive any bidding. 
Under CCPA, advertisers bidding behavior significantly changes when users opt-in to data processing and sharing.
However, we note that the difference in advertisers behavior is small,  i.e., effect size is less than 0.3, except for Arts and Regional personas where the effect size is medium.

\noindent
\textit{\textbf{Server-side data sharing.}} 
We evaluate reduction in server-side data sharing by analyzing  bidding from advertisers to which we do not leak data. 
Table \ref{table:didomi-table2} presents bids from advertisers to which we did not explicitly leak user data. 
Under GDPR, 3 personas bid higher than the control and 2 personas bid less than the control. 
However, the difference in bid values is less than 0.02, except for Science where it is 8 times higher than the control.
Under CCPA, 6 personas bid higher and 8 personas bid less than the control. 
For two personas i.e., Arts and Computers the bid values are 2.5 times higher than the control and for Kids persona the bid value is 4.5 times less than the control.
%

\noindent
\textit{\textbf{Client-side data sharing.}} 
We evaluate reduction in client-side data sharing by measuring cookie syncing by advertisers in network traffic.
Table \ref{table:didomi-table2} presents the cookie syncing participation of advertisers.  
Under GDPR, we note that there is difference between advertisers cookie syncing behavior for opt-out and opt-in. 
Specifically, we experience cookie syncing events in 6 persona when we opt-out but we experience substantial more cookie syncing when we opt-in. 
On average there are 3 and 223 cookie syncing events per persona when users opt-out and opt-in, respectively.
Under CCPA, advertisers engage in cookie syncing events on all 16 personas regardless of whether the user opts-out or opts-in. 
However, number of cookie syncing events substantially increases from 42 to 170 when users opt-out. 

We further investigate cookie syncing frequency of individual advertisers. 
It can be seen from Table \ref{tbl:Advertisers_cookie_syncing} that advertisers participate in as many as 31 and 211 cookie syncing events when we opt-out under GDPR and CCPA with Didomi, respectively.

\noindent
\textbf{\textit{Takeaway.}} 
Significant decreases in data utilization and sharing are observed when users choose to opt-out under both GDPR and CCPA regulations. The decline in data usage is more pronounced under CCPA in comparison to GDPR. Conversely, the decline in client-side data sharing is more notable under GDPR than CCPA. Despite the utilization of Didomi for obtaining consent, which notably curbs targeting, it doesn't entirely eradicate it. This is evidenced by the continued higher bids on certain user personas and the involvement of advertising in cookie synchronization. Consequently, while achieving GDPR compliance might be possible through Didomi, ensuring CCPA compliance on the same websites could be more challenging.

\subsection{OneTrust}
\label{appendix:onetrust}
\input{tables/new/onetrust-1.tex}

\input{tables/new/onetrust-2.tex}

\textit{\textbf{Data usage.}}
We evaluate reduction in data usage by analyzing advertisers bidding behavior. 
Table \ref{table:onetrust-table1} presents advertisers bidding on personas when users opt-out and opt-in through OneTrust under GDPR and CCPA.
We note that under GDPR, 6 personas bid higher than control and 5 personas bid lower than control.
Except for Home and Shopping personas where bid values substantially exceed when users opt-out, the difference between bid values as compared to the control is only 0.01.
We also note that advertisers did not return any bids for the arts persona. 
In contrast, under CCPA, except for 2 personas, i.e., Business and Society, 14 personas receive bid values that are higher than that of the control.

Next, we analyze if there is statistically significant difference between advertisers bidding patterns when users opt-out or opt-in under GDPR and CCPA. 
%
%
It can be seen in Table \ref{table:onetrust-table1} that under GDPR, for all personas, with the exception of Recreation persona, there is no statistically significant difference between advertisers bidding behavior. 
Under CCPA, for 8 personas here is no statistically significant difference between advertisers bidding behavior.
For the other 8 personas, however, advertisers have statistically significant different advertising behavior (with medium effect size for 6 personas).

\noindent
\textit{\textbf{Server-side data sharing.}} 
We evaluate reduction in server-side data sharing by analyzing  bidding from advertisers to which we do not leak data. 
Table \ref{table:onetrust-table2} presents bids from advertisers to which we did not explicitly leak data. 
Under GDPR, 5 personas bid higher than the control and 3 personas bid less than the control. 
However, the difference in bid values is at most 0.01, except for Home and Science, where the bid values are substantially higher.
Under CCPA, all personas receive higher bid values than the average bid values in the control, where the bid values for Science persona are 2.3 times higher than that of the control.
%

\noindent
\textit{\textbf{Client-side data sharing.}}
We evaluate reduction in client-side data sharing by measuring cookie syncing by advertisers in network traffic.
Table \ref{table:onetrust-table2} presents the cookie syncing participation of advertisers.
Under GDPR, we note that there is difference between advertisers cookie syncing behavior for opt-out and opt-in. 
Advertisers participate in cookie syncing across all personas in both cases when users opt-out and opt-in, however, when users opt-in, the number of cookie syncing event almost doubles from 14 to 27. 
Under CCPA, advertisers engage in cookie syncing events on all 16 personas regardless of whether the user opts-out or opts-in. 
Surprisingly, we notice 15 more cookie syncing events when users opt-out of data sharing/selling under CCPA.

We further investigate cookie syncing frequency of individual advertisers. 
It can be seen from Table \ref{tbl:Advertisers_cookie_syncing} that advertisers participate in as many as 101 and 419 cookie syncing events when we opt-out under GDPR and CCPA with OneTrust, respectively.

\noindent
\textbf{\textit{Takeaway.}} 
Discrepancies in the behavior of advertisers were observed in relation to GDPR and CCPA when users choose to opt-out via OneTrust. Notably, opting out did not result in a statistically notable variation in data utilization under GDPR; however, it did lead to such a difference under CCPA. There was a greater occurrence of both server-side and client-side data sharing under CCPA in comparison to GDPR. Surprisingly, there was an increased synchronization of cookies, signifying the sharing of client-side data, under CCPA as opposed to GDPR. Consequently, it can be inferred that ensuring compliance with both GDPR and CCPA might not be guaranteed on the same websites.

\subsection{Quantcast}
\label{appendix:quantcast}

\input{tables/new/quantcast-1.tex}

\input{tables/new/quantcast-2.tex}

\noindent
\textbf{\textit{Data Usage.}} 
We evaluate reduction in data usage by analyzing advertisers bidding behavior. 
Table \ref{table:quantcast-table1} presents advertisers bidding behavior for Quantcast.
We note that under both GDPR and CCPA, all personas bid higher than control and in 9 such personas under GDPR and 1 personas under CCPA the bid value is higher than the average and standard deviation of the control (i.e., 0.08 std. under GDPR and 0.18 std. under CCPA). 
For Games, Health, and Science personas the bid values are higher than twice the bid value for the control persona under GDPR.
For Health persona, the bid values are 3 times higher than that of the control under CCPA.

Next, we analyze if there is statistically significant difference between advertisers bidding patterns when users opt-out or opt-in to data processing and sharing/selling.
It can be seen in Table \ref{table:quantcast-table1} that under GDPR, for 11 personas advertisers behavior does not significantly changes regardless of whether users opt-out or opt-in. 
For 5 personas there are statistically significant differences in advertisers bidding, however, for all the cases the effect size is small. 
Under CCPA, for 14 personas advertisers behavior does not significantly changes regardless of whether users opt-out or opt-in. 
For Adult and Games personas there are statistically significant changes in advertisers behavior but the effect size is small.

\noindent
\textit{\textbf{Server-side data sharing.}} 
We evaluate reduction in server-side data sharing by analyzing bidding from advertisers to which we do not leak data. 
Table \ref{table:quantcast-table2} presents bids from advertisers to which we did not explicitly leak data.
Under GDPR, for 10 personas advertisers bid higher than the control and in 2 such cases the bid values are more than 2 times higher than that of the control. 
Only 4 personas receive bid values that are less than that of the control but the difference is only 0.04 at max. 
%
Under CCPA, 13 personas receive higher bid values, where bid values for Health persona are 3 times higher than that of the control. 
Only 2 personas receive bid values that are less than that of the control but the difference is only 0.01.

\noindent
\textit{\textbf{Client-side data sharing.}} 
We evaluate reduction in client-side data sharing by measuring cookie syncing by advertisers in network traffic.
Table \ref{table:quantcast-table2} presents the cookie syncing participation of advertisers for Quantcast. 
Under GDPR, we note that there is substantial difference between advertisers cookie syncing behavior for opt-out and opt-in. 
Specifically, we experience cookie syncing events in all persona when we opt-out but we experience substantial more cookie syncing when we opt-in. 
On average there are 45 and 169 cookie syncing events per persona when users opt-out and opt-in, respectively.
Under CCPA, advertisers engage in cookie syncing events on all 16 personas, except for Kids persona when users opt-in, regardless of whether the user opts-out or opts-in. 
However, on average the difference in cookie syncing events between opt-out and opt-in is not significantly large as compared to GDPR. 
On average there are 130 and 174 cookie syncing events per persona when users opt-out and opt-in, respectively.

We further investigate cookie syncing frequency of individual advertisers. 
It can be seen from Table \ref{tbl:Advertisers_cookie_syncing} that advertisers participate in as many as 170 and 317 cookie syncing events when we opt-out under GDPR and CCPA with Quantcast, respectively.

\noindent
\textbf{\textit{Takeaway.}} 
Significant decreases in data usage and sharing are not readily apparent when users choose to opt-out. Within the framework of GDPR, there are noticeable shifts in advertisers' bidding patterns for 5 different user personas, although these changes are only minor in terms of impact. When users opt-in under GDPR regulations, a notable increase in cookie syncing events becomes evident. This suggests that ensuring compliance with both GDPR and CCPA on the same websites might not be assured.

\subsection{NAI}
\label{appendix:nai}

\input{tables/new/nai-1.tex}

\input{tables/new/nai-2.tex}

We also analyze advertisers bidding behavior by exercising advertisers offered opt-out controls.
Specifically, we opt-out through National Advertising Initiative (NAI) -- an advertising consortium -- provided controls to opt-out of targeted advertising \cite{nai_opt_out}.
Similar to state-enforced regulations, i.e., GDPR and CCPA, we evaluate whether opt-out through NAI reduces targeting, whether the reduction is statistically different, and whether advertisers share data without user consent.

\textit{\textbf{Data usage.}}
We evaluate reduction in data usage by analyzing advertisers bidding behavior. 
Table \ref{table:nai-table1} presents advertisers bidding on personas when users opt-out and opt-in through NAI under GDPR and CCPA.
We note that under GDPR, all 16 personas bid higher than that of the control.
Note that advertisers do bid for control personas but with 0 values. 
Under CCPA, 5 personas bid higher than the control and 9 personas bid lower than control. 
For Health persona, the bid values are 3.14 times higher than the control. 
%
%
It can also be seen in Table \ref{table:nai-table1} that under both GDPR and CCPA, for all personas there is no statistically significant difference between advertisers bidding behavior.

\noindent
\textit{\textbf{Server-side data sharing.}} 
We evaluate reduction in server-side data sharing by analyzing bidding from advertisers to which we do not leak data. 
Table \ref{table:nai-table2} presents bids from advertisers to which we did not explicitly leak data.
Under GDPR, 12 personas receive bids with 0 values but 4 personas still receive higher bids than the control when users opt-out. 
Under CCPA, however, all 16 personas receive lower bids than the control.

\noindent
\textit{\textbf{Client-side data sharing.}} 
We evaluate reduction in client-side data sharing by measuring cookie syncing by advertisers in network traffic.
Table \ref{table:nai-table2} presents the cookie syncing participation of advertisers for NAI. 
Under GDPR, there is no substantial difference between advertisers cookie syncing behavior for opt-out and opt-in. 
In fact, there are 3 more cookie syncing events on average when users opt-out
Under CCPA, advertisers engage in cookie syncing events on all personas regardless of whether the user opts-out or opts-in. 
However, when users opt-in, the number of cookie syncing events on average doubles from 26 to 48.

We further investigate cookie syncing frequency of individual advertisers. 
It can be seen from Table \ref{tbl:Advertisers_cookie_syncing} that advertisers participate in as many as 107 and 190 cookie syncing events when we opt-out under GDPR and CCPA with NAI, respectively.


\textbf{\textit{Takeaway.}}  
The utilization of advertisers' data remains relatively stable. Nevertheless, advertisers submit lower bids under CCPA in contrast to GDPR. Correspondingly, we have observed a marked decrease in data sharing, both on the server-side and client-side, under CCPA. This suggests that achieving compliance with both GDPR and CCPA might not be guaranteed on identical websites.

\section{Advertisers Bidding Behavior with pre-opt-out}
\label{appendix:pre-opt-out}
Under GDPR processing personal data is prohibited, unless the data subject has consented to the processing (Article 6). 
However, under CCPA, data selling and sharing should immediately stop once consumers opt-out (Section 798.120 (a), Section 7013 (a)). 
%
Thus to eliminate the impact of data collection and sharing prior to opting-out, we conduct additional experiments where we opt-out prior to simulating personas. 


Table \ref{table:GDPR-bidding-mean-pre-opt-out} and Table \ref{table:CCPA-bidding-mean-pre-opt-out} present the ad bidding under GDPR and CCPA. Under GDPR, we note that advertisers bid higher for most personas than control across all four CMPs. In several instances the bid values are even higher than the sum of average and standard deviation of the bid values in control persona. Under CCPA, however, we note varying trends across CMPs. For Cookiebot, OneTrust, and Quantcast 16, 7, and 4 personas receive  higher bid values from advertisers despite opting out, respectively. In the case of Didomi, only 1 persona receives higher bid values.


Table \ref{table:GDPR-syncing-pre-opt-out} and Table \ref{table:CCPA-syncing-pre-opt-out} present the cookie syncing events from advertisers under GDPR and CCPA. We note that advertisers participate in cookie syncing events despite users opting out under both GDPR and CCPA.

\textbf{\textit{Takeaway.}} Similar to post opt-out, we note that under GDPR advertisers continue to use data even when we opt-out prior to collecting bids. Under CCPA, as compared to GDPR, less number of personas receive higher bid values than that of the control. However, there are still several personas where advertisers continue to bid higher than the control. In the case of client side data sharing, we did not notice any reduction in cookie syncing under both GDPR and CCPA.

\input{docs/pre-opt-out_syncing_gdpr.tex}

\input{docs/pre-opt-out_syncing_ccpa.tex}

\input{docs/pre-opt-out.tex}

%% file: tables/new/didomi-1.tex
\begin{table}[!t]
    \scriptsize
    \centering
    \caption{Ad bidding under GDPR and CCPA after opt-out (Out) and opt-in (In) with Didomi. Avg. column represents the mean of all bid value. \textcolor{highbids}{Light red} and \textcolor{lowbids}{Light blue} indicate bid values that are higher and lower than Control's avg., respectively. \textcolor{higherbids}{Dark red} and \textcolor{lowerbids}{Dark blue} indicate bid values that are Control's avg. $\pm$ std., respectively. Column p-val. and Eff. represent p-value and effect size, respectively}
    \adjustbox{max width=\columnwidth}{
    \begin{tabular}{l?{0.3mm}cc|cc?{0.3mm}cc|cc} 
        \toprule
                        & \multicolumn{4}{c?{0.3mm}}{\textbf{GDPR}}                                                  & \multicolumn{4}{c}{\textbf{CCPA}}                                                   \\ 
    
                        & \textbf{Out} & \textbf{In} & \multicolumn{2}{c?{0.3mm}}{\textbf{Stat. Test}} & \textbf{Out} & \textbf{In} & \multicolumn{2}{c}{\textbf{Stat. Test}}  \\ 
    
    \textbf{Persona}    & \textbf{Avg.}    & \textbf{Avg.}   & \textbf{p-val.} & \textbf{Eff.}        & \textbf{Avg.}    & \textbf{Avg.}   & \textbf{p-val.} & \textbf{Eff.}         \\ 
    \toprule
\textbf{Adult}      & --   & 0.11 & -- & --      & 0.17 {\cellcolor{lowbids}}  & 0.17 & 7.5E-7 & 0.20  \\
\textbf{Arts}       & 0.05 & 0.16 {\cellcolor{highbids}} & 6.00 & 0.01     & 0.39 {\cellcolor{higherbids}}& 0.18 {\cellcolor{highbids}} & 6.3E-18 & 0.31  \\
\textbf{Business}   & --   & 0.11 & -- & --      & 0.16 {\cellcolor{lowbids}}& 0.19 {\cellcolor{highbids}} & 3.8E-12 & 0.26  \\
\textbf{Computers}  & 0.05 & 0.11 & 6.31 & 0.01     & 0.45 {\cellcolor{higherbids}}& 0.15 {\cellcolor{lowbids}}& 4.2E-17 & 0.29  \\
\textbf{Games}      & 0.03 {\cellcolor{lowerbids}}  & 0.12 {\cellcolor{highbids}} & 2.52 & 0.03     & 0.15 {\cellcolor{lowbids}}& 0.16 {\cellcolor{lowbids}}& 3.6E-10 & 0.25  \\
\textbf{Health}     & --   & 0.11 & -- & --      & 0.22 {\cellcolor{highbids}}  & 0.20 {\cellcolor{highbids}} & 1.9E-11 & 0.23  \\
\textbf{Home}       & 0.05 & 0.12 {\cellcolor{highbids}} & 7.44 & 3E-3 & 0.23 {\cellcolor{highbids}}  & 0.16 {\cellcolor{lowbids}}& 9.2E-14 & 0.29  \\
\textbf{Kids}       & 0.06 {\cellcolor{higherbids}}  & 0.14 {\cellcolor{highbids}} & 7.52 & 3E-3 & 0.18 & 0.18 {\cellcolor{highbids}} & 2.2E-5 & 0.15  \\
\textbf{News}       & --   & 0.09 {\cellcolor{lowbids}}& -- & --      & 0.17 {\cellcolor{lowbids}}& 0.17 & 4E-10 & 0.23  \\
\textbf{Recreation} & 0.06 {\cellcolor{higherbids}}  & 0.11 & 5.62 & 0.01     & 0.16 {\cellcolor{lowbids}}& 0.27 {\cellcolor{highbids}} & 3.9E-12 & 0.27  \\
\textbf{Reference}  & --   & 0.13 {\cellcolor{highbids}} & -- & --     & 0.17 {\cellcolor{lowbids}}& 0.16 {\cellcolor{lowbids}}& 1.7E-9 & 0.23  \\
\textbf{Regional}   & --   & 0.13 {\cellcolor{highbids}} & -- & --     & 0.23 {\cellcolor{highbids}}  & 0.14 {\cellcolor{lowbids}}& 1.9E-27 & 0.37  \\
\textbf{Science}    & 0.41 {\cellcolor{higherbids}}  & 0.12 {\cellcolor{highbids}} & 1.12 & 0.04  & 0.18 & 0.23 {\cellcolor{highbids}} & 8.7E-14 & 0.27  \\
\textbf{Shopping}   & 0.03 {\cellcolor{lowerbids}}  & 0.10 {\cellcolor{lowbids}}& 7.23 & 4E-3 & 0.20 {\cellcolor{highbids}}  & 0.18 {\cellcolor{highbids}} & 1.3E-11 & 0.23  \\
\textbf{Society}    & 0.05 & 0.12 {\cellcolor{highbids}} & 6.59 & 0.01     & 0.24 {\cellcolor{highbids}}  & 0.17 & 2.4E-12 & 0.26  \\
\textbf{Sports}     & 0.05 & 0.13 {\cellcolor{highbids}} & 7.98 & 1.08E-4 & 0.17 {\cellcolor{lowbids}}& 0.18 {\cellcolor{highbids}} & 2.5E-12 & 0.23  \\
\midrule
\textbf{Control}    & 0.05 & 0.11 & 6.46 & 0.01     & 0.18 & 0.17 & 1.3E-20 & 0.33               \\
\bottomrule
\end{tabular}
}
\vspace{-1em}
\label{table:didomi-table1}
\end{table}

%% file: tables/new/didomi-2.tex
\begin{table}[!t]
    \scriptsize
    \centering
    \caption{Ad bidding and cookie syncing under GDPR and CCPA after opt-out (Out) and opt-in (In) with Didomi. Avg. column represents the mean of all bid value from advertisers who did not bid or appear when we simulated personas but appeared and bid after we opt-out. Out and In under C-Sync. represent number of cookie syncing events after opt-out and opt-in, respectively.}
    \begin{tabular}{l?{0.3mm}c|cc?{0.3mm}c|cc} 
    \toprule
                        & \multicolumn{3}{c?{0.3mm}}{\textbf{GDPR}}                              & \multicolumn{3}{c}{\textbf{CCPA}}                               \\ 
    
                        & \textbf{Out} & \multicolumn{2}{c?{0.3mm}}{\textbf{C-Sync.}} & \textbf{Out} & \multicolumn{2}{l}{\textbf{C-Sync.}}  \\ 
    
    \textbf{Persona}    & \textbf{Avg.}    & \textbf{Out} & \textbf{In}           & \textbf{Avg.}    & \textbf{Out} & \textbf{In}            \\ 
\toprule
\textbf{Adult}      & --   & 0  & 205 & 0.12 {\cellcolor{lowbids}}& 36 & 136  \\
\textbf{Arts}       & 0.05 & 1  & 209 & 0.45 {\cellcolor{higherbids}}& 43 & 163  \\
\textbf{Business}   & --   & 0  & 232 & 0.15 {\cellcolor{lowbids}}& 21 & 129  \\
\textbf{Computers}  & 0.05 & 15 & 276 & 0.45 {\cellcolor{higherbids}}& 38 & 161  \\
\textbf{Games}      & 0.03 {\cellcolor{lowerbids}}& 17 & 247 & 0.15 {\cellcolor{lowbids}}& 30 & 178  \\
\textbf{Health}     & --   & 0  & 223 & 0.16 {\cellcolor{lowbids}}& 86 & 222  \\
\textbf{Home}       & 0.05 & 0  & 153 & 0.23 {\cellcolor{highbids}}& 19 & 155  \\
\textbf{Kids}       & 0.06 {\cellcolor{higherbids}}& 2  & 252 & 0.04 {\cellcolor{lowbids}}& 65 & 169  \\
\textbf{News}       & --   & 0  & 251 & 0.18 & 31 & 162  \\
\textbf{Recreation} & 0.06 {\cellcolor{higherbids}}& 2  & 208 & 0.14 {\cellcolor{lowbids}}& 47 & 232  \\
\textbf{Reference}  & --   & 0  & 221 & 0.17 {\cellcolor{lowbids}}& 32 & 143  \\
\textbf{Regional}   & --   & 0  & 203 & 0.22 {\cellcolor{highbids}}& 36 & 144  \\
\textbf{Science}    & 0.41 {\cellcolor{higherbids}}& 0  & 246 & 0.16 {\cellcolor{lowbids}}& 52 & 219  \\
\textbf{Shopping}   & 0.03 {\cellcolor{lowerbids}}& 0  & 223 & 0.19 {\cellcolor{highbids}}& 69 & 185  \\
\textbf{Society}    & 0.05 & 0  & 204 & 0.24 {\cellcolor{highbids}}& 31 & 161  \\
\textbf{Sports}     & 0.05 & 3  & 218 & 0.18 & 37 & 162  \\
\midrule
\textbf{Control}    & 0.05 & -- & --  & 0.18 & -- & --      \\ 
\midrule
\textbf{Average} & -- & 3 & 223 & -- & 42 & 170 \\

\bottomrule
\end{tabular}
\vspace{-2em}
\label{table:didomi-table2}
\end{table}

%% file: tables/new/onetrust-1.tex
\begin{table}[!t]
    \scriptsize
    \centering
    \caption{Ad bidding under GDPR and CCPA after opt-out (Out) and opt-in (In) with OneTrust. Avg. column represents the mean of all bid value. \textcolor{highbids}{Light red} and \textcolor{lowbids}{Light blue} indicate bid values that are higher and lower than Control's avg., respectively. \textcolor{higherbids}{Dark red} and \textcolor{lowerbids}{Dark blue} indicate bid values that are Control's avg. $\pm$ std., respectively. Column p-val. and Eff. represent p-value and effect size, respectively}
    \adjustbox{max width=\columnwidth}{
    \begin{tabular}{l?{0.3mm}cc|cc?{0.3mm}cc|cc} 
        \toprule
                        & \multicolumn{4}{c?{0.3mm}}{\textbf{GDPR}}                                                  & \multicolumn{4}{c}{\textbf{CCPA}}                                                   \\ 
    
                        & \textbf{Out} & \textbf{In} & \multicolumn{2}{c?{0.3mm}}{\textbf{Stat. Test}} & \textbf{Out} & \textbf{In} & \multicolumn{2}{c}{\textbf{Stat. Test}}  \\ 
    
    \textbf{Persona}    & \textbf{Avg.}    & \textbf{Avg.}   & \textbf{p-val.} & \textbf{Eff.}        & \textbf{Avg.}    & \textbf{Avg.}   & \textbf{p-val.} & \textbf{Eff.}         \\ 
    \toprule
\textbf{Adult}      & 0.02 & 0.01 {\cellcolor{lowerbids}}& 8.00 & 0 & 1.09 {\cellcolor{highbids}}& 0.46 {\cellcolor{lowbids}}& 2.3E-6 & 0.35  \\
\textbf{Arts}       & --   & 0.04 {\cellcolor{higherbids}}& -- & -- & 1.06 {\cellcolor{highbids}}& 1.14 {\cellcolor{highbids}}& 4.63     & 0.05  \\
\textbf{Business}   & 0.01 {\cellcolor{lowerbids}}& 0.06 {\cellcolor{higherbids}}& 0.28 & 0.42 & 0.61 {\cellcolor{lowbids}}& 0.19 {\cellcolor{lowbids}}& 5.5E-11 & 0.50  \\
\textbf{Computers}  & 0.02 & 0.01 {\cellcolor{lowerbids}}& 2.01 & 0.28& 1.02 {\cellcolor{highbids}}& 1.40 {\cellcolor{highbids}}& 4.50     & 0.05  \\
\textbf{Games}      & 0.02 & 0.03 {\cellcolor{higherbids}}& 8.00 & 0 & 0.99 {\cellcolor{highbids}}& 0.81 {\cellcolor{highbids}}& 0.47     & 0.11  \\
\textbf{Health}     & 0.03 {\cellcolor{higherbids}}& 0.15 {\cellcolor{higherbids}}& 0.11 & 0.78 & 0.98 {\cellcolor{highbids}}& 0.70 {\cellcolor{highbids}}& 7.7E-6 & 0.31  \\
\textbf{Home}       & 0.31 {\cellcolor{higherbids}}& 0.05 {\cellcolor{higherbids}}& 2.52 & 0.37 & 1.03 {\cellcolor{highbids}}& 1.61 {\cellcolor{highbids}}& 2.23     & 0.10  \\
\textbf{Kids}       & 0.01 {\cellcolor{lowerbids}}& 0.02 & 0.62 & 0.44 & 0.91 {\cellcolor{highbids}}& 0.66 {\cellcolor{lowbids}}& 1.2E-4 & 0.23  \\
\textbf{News}       & 0.03 {\cellcolor{higherbids}}& 0.46 {\cellcolor{higherbids}}& 0.32 & 0.64 & 0.95 {\cellcolor{highbids}}& 0.75 {\cellcolor{highbids}}& 0.66     & 0.10  \\
\textbf{Recreation} & 0.02 & 0.14 {\cellcolor{higherbids}}& 0.02 & 0.81 & 1.03 {\cellcolor{highbids}}& 0.46 {\cellcolor{lowbids}}& 3.6E-11 & 0.39  \\
\textbf{Reference}  & 0.03 {\cellcolor{higherbids}}& 0.20 {\cellcolor{higherbids}}& 3.07 & 0.31 & 1.10 {\cellcolor{highbids}}& 0.72 {\cellcolor{highbids}}& 3.7E-5 & 0.27  \\
\textbf{Regional}   & 0.01 {\cellcolor{lowerbids}}& 0.02 & 2.43 & 0.24 & 0.91 {\cellcolor{highbids}}& 0.35 {\cellcolor{lowbids}}& 6E-11 & 0.43  \\
\textbf{Science}    & 0.03 {\cellcolor{higherbids}}& 0.09 {\cellcolor{higherbids}}& 1.75 & 0.42 & 1.05 {\cellcolor{highbids}}& 0.68 {\cellcolor{lowbids}}& 0.16     & 0.17  \\
\textbf{Shopping}   & 1.10 {\cellcolor{higherbids}}& 0.13 {\cellcolor{higherbids}}& 3.84 & 0.35 & 1.00 {\cellcolor{highbids}}& 0.77 {\cellcolor{highbids}}& 0.57     & 0.10  \\
\textbf{Society}    & 0.01 {\cellcolor{lowerbids}}& 0.02 & 2.23 & 0.25 & 0.85 {\cellcolor{lowbids}}& 0.20 {\cellcolor{lowbids}}& 1.8E-5 & 0.37  \\
\textbf{Sports}     & 0.01 {\cellcolor{lowerbids}}& 0.11 {\cellcolor{higherbids}}& 0.16 & 0.47 & 0.94 {\cellcolor{highbids}}& 0.77 {\cellcolor{highbids}}& 0.61     & 0.10  \\
\midrule
\textbf{Control}    & 0.02 & 0.02 & 8.00 & 0 & 0.88 & 0.69 & 1.94     & 0.09  \\

\bottomrule
\end{tabular}
}
\vspace{-1em}
\label{table:onetrust-table1}
\end{table}

%% file: tables/new/onetrust-2.tex
\begin{table}[!t]
    \scriptsize
    \centering
    \caption{Ad bidding and cookie syncing under GDPR and CCPA after opt-out (Out) and opt-in (In) with OneTrust. Avg. column represents the mean of all bid value from advertisers who did not bid or appear when we simulated personas but appeared and bid after we opt-out. Out and In under C-Sync. represent number of cookie syncing events after opt-out and opt-in, respectively.}
    \begin{tabular}{l?{0.3mm}c|cc?{0.3mm}c|cc} 
    \toprule
                        & \multicolumn{3}{c?{0.3mm}}{\textbf{GDPR}}                              & \multicolumn{3}{c}{\textbf{CCPA}}                               \\ 
    
                        & \textbf{Out} & \multicolumn{2}{c?{0.3mm}}{\textbf{C-Sync.}} & \textbf{Out} & \multicolumn{2}{l}{\textbf{C-Sync.}}  \\ 
    
    \textbf{Persona}    & \textbf{Avg.}    & \textbf{Out} & \textbf{In}           & \textbf{Avg.}    & \textbf{Out} & \textbf{In}            \\ 
\toprule
\textbf{Adult}      & 0.02 & 19 & 20 & 1.64 {\cellcolor{highbids}}& 85  & 128  \\
\textbf{Arts}       & --   & 0  & 22 & 1.43 {\cellcolor{highbids}}& 53  & 71   \\
\textbf{Business}   & 0.01 {\cellcolor{lowerbids}}& 24 & 49 & 1.29 {\cellcolor{highbids}}& 125 & 77   \\
\textbf{Computers}  & 0.02 & 29 & 38 & 1.78 {\cellcolor{highbids}}& 86  & 84   \\
\textbf{Games}      & 0.02 & 16 & 22 & 1.50 {\cellcolor{highbids}}& 95  & 58   \\
\textbf{Health}     & 0.03 {\cellcolor{higherbids}}& 1  & 25 & 1.37 {\cellcolor{highbids}}& 115 & 120  \\
\textbf{Home}       & 0.31 {\cellcolor{higherbids}}& 15 & 8  & 1.63 {\cellcolor{highbids}}& 110 & 56   \\
\textbf{Kids}       & 0.02 & 12 & 27 & 1.95 {\cellcolor{highbids}}& 124 & 116  \\
\textbf{News}       & 0.02 & 8  & 27 & 1.76 {\cellcolor{highbids}}& 98  & 47   \\
\textbf{Recreation} & 0.02 & 7  & 13 & 1.55 {\cellcolor{highbids}}& 86  & 77   \\
\textbf{Reference}  & 0.03 {\cellcolor{higherbids}}& 11 & 45 & 1.86 {\cellcolor{highbids}}& 67  & 89   \\
\textbf{Regional}   & 0.01 {\cellcolor{lowerbids}}& 33 & 30 & 1.36 {\cellcolor{highbids}}& 85  & 53   \\
\textbf{Science}    & 0.03 {\cellcolor{higherbids}}& 1  & 13 & 2.05 {\cellcolor{higherbids}}& 114 & 106  \\
\textbf{Shopping}   & 1.10 {\cellcolor{higherbids}}& 14 & 14 & 1.77 {\cellcolor{highbids}}& 136 & 76   \\
\textbf{Society}    & 0.01 {\cellcolor{lowerbids}}& 10 & 21 & 1.83 {\cellcolor{highbids}}& 89  & 87   \\
\textbf{Sports}     & 0.02 & 20 & 54 & 1.51 {\cellcolor{highbids}}& 74  & 57   \\
\midrule
\textbf{Control}    & 0.02 & -- & -- & 0.88 & --  & --      \\ 
\midrule
\textbf{Average} & -- & 14 & 27 & -- & 96 & 81 \\

\bottomrule
\end{tabular}
\vspace{-1em}
\label{table:onetrust-table2}
\end{table}

%% file: tables/new/quantcast-1.tex
\begin{table}[!t]
    \scriptsize
    \centering
    \caption{Ad bidding under GDPR and CCPA after opt-out (Out) and opt-in (In) with Quantcast. Avg. column represents the mean of all bid value. \textcolor{highbids}{Light red} and \textcolor{lowbids}{Light blue} indicate bid values that are higher and lower than Control's avg., respectively. \textcolor{higherbids}{Dark red} and \textcolor{lowerbids}{Dark blue} indicate bid values that are Control's avg. $\pm$ std., respectively. Column p-val. and Eff. represent p-value and effect size, respectively}
    \adjustbox{max width=\columnwidth}{
    \begin{tabular}{l?{0.3mm}cc|cc?{0.3mm}cc|cc} 
        \toprule
                        & \multicolumn{4}{c?{0.3mm}}{\textbf{GDPR}}                                                  & \multicolumn{4}{c}{\textbf{CCPA}}                                                   \\ 
    
                        & \textbf{Out} & \textbf{In} & \multicolumn{2}{c?{0.3mm}}{\textbf{Stat. Test}} & \textbf{Out} & \textbf{In} & \multicolumn{2}{c}{\textbf{Stat. Test}}  \\ 
    
    \textbf{Persona}    & \textbf{Avg.}    & \textbf{Avg.}   & \textbf{p-val.} & \textbf{Eff.}        & \textbf{Avg.}    & \textbf{Avg.}   & \textbf{p-val.} & \textbf{Eff.}         \\ 
    \toprule
\textbf{Adult}      & 0.22 {\cellcolor{highbids}}& 0.15 & 0.31     & 0.07 & 0.25 {\cellcolor{highbids}}& 0.16 {\cellcolor{lowbids}}& 0.01 & 0.12  \\
\textbf{Arts}       & 0.22 {\cellcolor{highbids}}& 0.14 {\cellcolor{lowbids}}& 0.27     & 0.07 & 0.30 {\cellcolor{highbids}}& 0.18 & 5.17 & 0.02  \\
\textbf{Business}   & 0.19 {\cellcolor{highbids}}& 0.15 & 1E-3 & 0.13 & 0.20 {\cellcolor{highbids}}& 0.21 {\cellcolor{highbids}}& 1.84 & 0.04 \\
\textbf{Computers}  & 0.28 {\cellcolor{higherbids}}& 0.16 {\cellcolor{highbids}}& 0.81     & 0.06 & 0.31 {\cellcolor{highbids}}& 0.33 {\cellcolor{higherbids}}& 4.98 & 0.02  \\
\textbf{Games}      & 0.43 {\cellcolor{higherbids}}& 0.14 {\cellcolor{lowbids}}& 0.08     & 0.09 & 0.23 {\cellcolor{highbids}}& 0.23 {\cellcolor{highbids}}& 0.01 & 0.11  \\
\textbf{Health}     & 0.41 {\cellcolor{higherbids}}& 0.15 & 1.98     & 0.04 & 0.51 {\cellcolor{higherbids}}& 0.38 {\cellcolor{higherbids}}& 0.41 & 0.07  \\
\textbf{Home}       & 0.23 {\cellcolor{highbids}}& 0.14 {\cellcolor{lowbids}}& 4.20     & 0.02 & 0.21 {\cellcolor{highbids}}& 0.28 {\cellcolor{highbids}}& 0.06 & 0.09  \\
\textbf{Kids}       & 0.30 {\cellcolor{higherbids}}& 0.14 {\cellcolor{lowbids}}& 2.76     & 0.03 & 0.23 {\cellcolor{highbids}}& 0.19 {\cellcolor{highbids}}& 5.16 & 0.02  \\
\textbf{News}       & 0.22 {\cellcolor{highbids}}& 0.14 {\cellcolor{lowbids}}& 0.02     & 0.11 & 0.20 {\cellcolor{highbids}}& 0.18 & 0.62 & 0.07  \\
\textbf{Recreation} & 0.31 {\cellcolor{higherbids}}& 0.15 & 0.09     & -- & 0.27 {\cellcolor{highbids}}& 0.21 {\cellcolor{highbids}}& 2.81 & 0.03  \\
\textbf{Reference}  & 0.26 {\cellcolor{higherbids}}& 0.14 {\cellcolor{lowbids}}& 5.86     & 0.01 & 0.22 {\cellcolor{highbids}}& 0.17 {\cellcolor{lowbids}}& 0.42 & 0.07  \\
\textbf{Regional}   & 0.27 {\cellcolor{higherbids}}& 0.16 {\cellcolor{highbids}}& 2.2E-4 & 0.14 & 0.24 {\cellcolor{highbids}}& 0.19 {\cellcolor{highbids}}& 5.86 & 0.01  \\
\textbf{Science}    & 0.36 {\cellcolor{higherbids}}& 0.15 & 1.38     & 0.04 & 0.23 {\cellcolor{highbids}}& 0.15 {\cellcolor{lowbids}}& 0.06 & 0.10 \\
\textbf{Shopping}   & 0.18 {\cellcolor{highbids}}& 0.12 {\cellcolor{lowbids}}& 4.E-3 & 0.14 & 0.22 {\cellcolor{highbids}}& 0.21 {\cellcolor{highbids}}& 4.00 & 0.02  \\
\textbf{Society}    & 0.30 {\cellcolor{higherbids}}& 0.16 {\cellcolor{highbids}}& 1.48     & 0.05 & 0.18 {\cellcolor{highbids}}& 0.16 {\cellcolor{lowbids}}& 2.02 & 0.04  \\
\textbf{Sports}     & 0.21 {\cellcolor{highbids}}& 0.16 {\cellcolor{highbids}}& 1.E-3 & 0.13 & 0.23 {\cellcolor{highbids}}& 0.18 & 2.19 & 0.04  \\
\midrule
\textbf{Control}    & 0.16 & 0.15 & 3.99     & 0.03 & 0.17 & 0.18 & 2.61 & 0.04 \\

\bottomrule
\end{tabular}
}
\vspace{-1em}
\label{table:quantcast-table1}
\end{table}

%% file: tables/new/quantcast-2.tex
\begin{table}[!t]
    \scriptsize
    \centering
    \caption{Ad bidding and cookie syncing under GDPR and CCPA after opt-out (Out) and opt-in (In) with Quantcast. Avg. column represents the mean of all bid value from advertisers who did not bid or appear when we simulated personas but appeared and bid after we opt-out. Out and In under C-Sync. represent number of cookie syncing events after opt-out and opt-in, respectively.}
    \begin{tabular}{l?{0.3mm}c|cc?{0.3mm}c|cc} 
    \toprule
                        & \multicolumn{3}{c?{0.3mm}}{\textbf{GDPR}}                              & \multicolumn{3}{c}{\textbf{CCPA}}                               \\ 
    
                        & \textbf{Out} & \multicolumn{2}{c?{0.3mm}}{\textbf{C-Sync.}} & \textbf{Out} & \multicolumn{2}{l}{\textbf{C-Sync.}}  \\ 
    
    \textbf{Persona}    & \textbf{Avg.}    & \textbf{Out} & \textbf{In}           & \textbf{Avg.}    & \textbf{Out} & \textbf{In}            \\ 
\toprule
\textbf{Adult}      & 0.19 {\cellcolor{highbids}}& 54 & 204 & 0.20 {\cellcolor{highbids}}& 128 & 161  \\
\textbf{Arts}       & 0.16 & 50 & 210 & 0.29 {\cellcolor{highbids}}& 108 & 149  \\
\textbf{Business}   & 0.14 {\cellcolor{lowbids}}& 43 & 154 & 0.19 {\cellcolor{highbids}}& 118 & 179  \\
\textbf{Computers}  & 0.28 {\cellcolor{higherbids}}& 52 & 192 & 0.21 {\cellcolor{highbids}}& 122 & 125  \\
\textbf{Games}      & 0.44 {\cellcolor{higherbids}}& 61 & 180 & 0.24 {\cellcolor{highbids}}& 120 & 184  \\
\textbf{Health}     & 0.17 {\cellcolor{highbids}}& 37 & 141 & 0.52 {\cellcolor{higherbids}}& 184 & 252  \\
\textbf{Home}       & 0.16 & 45 & 135 & 0.18 {\cellcolor{highbids}}& 101 & 210  \\
\textbf{Kids}       & 0.25 {\cellcolor{higherbids}}& 52 & 175 & 0.23 {\cellcolor{highbids}}& 136 & --   \\
\textbf{News}       & 0.20 {\cellcolor{highbids}}& 36 & 163 & 0.17 & 89  & 179  \\
\textbf{Recreation} & 0.18 {\cellcolor{highbids}}& 42 & 176 & 0.16 {\cellcolor{lowbids}}& 137 & 160  \\
\textbf{Reference}  & 0.18 {\cellcolor{highbids}}& 46 & 191 & 0.22 {\cellcolor{highbids}}& 145 & 142  \\
\textbf{Regional}   & 0.13 {\cellcolor{lowbids}}& 49 & 138 & 0.23 {\cellcolor{highbids}}& 163 & 215  \\
\textbf{Science}    & 0.37 {\cellcolor{higherbids}}& 40 & 191 & 0.23 {\cellcolor{highbids}}& 146 & 170  \\
\textbf{Shopping}   & 0.15 {\cellcolor{lowbids}}& 47 & 139 & 0.22 {\cellcolor{highbids}}& 189 & 158  \\
\textbf{Society}    & 0.12 {\cellcolor{lowbids}}& 35 & 126 & 0.16 {\cellcolor{lowbids}}& 109 & 166  \\
\textbf{Sports}     & 0.18 {\cellcolor{highbids}}& 37 & 189 & 0.19 {\cellcolor{highbids}}& 83  & 162  \\
\midrule
\textbf{Control}    & 0.16 & -- & --  & 0.17 & --  & --          \\ 
\midrule
\textbf{Average} & -- & 45 & 169 & -- & 130 & 174 \\

\bottomrule
\end{tabular}
\vspace{-2em}
\label{table:quantcast-table2}
\end{table}

%% file: tables/new/nai-1.tex
\begin{table}[!t]
    \scriptsize
    \centering
    \caption{Ad bidding under GDPR and CCPA after opt-out (Out) and opt-in (In) with NAI. Avg. column represents the mean of all bid value. \textcolor{highbids}{Light red} and \textcolor{lowbids}{Light blue} indicate bid values that are higher and lower than Control's avg., respectively. \textcolor{higherbids}{Dark red} and \textcolor{lowerbids}{Dark blue} indicate bid values that are Control's avg. $\pm$ std., respectively. Column p-val. and Eff. represent p-value and effect size, respectively}
    \adjustbox{max width=\columnwidth}{
    \begin{tabular}{l?{0.3mm}cc|cc?{0.3mm}cc|cc} 
        \toprule
                        & \multicolumn{4}{c?{0.3mm}}{\textbf{GDPR}}                                                  & \multicolumn{4}{c}{\textbf{CCPA}}                                                   \\ 
    
                        & \textbf{Out} & \textbf{In} & \multicolumn{2}{c?{0.3mm}}{\textbf{Stat. Test}} & \textbf{Out} & \textbf{In} & \multicolumn{2}{c}{\textbf{Stat. Test}}  \\ 
    
    \textbf{Persona}    & \textbf{Avg.}    & \textbf{Avg.}   & \textbf{p-val.} & \textbf{Eff.}        & \textbf{Avg.}    & \textbf{Avg.}   & \textbf{p-val.} & \textbf{Eff.}         \\  
    \toprule
\textbf{Adult}      & 1.4E-3 {\cellcolor{higherbids}}& 3.9E-3 {\cellcolor{highbids}}& 6.88 & 4E-3 & 0.10 {\cellcolor{highbids}}& 0.07 {\cellcolor{highbids}}& 4.89 & 0.03      \\
\textbf{Arts}       & 1.3E-3 {\cellcolor{higherbids}}& 1.6E-3 {\cellcolor{lowbids}}& 4.50 & 0.02     & 0.06 {\cellcolor{lowbids}}& 0.07 {\cellcolor{highbids}}& 0.13 & 0.15      \\
\textbf{Business}   & 1.9E-3 {\cellcolor{higherbids}}& 0.01 {\cellcolor{highbids}}    & 2.72 & 0.04     & 0.02 {\cellcolor{lowbids}}& 0.08 {\cellcolor{highbids}}& 0.23 & 0.14      \\
\textbf{Computers}  & 4.1E-3 {\cellcolor{higherbids}}& 0.01 {\cellcolor{highbids}}    & 5.99 & 0.01    & 0.03 {\cellcolor{lowbids}}& 0.03 {\cellcolor{lowbids}}& 2.17 & 0.07     \\
\textbf{Games}      & 4.4E-4 {\cellcolor{higherbids}}& 0.00 {\cellcolor{lowbids}}   & 2.08 & 0.02     & 0.09 {\cellcolor{highbids}}& 0.03 {\cellcolor{lowbids}}& 0.22 & 0.14      \\
\textbf{Health}     & 1E-3 {\cellcolor{higherbids}}& 0.01 {\cellcolor{highbids}}    & 6.94 & 0.01     & 0.16 {\cellcolor{highbids}}& 0.20 {\cellcolor{higherbids}}& 1.90 & 0.08      \\
\textbf{Home}       & 0.00     & 0.00     {\cellcolor{lowbids}}& --   & --       & 0.03 {\cellcolor{lowbids}}& 0.05 {\cellcolor{highbids}}& 6.51 & 0.02     \\
\textbf{Kids}       & 7.3E-4 {\cellcolor{higherbids}}& 4.3E-3 {\cellcolor{highbids}}& 6.48 & 0.01     & 0.07 & 0.08 {\cellcolor{highbids}}& 1.89 & 0.08      \\
\textbf{News}       & 4.1E-4 {\cellcolor{higherbids}}& 1.3E-3 {\cellcolor{lowbids}}& 5.48 & 0.01     & 0.03 {\cellcolor{lowbids}}& 0.06 {\cellcolor{highbids}}& 0.60 & 0.11      \\
\textbf{Recreation} & 10E-4 {\cellcolor{higherbids}}& 2.3E-3 {\cellcolor{lowbids}}& 7.02 & 0.01     & 0.02 {\cellcolor{lowbids}}& 0.08 {\cellcolor{highbids}}& 0.16 & 0.17      \\
\textbf{Reference}  & 2.3E-3 {\cellcolor{higherbids}}& 1.9E-4 {\cellcolor{lowbids}}& 1.58 & 0.05     & 0.02 {\cellcolor{lowbids}}& 0.07 {\cellcolor{highbids}}& 0.05 & 0.18     \\
\textbf{Regional}   & 1.8E-3 {\cellcolor{higherbids}}& 2E-4 {\cellcolor{lowbids}}& 1.86 & 0.04    & 0.22 {\cellcolor{higherbids}}& 0.09 {\cellcolor{highbids}}& 7.89 & 1E-3  \\
\textbf{Science}    & 5E-4 {\cellcolor{higherbids}}& 0.01   {\cellcolor{highbids}}  & 4.19 & 0.02    & 0.02 {\cellcolor{lowbids}}& 0.03 {\cellcolor{lowbids}}& 3.56 & 0.04     \\
\textbf{Shopping}   & 2.5E-3 {\cellcolor{higherbids}}& 1.1E-3 {\cellcolor{lowbids}}& 4.86 & 0.02     & 0.07 & 0.07 {\cellcolor{highbids}}& 7.94 & 1E-3  \\
\textbf{Society}    & 0.00     & 0.01  {\cellcolor{highbids}}   & 1.42 & 0.04     & 0.05 {\cellcolor{lowbids}}& 0.11 {\cellcolor{highbids}}& 0.23 & 0.15      \\
\textbf{Sports}     & 0.01     {\cellcolor{higherbids}}& 0.01   {\cellcolor{highbids}}  & 3.44 & 0.03    & 0.09 {\cellcolor{highbids}}& 0.14 {\cellcolor{highbids}}& 0.24 & 0.15      \\
\midrule
\textbf{Control}    & 0.00     & 3.8E-3 & 1.05 & 0.05     & 0.07 & 0.05 & 2.02 & 0.07             \\
\bottomrule
\end{tabular}
}
\vspace{-1em}
\label{table:nai-table1}
\end{table}

%% file: tables/new/nai-2.tex
\begin{table}[!t]
    \scriptsize
    \centering
    \caption{Ad bidding and cookie syncing under GDPR and CCPA after opt-out (Out) and opt-in (In) with NAI. Avg. column represents the mean of all bid value from advertisers who did not bid or appear when we simulated personas but appeared and bid after we opt-out. Out and In under C-Sync. represent number of cookie syncing events after opt-out and opt-in, respectively.}
    \begin{tabular}{l?{0.3mm}c|cc?{0.3mm}c|cc} 
    \toprule
                        & \multicolumn{3}{c?{0.3mm}}{\textbf{GDPR}}                              & \multicolumn{3}{c}{\textbf{CCPA}}                               \\ 
    
                        & \textbf{Out} & \multicolumn{2}{c?{0.3mm}}{\textbf{C-Sync.}} & \textbf{Out} & \multicolumn{2}{l}{\textbf{C-Sync.}}  \\ 
    
    \textbf{Persona}    & \textbf{Avg.}    & \textbf{Out} & \textbf{In}           & \textbf{Avg.}    & \textbf{Out} & \textbf{In}            \\ 
\toprule
\textbf{Adult}      & 1.4E-3 {\cellcolor{higherbids}}& 7  & 0  & 2.3E-3 {\cellcolor{lowerbids}}& 19 & 23  \\
\textbf{Arts}       & 0.00     & 13 & 0  & 1.3E-3 {\cellcolor{lowerbids}}& 5  & 86  \\
\textbf{Business}   & 0.00     & 22 & 21 & 0.00     {\cellcolor{lowerbids}}& 18 & 27  \\
\textbf{Computers}  & 4.1E-3 {\cellcolor{higherbids}}& 11 & 19 & 0.00     {\cellcolor{lowerbids}}& 20 & 33  \\
\textbf{Games}      & 0.00     & 15 & 0  & 0.00     {\cellcolor{lowerbids}}& 10 & 20  \\
\textbf{Health}     & 1E-3 {\cellcolor{higherbids}}& 5  & 5  & 0.01     {\cellcolor{lowerbids}}& 57 & 71  \\
\textbf{Home}       & 0.00     & 0  & 0  & 0.00     {\cellcolor{lowerbids}}& 18 & 31  \\
\textbf{Kids}       & 0.00     & 7  & 14 & 0.00     {\cellcolor{lowerbids}}& 38 & 67  \\
\textbf{News}       & 0.00     & 5  & 2  & 1.9E-3 {\cellcolor{lowerbids}}& 16 & 18  \\
\textbf{Recreation} & 0.00     & 9  & 6  & 2.3E-3 {\cellcolor{lowerbids}}& 19 & 49  \\
\textbf{Reference}  & 1.4E-3 {\cellcolor{higherbids}}& 20 & 13 & 4E-3 {\cellcolor{lowerbids}}& 25 & 27  \\
\textbf{Regional}   & 0.00     & 13 & 14 & 0.03     {\cellcolor{lowerbids}}& 22 & 49  \\
\textbf{Science}    & 0.00     & 5  & 0  & 0.00     {\cellcolor{lowerbids}}& 54 & 22  \\
\textbf{Shopping}   & 0.00     & 23 & 18 & 0.00     {\cellcolor{lowerbids}}& 27 & 74  \\
\textbf{Society}    & 0.00     & 0  & 0  & 0.00     {\cellcolor{lowerbids}}& 17 & 93  \\
\textbf{Sports}     & 0.00     & 1  & 0  & 3.5E-3 {\cellcolor{lowerbids}}& 46 & 78  \\
\midrule
\textbf{Control}    & 0.00     & -- & -- & 0.07     & -- & --     \\ 
\midrule
\textbf{Average} & -- & 10 & 7 & -- & 26 & 48 \\

\bottomrule
\end{tabular}
\vspace{-1em}
\label{table:nai-table2}
\end{table}

%% file: docs/pre-opt-out_syncing_gdpr.tex
\begin{table}
    \scriptsize
    \centering
    \caption{Cookie syncing events by advertisers under GDPR in Germany after Pre-opt-out. Evt. column represents the number of cookie syncing events from advertisers, respectively.}

\begin{tabular}{l?{0.3mm}l?{0.3mm}l?{0.3mm}l?{0.3mm}l|l} 
    \toprule
                    & \multicolumn{5}{c}{\textbf{GDPR}}                                                                                                                                                                                                                                                                                                                                              \\ 

     & \multicolumn{1}{c?{0.3mm}}{\textbf{Cookiebot}} & \multicolumn{1}{c?{0.3mm}}{\textbf{Didomi}}& \multicolumn{1}{c?{0.3mm}}{\textbf{Onetrust}}&
     \multicolumn{1}{c?{0.3mm}}{\textbf{Quantcast}}& \multicolumn{1}{c}{\textbf{NAI}}      \\  

    \textbf{Persona}      & \multicolumn{1}{c?{0.3mm}}{\textbf{Evt.}}  & \multicolumn{1}{c?{0.3mm}}{\textbf{Evt.}}  & \multicolumn{1}{c?{0.3mm}}{\textbf{Evt.}}  & \multicolumn{1}{c?{0.3mm}}{\textbf{Evt.}}  & \multicolumn{1}{c}{\textbf{Evt.}} \\ 
    \toprule
    \textbf{Adult}      & -- & 6  & 19 & 27 & 16  \\
    \textbf{Arts}       & -- & -- & 29 & 24 & 17  \\
    \textbf{Business}   & -- & -- & 46 & 26 & 11  \\
    \textbf{Computers}  & -- & 9  & 20 & 37 & 29  \\
    \textbf{Games}      & -- & 8  & 30 & 20 & 15  \\
    \textbf{Health}     & -- & 13 & 21 & 12 & 5   \\
    \textbf{Home}       & -- & -- & 46 & 27 & 6   \\
    \textbf{Kids}       & -- & 15 & 34 & 41 & 13  \\
    \textbf{News}       & -- & -- & 21 & 25 & 13  \\
    \textbf{Recreation} & -- & -- & 4  & 23 & 5   \\
    \textbf{Reference}  & -- & -- & 34 & 31 & 22  \\
    \textbf{Regional}   & -- & -- & 34 & 37 & 21  \\
    \textbf{Science}    & -- & -- & 16 & 16 & 10  \\
    \textbf{Shopping}   & -- & 7  & 41 & 25 & 16  \\
    \textbf{Society}    & -- & -- & 54 & 21 & 7   \\
    \textbf{Sports}     & -- & -- & 41 & 27 & 1    \\
\midrule
\textbf{Average} & -- & 4 & 31 & 26 & 13         \\

\bottomrule
\end{tabular}
\label{table:GDPR-syncing-pre-opt-out}
\end{table}

%% file: docs/pre-opt-out_syncing_ccpa.tex
\begin{table}
    \scriptsize
    \centering
    \caption{Cookie syncing events by advertisers under CCPA in California after Pre-opt-out. Evt. column represents the number of cookie syncing events from advertisers, respectively.}

\begin{tabular}{l?{0.3mm}l?{0.3mm}l?{0.3mm}l?{0.3mm}l|l} 
    \toprule
                    & \multicolumn{5}{c}{\textbf{CCPA}}                                                                                                                                                                                                                                                                                                                                              \\ 

     & \multicolumn{1}{c?{0.3mm}}{\textbf{Cookiebot}} & \multicolumn{1}{c?{0.3mm}}{\textbf{Didomi}}& \multicolumn{1}{c?{0.3mm}}{\textbf{Onetrust}}&
     \multicolumn{1}{c?{0.3mm}}{\textbf{Quantcast}}& \multicolumn{1}{c}{\textbf{NAI}}      \\  

    \textbf{Persona}      & \multicolumn{1}{c?{0.3mm}}{\textbf{Evt.}}  & \multicolumn{1}{c?{0.3mm}}{\textbf{Evt.}}  & \multicolumn{1}{c?{0.3mm}}{\textbf{Evt.}}  & \multicolumn{1}{c?{0.3mm}}{\textbf{Evt.}}  & \multicolumn{1}{c}{\textbf{Evt.}} \\ 
    \toprule
\textbf{Adult}      & 17 & 39 & 107 & 125 & 31  \\
\textbf{Arts}       & 27 & 35 & 90  & 90  & 24  \\
\textbf{Business}   & 23 & 36 & 77  & 94  & 27  \\
\textbf{Computers}  & 16 & 34 & 79  & 73  & 32  \\
\textbf{Games}      & 15 & 45 & 109 & 96  & 25  \\
\textbf{Health}     & 13 & 41 & 134 & 89  & 71  \\
\textbf{Home}       & -- & 37 & 46  & 104 & 29  \\
\textbf{Kids}       & 26 & 57 & 111 & 90  & 59  \\
\textbf{News}       & 37 & 34 & 100 & 66  & 31  \\
\textbf{Recreation} & 21 & 57 & 61  & 83  & 45  \\
\textbf{Reference}  & -- & 35 & 64  & 95  & 30  \\
\textbf{Regional}   & 18 & 32 & 106 & 137 & 43  \\
\textbf{Science}    & -- & 46 & 71  & 86  & 49  \\
\textbf{Shopping}   & 13 & 37 & 62  & 98  & 50  \\
\textbf{Society}    & 47 & 54 & 94  & 91  & 56  \\
\textbf{Sports}     & 18 & 34 & 112 & 107 & 31   \\
\midrule
\textbf{Average} & 18 & 41 & 89 & 95 & 40         \\

\bottomrule
\end{tabular}
\label{table:CCPA-syncing-pre-opt-out}
\end{table}

%% file: docs/pre-opt-out.tex
\begin{table*}
    \scriptsize
    \centering
    \caption{Ad bidding under GDPR in Germany. Avg. column represents the mean of all bid value. Std. represents the standard deviation of all bid value. Pre-opt-out represents the bids when we opt-out the usage of user data before we stimulate the persona, respectively.   \textcolor{highbids}{Light red} and \textcolor{lowbids}{Light blue} indicate bid values that are higher and lower than Control's avg., respectively. \textcolor{higherbids}{Dark red} and \textcolor{lowerbids}{Dark blue} indicate bid values that are Control's avg. $\pm$ std., respectively. Column p-val. and Eff. represent p-value and effect size, respectively}

\begin{tabular}{l?{0.3mm}ll?{0.3mm}ll?{0.3mm}ll?{0.3mm}ll|ll} 
    \toprule
                    & \multicolumn{10}{c}{\textbf{GDPR}}                                                                                                                                                                                                                                                                                                                                              \\ 

     & \multicolumn{2}{c?{0.3mm}}{\textbf{Cookiebot}} & \multicolumn{2}{c?{0.3mm}}{\textbf{Didomi}}& \multicolumn{2}{c?{0.3mm}}{\textbf{Onetrust}}&
     \multicolumn{2}{c?{0.3mm}}{\textbf{Quantcast}}& \multicolumn{2}{c}{\textbf{NAI}}      \\  

    \textbf{Persona}     & \multicolumn{1}{c}{\textbf{Avg.}} & \multicolumn{1}{c?{0.3mm}}{\textbf{Std.}} & \multicolumn{1}{c}{\textbf{Avg.}} & \multicolumn{1}{c?{0.3mm}}{\textbf{Std.}} & \multicolumn{1}{c}{\textbf{Avg.}} & \multicolumn{1}{c?{0.3mm}}{\textbf{Std.}} & \multicolumn{1}{c}{\textbf{Avg.}} & \multicolumn{1}{c?{0.3mm}}{\textbf{Std.}} & \multicolumn{1}{c}{\textbf{Avg.}} & \multicolumn{1}{c}{\textbf{Std.}} \\ 
    \toprule
\textbf{Adult}      & 0.37   {\cellcolor{highbids}}                                                                                                        & 0.20                                           & 0.02        {\cellcolor{higherbids}}                                                                                                                                                                       & 0.00                                            & 0.08                                          {\cellcolor{highbids}}                                                                 & 0.04                                           & 0.58                                           {\cellcolor{higherbids}}                                                                                                                                    & 1.99                                           & 3.3E-4 {\cellcolor{lowbids}}                                                                 & 1.8E-3                                           \\
\textbf{Arts}       & 0.37       {\cellcolor{highbids}}                                                                                                    & 0.20                                           & 0.11         {\cellcolor{higherbids}}                                                                                                                                                                      & 0.08                                            & 0.16                                           {\cellcolor{higherbids}}                                                                                                                                    & 0.22                                           & 0.16                                          {\cellcolor{highbids}}                                                                 & 0.08                                           & 4.6E-3 {\cellcolor{higherbids}}                                                                                                                                    & 0.03                                           \\
\textbf{Business}   & 0.37    {\cellcolor{highbids}}                                                                                                       & 0.20                                           & --                                             & --                                              & 0.09                                           {\cellcolor{highbids}}                                                                 & 0.05                                           & 0.58                                          {\cellcolor{higherbids}}                                                                                                                                    & 2.05                                           & 0.04                                           {\cellcolor{higherbids}}                                                                                                                                    & 0.27                                           \\
\textbf{Computers}  & 0.38        {\cellcolor{highbids}}                                                                                                   & 0.20                                           & 0.01       & 0.00                                            & 0.05                                          {\cellcolor{lowbids}}                                                               & 0.03                                           & 0.16                                          {\cellcolor{highbids}}                                                                 & 0.07                                           & 0.01                                           {\cellcolor{higherbids}}                                                                                                                                    & 0.06                                           \\
\textbf{Games}      & 0.37     {\cellcolor{highbids}}                                                                                                      & 0.20                                           & 0.01       & 0.00                                            & 0.07                                                                                                       & 0.05                                           & 0.16                                          {\cellcolor{highbids}}                                                                 & 0.07                                           & 6E-4 {\cellcolor{highbids}}                                                                 & 2.9E-3                                           \\
\textbf{Health}     & 0.37         {\cellcolor{highbids}}                                                                                                  & 0.20                                           & 0.13     {\cellcolor{higherbids}}              & 0.12                                            & 0.11                                          {\cellcolor{highbids}}                                                                 & 0.05                                           & 0.16                                          {\cellcolor{highbids}}                                                                 & 0.07                                           & 0.01                                           {\cellcolor{higherbids}}                                                                                                                                    & 0.08                                           \\
\textbf{Home}       & 0.37    {\cellcolor{highbids}}                                                                                                       & 0.20                                           & --                                             & --                                                                                        & 0.08                                          {\cellcolor{highbids}}                                                                 & 0.05                                           & 0.23      {\cellcolor{higherbids}}                                                                                                                                                                         & 0.93                                           & 1.7E-3 {\cellcolor{highbids}}                                                                 & 0.01 \\
\textbf{Kids}       & 0.37      {\cellcolor{highbids}}                                                                                                     & 0.20                                           & 0.01                                                                                                        & 0.00                                            & 0.13                                          {\cellcolor{highbids}}                                                                 & 0.04                                           & 0.60                                           {\cellcolor{higherbids}}                                                                                                                                    & 2.11                                           & 5E-3                                           {\cellcolor{higherbids}}                                                                                                                                    & 0.03                                           \\
\textbf{News}       & 0.37           {\cellcolor{highbids}}                                                                                                & 0.20                                           & 0.06      {\cellcolor{higherbids}}                                                                                                                                                                         & 0.04                                            & 0.08                                          {\cellcolor{highbids}}                                                                 & 0.02                                           & 0.55                                           {\cellcolor{higherbids}}                                                                                                                                    & 1.97                                           & 3E-4 {\cellcolor{lowbids}}                                                                                                   & 1.6E-3                                           \\
\textbf{Recreation} & 0.37    {\cellcolor{highbids}}                                                                                                       & 0.20                                           & --                                             & --                                              & 0.02                                          {\cellcolor{lowbids}}                                                               & 0.00                                           & 0.64                                           {\cellcolor{higherbids}}                                                                                                                                    & 2.21                                           & 3.6E-4 {\cellcolor{lowbids}}                                                               & 2.7E-3                                           \\
\textbf{Reference}  & 0.37        {\cellcolor{highbids}}                                                                                                   & 0.20                                           & --                                             & --                                              & 0.06                                          {\cellcolor{lowbids}}                                                               & 0.06                                           & 0.16                                          {\cellcolor{highbids}}                                                                 & 0.07                                           & 1.4E-3                                          {\cellcolor{highbids}}                                                                 & 0.01                                           \\
\textbf{Regional}   & 0.37        {\cellcolor{highbids}}                                                                                                   & 0.20                                           & --                                             & --                                              & 0.04                                          {\cellcolor{lowbids}}                                                               & 0.05                                           & 0.15                                          {\cellcolor{highbids}}                                                                 & 0.07                                           & 0.01                                           {\cellcolor{higherbids}}                                                                                                                                    & 0.03                                           \\
\textbf{Science}    & 0.36        {\cellcolor{highbids}}                                                                                                   & 0.21                                           & --                                             & --                                              & 0.12                                           {\cellcolor{highbids}}                                                                 & 0.01                                           & 0.55                                          {\cellcolor{higherbids}}                                                                                                                                    & 2.01                                           & 0.01             {\cellcolor{higherbids}}                                                                                                                                                                  & 0.07                                           \\
\textbf{Shopping}   & 0.37        {\cellcolor{highbids}}                                                                                                   & 0.20                                           & 0.02                                           {\cellcolor{higherbids}}                                                                                                                                    & 0.00                                            & 0.17                                           {\cellcolor{higherbids}}                                                                                                                                    & 0.10                                           & 0.60                                           {\cellcolor{higherbids}}                                                                                                                                    & 2.17                                           & 0.01                                           {\cellcolor{higherbids}}                                                                                                                                    & 0.09                                           \\
\textbf{Society}    & 0.38           {\cellcolor{highbids}}                                                                                                & 0.19                                           & --                                             & --                                              & 0.14                                          {\cellcolor{highbids}}                                                                 & 0.19                                           & 0.16                                          {\cellcolor{highbids}}                                                                 & 0.07                                           & 0.01              {\cellcolor{higherbids}}                                                                                                                                                                 & 0.08                                           \\
\textbf{Sports}     & 0.37         {\cellcolor{highbids}}                                                                                                  & 0.20                                           & --                                             & --                                              & 0.07                                                                                                           & 0.10                                           & 0.16                                          {\cellcolor{highbids}}                                                                 & 0.07                                           & 0.01                   {\cellcolor{higherbids}}                                                                                                                                                            & 0.08                                           \\
\midrule
\textbf{Control}    & 0.31                                           & 0.23                                           & 0.01                                           & 3.6E-4  & 0.07                                           & 0.08                                           & 0.14                                           & 0.08                                          & 3.7E-4                                           & 2.7E-3                                           \\

\bottomrule
\end{tabular}
\label{table:GDPR-bidding-mean-pre-opt-out}
\end{table*}

\begin{table*}
    \scriptsize
    \centering
    \caption{Ad bidding under CCPA in California. Avg. column represents the mean of all bid value. Std. represents the standard deviation of all bid value. Pre-opt-out represents the bids when we opt-out the usage of user data before we stimulate the persona, respectively.   \textcolor{highbids}{Light red} and \textcolor{lowbids}{Light blue} indicate bid values that are higher and lower than Control's avg., respectively. \textcolor{higherbids}{Dark red} and \textcolor{lowerbids}{Dark blue} indicate bid values that are Control's avg. $\pm$ std., respectively. Column p-val. and Eff. represent p-value and effect size, respectively}

\begin{tabular}{l?{0.3mm}ll?{0.3mm}ll?{0.3mm}ll?{0.3mm}ll|ll} 
    \toprule
                    & \multicolumn{10}{c}{\textbf{CCPA}}                                                                                                                                                                                                                                                                                                                                              \\ 

     & \multicolumn{2}{c?{0.3mm}}{\textbf{Cookiebot}} & \multicolumn{2}{c?{0.3mm}}{\textbf{Didomi}}& \multicolumn{2}{c?{0.3mm}}{\textbf{Onetrust}}&
     \multicolumn{2}{c?{0.3mm}}{\textbf{Quantcast}}& \multicolumn{2}{c}{\textbf{NAI}}      \\  

    \textbf{Persona}     & \multicolumn{1}{c}{\textbf{Avg.}} & \multicolumn{1}{c?{0.3mm}}{\textbf{Std.}} & \multicolumn{1}{c}{\textbf{Avg.}} & \multicolumn{1}{c?{0.3mm}}{\textbf{Std.}} & \multicolumn{1}{c}{\textbf{Avg.}} & \multicolumn{1}{c?{0.3mm}}{\textbf{Std.}} & \multicolumn{1}{c}{\textbf{Avg.}} & \multicolumn{1}{c?{0.3mm}}{\textbf{Std.}} & \multicolumn{1}{c}{\textbf{Avg.}} & \multicolumn{1}{c}{\textbf{Std.}} \\ 
    \toprule
\textbf{Adult}      & 0.38 {\cellcolor{highbids}}                                                                                              & 0.21                               & 0.13 {\cellcolor{lowbids}}                               & 0.08                                & 0.38   {\cellcolor{lowbids}}                             & 0.68                               & 0.18     {\cellcolor{lowbids}}                           & 0.26                               & 0.07                                                                                            & 0.12                               \\ 

\textbf{Arts}       & 0.37   {\cellcolor{highbids}}                                                                                            & 0.21                               & 0.11  {\cellcolor{lowbids}}                              & 0.10                                & 1.05     {\cellcolor{highbids}}                                                                                          & 1.36                               & 0.22 {\cellcolor{highbids}}                                                                                              & 0.65                               & 0.05 {\cellcolor{lowbids}}                               & 0.13                               \\ 

\textbf{Business}   & 0.36  {\cellcolor{highbids}}                                                                                             & 0.22                               & 0.10  {\cellcolor{lowbids}}                              & 0.07                                & 1.26  {\cellcolor{highbids}}                                                                                             & 1.49                               & 0.17 {\cellcolor{lowbids}}                               & 0.11                               & 0.05  {\cellcolor{lowbids}}                              & 0.12                               \\ 

\textbf{Computers}  & 0.36   {\cellcolor{highbids}}                                                                                            & 0.20                               & 0.11  {\cellcolor{lowbids}}                              & 0.10                                & 1.12  {\cellcolor{highbids}}                                                                                             & 1.33                               & 0.32  {\cellcolor{highbids}}                                                                                             & 1.66                               & 0.05    {\cellcolor{lowbids}}                            & 0.15                               \\ 

\textbf{Games}      & 0.37   {\cellcolor{highbids}}                                                                                            & 0.20                               & 0.15                               & 0.11                                & 0.73    {\cellcolor{lowbids}}                            & 1.07                               & 0.19    {\cellcolor{lowbids}}                            & 0.19                               & 0.04   {\cellcolor{lowbids}}                             & 0.11                               \\ 

\textbf{Health}     & 0.38   {\cellcolor{highbids}}                                                                                            & 0.20                               & 0.09    {\cellcolor{lowbids}}                            & 0.08                                & 0.61  {\cellcolor{lowbids}}                              & 0.86                               & 0.44 {\cellcolor{highbids}}                                                                                              & 0.86                               & 0.52    {\cellcolor{higherbids}}                                                                                                                                                              & 1.25                               \\ 

\textbf{Home}       & 0.37   {\cellcolor{highbids}}                                                                                            & 0.20                               & 0.16  {\cellcolor{highbids}}                                                                                             & 0.33                                & 0.80     {\cellcolor{highbids}}                                                                                          & 1.12                               & 0.18   {\cellcolor{lowbids}}                             & 0.17                               & 0.06    {\cellcolor{lowbids}}                            & 0.14                               \\ 

\textbf{Kids}       & 0.38   {\cellcolor{highbids}}                                                                                            & 0.19                               & 0.06  {\cellcolor{lowbids}}                              & 0.06                                & 0.59    {\cellcolor{lowbids}}                            & 0.96                               & 0.17  {\cellcolor{lowbids}}                              & 0.12                               & 0.10   {\cellcolor{highbids}}                                                                                            & 0.19                               \\ 

\textbf{News}       & 0.37  {\cellcolor{highbids}}                                                                                             & 0.19                               & 0.12  {\cellcolor{lowbids}}                              & 0.09                                & 0.66  {\cellcolor{lowbids}}                              & 1.10                               & 0.17   {\cellcolor{lowbids}}                             & 0.10                               & 0.04  {\cellcolor{lowbids}}                              & 0.11                               \\ 

\textbf{Recreation} & 0.36   {\cellcolor{highbids}}                                                                                            & 0.20                               & 0.10  {\cellcolor{lowbids}}                              & 0.09                                & 0.50   {\cellcolor{lowbids}}                             & 0.82                               & 0.17   {\cellcolor{lowbids}}                             & 0.17                               & 0.17 {\cellcolor{highbids}}                                                                                              & 0.60                               \\ 

\textbf{Reference}  & 0.37  {\cellcolor{highbids}}                                                                                             & 0.20                               & 0.09    {\cellcolor{lowbids}}                            & 0.09                                & 0.67   {\cellcolor{lowbids}}                             & 1.10                               & 0.16   {\cellcolor{lowbids}}                             & 0.11                               & 0.02    {\cellcolor{lowbids}}                            & 0.04                               \\ 

\textbf{Regional}   & 0.37   {\cellcolor{highbids}}                                                                                            & 0.22                               & 0.10   {\cellcolor{lowbids}}                             & 0.07                                & 0.50   {\cellcolor{lowbids}}                             & 0.78                               & 0.22    {\cellcolor{highbids}}                                                                                           & 1.04                               & 0.06   {\cellcolor{lowbids}}                             & 0.14                               \\ 

\textbf{Science}    & 0.38    {\cellcolor{highbids}}                                                                                           & 0.20                               & 0.13   {\cellcolor{lowbids}}                             & 0.15                                & 0.81    {\cellcolor{highbids}}                                                                                           & 1.17                               & 0.20      {\cellcolor{lowbids}}                          & 0.58                               & 0.05    {\cellcolor{lowbids}}                            & 0.12                               \\ 

\textbf{Shopping}   & 0.36  {\cellcolor{highbids}}                                                                                             & 0.20                               & 0.12     {\cellcolor{lowbids}}                           & 0.10                                & 0.86   {\cellcolor{highbids}}                                                                                            & 1.28                               & 0.21                               & 0.63                               & 0.05  {\cellcolor{lowbids}}                              & 0.11                               \\ 

\textbf{Society}    & 0.35                              & 0.21                               & 0.13    {\cellcolor{lowbids}}                            & 0.11                                & 0.31    {\cellcolor{lowbids}}                            & 0.37                               & 0.20    {\cellcolor{lowbids}}                            & 0.18                               & 0.03  {\cellcolor{lowbids}}                              & 0.07                               \\ 

\textbf{Sports}     & 0.37    {\cellcolor{highbids}}                                                                                           & 0.22                               & 0.08     {\cellcolor{lowbids}}                           & 0.07                                & 0.75  {\cellcolor{highbids}}                                                                                             & 1.04                               & 0.16    {\cellcolor{lowbids}}                            & 0.10                               & 0.10   {\cellcolor{highbids}}                                                                                            & 0.19                               \\ 
\midrule
\textbf{Control}    & 0.35                               & 0.21                               & 0.15                               & 0.14                                & 0.74                               & 1.10                               & 0.21                               & 0.82                               & 0.07                               & 0.17                               \\
\bottomrule
\end{tabular}
\label{table:CCPA-bidding-mean-pre-opt-out}
\end{table*}